\documentclass[reprint,amsmath,amssymb,aps,nofootinbib,prd]{revtex4-2}

\usepackage{graphicx}
\usepackage{dcolumn}
\usepackage{bm}
\usepackage[dvipsnames]{xcolor}
\usepackage[normalem]{ulem}
\usepackage{natbib}
\usepackage{enumerate}
\usepackage{longtable}

\newcommand{\aap}{Astron.\ Astrophys.}
\newcommand{\mnras}{Mon.\ Not.\ R.\ Astron.\ Soc.}
\newcommand{\physrep}{Phys.\ Rep.}
\newcommand{\apjl}{Astrophys.\ J.\ Lett.}

\newcommand{\araa}{Ann.\ Rev.\ Astron.\ Astrophys.}

\newcommand{\jcap}{J. Cosmology \& Astropaticles.}
\newcommand{\nphysa}{Nuc. Phys. A}

\newcommand{\Msun}{$M_\odot$}
\newcommand{\Mmax}{M_\text{\textsc{tov}}}
\newcommand{\Rmax}{R_\text{\textsc{tov}}}
\newcommand{\csmax}{c_\text{s\textsc{tov}}}
\newcommand{\rhomax}{\rho_\text{\textsc{tov}}}
\newcommand{\Pmax}{P_\text{\textsc{tov}}}
\newcommand{\Rhalf}{R_\text{1/2}}
\newcommand{\ratio}{r_\text{1/2}}

\usepackage{moreverb} 
\immediate\write18{texcount -inc -incbib 
-sum DDOMasses_05.tex > /tmp/wordcount.tex}

\immediate\write18{texcount -char -freq
 DDOMasses_05.tex > /tmp/charcount.tex}

\begin{document}
 

\title[
NS $M-R$ and EoS are three-parametric]{
{A three-parameter characterization of neutron stars' mass$-$radius relation and equation of state}}%

\author{Dmitry D. Ofengeim$^1$}\email{Corresponding Author: ddofengeim@gmail.com}
\author{Peter S. Shternin$^2$}
\author{Tsvi Piran$^1$}%

\affiliation{$^1$Racah Institute of Physics, The Hebrew University, Jerusalem 91904, Israel}
\affiliation{$^2$Ioffe Institute, Politekhnicheskaya 26, St. Petersburg, 194021, Russia}

\date{\today}

\begin{abstract}

Numerous models of neutron star (NS) equation of state (EoS) exist based on different superdense-matter physics approaches. Nevertheless, some NS properties show universal (EoS-independent) relations. Here, we propose a novel class of such universalities. Despite different physics inputs, a wide class of realistic nucleonic, hyperonic, and hybrid EoS models can be accurately described using only three parameters. For a given EoS, these are the mass and radius of the maximum-mass NS (or pressure and density in its center) and the radius of a half-maximum-mass star. With such a parametrization, we build universal analytic expressions for mass-radius and pressure-density relations. They form a semi-analytic mapping from the mass-radius relation to the EoS in NS cores (the so-called inverse Oppenheimer-Volkoff mapping). This mapping simplifies the process of inferring the EoS from observations of NS masses and radii. Applying it to current NS observations we set new limits on the high-density end of the EoS.

\end{abstract}

\maketitle

\section{Introduction}
\label{sec:intro}

One of the most intriguing problems of modern physics is to unveil the mystery of the neutron star (NS) equation of state (EoS, pressure $P$ $-$ density $\rho$ relation). These extreme objects are denser than atomic nuclei, giving us a unique opportunity to study the superdense  matter through astrophysical observations~\cite{DegenaarSuleimanov2018}. One of the few ways to study it is to measure NS masses $M$ and radii $R$. As the star is in hydrostatic equilibrium, its EoS determines the $M-R$ relation. Inverting this relation, we can find the EoS. In the simplest case of negligible rotation, NS hydrostatics is described by the Tolman-Oppenheimer-Volkoff (TOV) equations \cite{Tolman1939,OppVol1939}. 

The mapping from the $P-\rho$ relation to the $M-R$ one is called the Oppenheimer-Volkoff (OV) mapping \cite{Lind1992}. It is a bijection, i.e., there exists inverse OV mapping (IOVM). Thus, unambiguous finding of EoS from $M$ and $R$ observations is, in principle, possible. Additionally, a maximum-mass NS (MMNS) exists regardless of the EoS model used. The MMNS characteristics, such as mass $\Mmax$, radius $\Rmax$, pressure $\Pmax$, density $\rhomax$ in the center, etc., are specific for a given EoS. While $\Mmax$ is not an upper limit for NS mass (a fast-spinning supermassive NS can have $M>\Mmax$), $\Pmax$ and $\rhomax$ for the true EoS are apparently the highest pressure and density can happen in the stationary object of the current Universe.

While the true NS EoS is unique, there are hundreds of various theoretical EoS models present in the literature \cite{BurgioFantina2018}. Each yields its own $M-R$, $M$ --- central $\rho$ and $P$ relations and, in particular, its own $\Mmax$, $\Rmax$, $\rhomax$, and $\Pmax$. Nevertheless, it has been known for a long time that some relations between various NS properties are universal across the EoS manifold (e.g. \cite{AnderssonKokkotasMNRAS1998,LattPrakApJ01,BejerHaenselAA2002,YagiYounes2017,Ofengeim2020,SotaniKumarPRD2021,SaesMendesPRD2022,KonstantinouMorsinkApJ2022,Cai2023ApJ}). Some of these relations, like the I-Love-Q relation \cite{YagiYounesSci2013,YagiYunesCQG2016}, originate from a unifying ability of relativistic gravity \cite{YagiYounes2017} and are very accurate for almost all existing models of EoS. Others are less precise and exist due to some common features of a subclass of EoSs, which are considered to be ``realistic'' (see, e.g., \cite{Legred+PRD2024}).

In this work, we present a new class of universal relations describing the fundamental NS features---their $M-R$ and $P-\rho$ relations (Sec.~\ref{sec:univ}). They are based on the idea that $\Mmax$, $\Rmax$, $\Pmax$, and $\rhomax$ set suitable universal scales for masses, radii, pressures, and densities of NSs. Using these relations we build an explicit semi-analytic IOVM (Sec.~\ref{sec:IOV}). Applying this relation to measured NS masses and radii, including the recent observations of PSR~J0437$-$4715~\cite{Choudhury2024}, we set new limits on the high-density EoS. The precision of the obtained constraints is higher than that of other results. Other possible applications of the universal relations are also considered (Sec.~\ref{sec:concl}).

\section{Universal relations}
\label{sec:univ}
The universal relations presented below were verified for a collection of $169$ EoSs based on various underlying physics-motivated models of low-energy hardronic interactions. Among those, $111$ pass the criterion $\Mmax\gtrsim2\,$\Msun\ (based on astrophysical measurements of NS masses \cite{0348paper2013,0740paper2021,Kandel2023ApJ}; see below). For $17$ models the maximal speed of sound (for $\rho\leqslant \rhomax$) is higher than the speed of light $c$. However, we keep all these models to obtain a more comprehensive cover of the $M-R$ and $P-\rho$ planes. Removing these EoS models doesn't change significantly the fits we present below. 
In this collection, $100$ models exhibit nucleonic composition of the NS core, $36$ have additional heavier baryons (hyperons and, in a few cases, $\Delta$-isobars), and $33$ have hybrid composition, in which  a transition to deconfined quark matter occurs at some density. See Table~\ref{tab:app:zooList} in Appendix~\ref{sec:app:EoS} for details.

\begin{figure}
    \centering
    \includegraphics[width=\columnwidth]{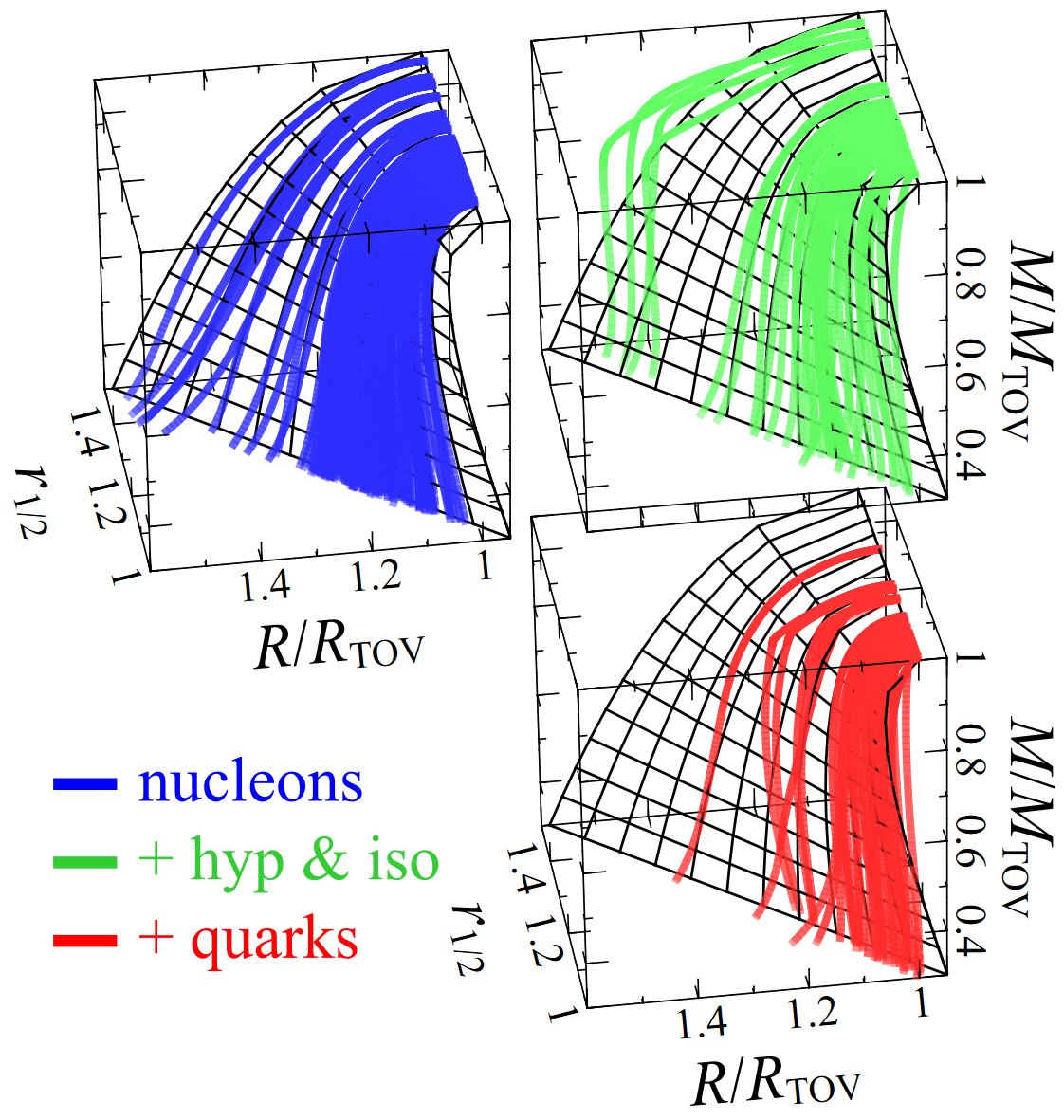}
    \caption{A geometrical view of the EoS manifold. The curves correspond to different EoS models:  nucleonic (blue), with hyperons and $\Delta$-isobars (green), and hybrid (red). Except for a few 
    outliers that correspond to obsolete EoSs with $\Mmax<2\,$\Msun, all EoSs lie on a single surface approximated by Eq.~(\ref{eq:MRfit}) (black grid).}
    \label{fig:3D}
\end{figure}
\begin{figure}
    \centering
    \includegraphics[width=\columnwidth]{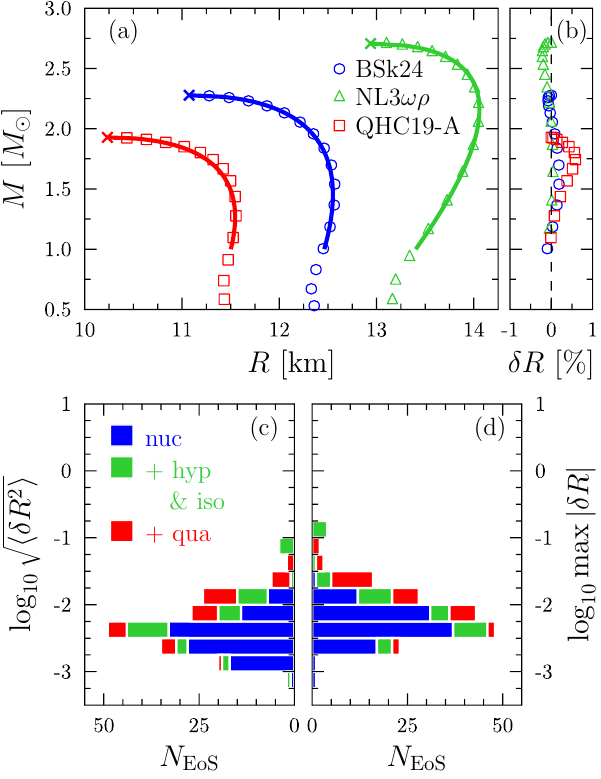}
    \caption{Demonstration of the approximations for $M-R$ curves [Eq.~(\ref{eq:MRfit}), panel (a)] for selected nucleonic (BSk24), hyperonic (NL3$\omega\rho$), and hybrid (QHC19-A) EoS models. Symbols stand for exact values for given models, lines show the approximations with $\Mmax$, $\Rmax$, $\rhomax$, $\Pmax$, and $\ratio$ parameters taken from the corresponding EoSs. Crosses mark the $(\Rmax,\Mmax)$ and $(\rhomax,\Pmax)$ points. Panel (b) shows relative errors of the approximations for these three EoSs. The histograms show the distributions of the relative root mean square error (c) and absolute value of maximum relative error (d) over the whole EoS collection.}
    \label{fig:demoMR}
\end{figure}

The $M-R$ curves can be described by the universal form:
\begin{multline}
\label{eq:MRfit}
    R/\Rmax = 1 + \left[ 2\left(\sqrt{2}-1\right)\ratio - a \right]\sqrt{1-M/\Mmax} \\
    + \left[ 2\left(\sqrt{2}-1\right)\ratio - 2 + a\sqrt{2} \right](1-M/\Mmax),
\end{multline}
where $a=0.492$. The maximum mass $\Mmax$, corresponding radius $\Rmax$, and the ratio $\ratio\equiv\Rhalf/\Rmax$ with $\Rhalf \equiv R(\Mmax/2)$, are the three parameters governing the manifold of NS $M-R$ curves. 
Geometrically, this means that the $M-R$ relations for different EoSs form a surface in the 3-dimensional space $(M/\Mmax,R/\Rmax,\ratio)$. This is an approximate statement (see  Fig.~\ref{fig:3D}). Several clear outliers correspond to obsolete models with severe softening around phase transition points and $\Mmax<2\,$\Msun. Moreover, this approximation is valid only for $M\gtrsim 1$\Msun. Below this mass, the form of $M-R$ curves is controlled by a larger number of parameters. 
Typical accuracy of Eq.~(\ref{eq:MRfit}) is demonstrated in Fig.~\ref{fig:demoMR}a,b for three particular examples of EoS. For the whole set of models, errors of this approximation rarely exceed $3\%$. For the outliers, maximal error is $\sim 10\%$ (see the histograms in Figs.~\ref{fig:demoMR}c,d).

\begin{figure*}
\centering
    \includegraphics[width=\textwidth]{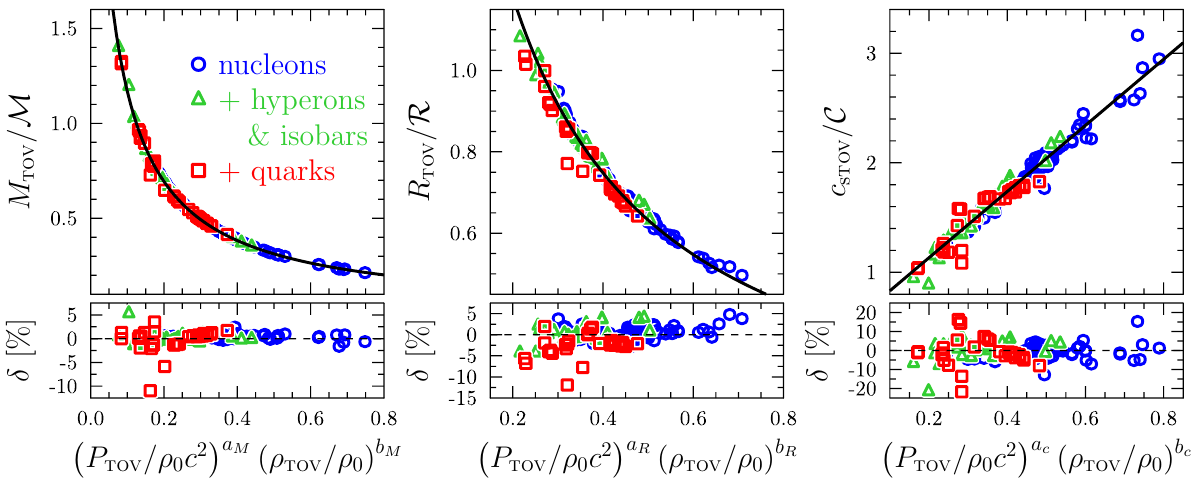}
\caption{\label{fig:corrs}
EoS-independent correlations of MMNS characteristics. Each symbol corresponds to a single EoS (nucleonic, with hyperons and isobars, or hybrid). The solid lines show the approximations~(\ref{eq:maxCorrs}). 
$\mathcal{R} = \sqrt{\Pmax/(G \rhomax^2)}$,
$\mathcal{M} = \rhomax \mathcal{R}^3$,
$\mathcal{C} = R_{\rm TOV}^\text{fit}\sqrt{G\rhomax}$, 
where $R_{\rm TOV}^\text{fit}$ is given by Eq.~(\ref{eq:maxCorrs-R}). The exponents $a_i$ and $b_i$, with $i = M,R,c$, are given in Table~\ref{tab:maxCorrCoeffs}. The bottom panels show relative errors of the fits.}
\end{figure*}

$\Mmax$ and $\Rmax$ are strongly correlated with other parameters describing MMNS, like $\rhomax$, $\Pmax$, and the speed of sound in the center $\csmax$ \cite{Ofengeim2020,Cai2023ApJ,OfShtPir2023}. Here we present updated fits to these correlations:
\begin{subequations}\label{eq:maxCorrs}
\begin{align}
    \Mmax &= \sqrt{\frac{\Pmax^3}{G^3 \rhomax^4}} \frac{1}{f_M(\Pmax,\rhomax)},
    \label{eq:maxCorrs-M}\\
    \Rmax &= \sqrt{\frac{\Pmax}{G \rhomax^2}} \frac{1}{f_R(\Pmax,\rhomax)}, 
    \label{eq:maxCorrs-R}\\
    \csmax &= \sqrt{\frac{\Pmax}{\rhomax}}\frac{f_c(\Pmax,\rhomax)}{f_R(\Pmax,\rhomax)} = c\ \zeta(\Pmax,\rhomax) \ . \label{eq:maxCorrs-c}
\end{align}
The dimensionless functions $f_M$, $f_R$, and $f_c$ have the generic form
\begin{equation}\label{eq:maxCorr-f}
    f_i = c_i \left(\frac{\Pmax}{\rho_0c^2}\right)^{a_i} \left(\frac{\rhomax}{\rho_0}\right)^{b_i} + d_i \ .
\end{equation}
\end{subequations}
The coefficients $a_i$, $b_i$, $c_i$, $d_i$ are listed in Table~\ref{tab:maxCorrCoeffs}, and $\rho_0 = 2.8\times 10^{14}\,\text{g}\,\text{cm}^{-3}$ is the nuclear saturation density. These correlations are illustrated in Fig.~\ref{fig:corrs}.

\begin{table}
	\centering
	\caption{\label{tab:maxCorrCoeffs} Coefficients in the fits (\ref{eq:maxCorrs}) of $\Mmax-\Pmax,\rhomax$, $\Rmax-\Pmax,\rhomax$, and $\csmax-\Pmax,\rhomax$ correlations. The last two columns show the relative root mean square and maximum relative  errors of the fits.}
   \renewcommand{\arraystretch}{1.4}
	\begin{tabular}{lcccccc}
		\hline\hline
		$f_i$  & $a_i$     & $b_i$     & $c_i$    & $d_i$     & rrms    & max error\\
		\hline
		$f_M$  & $1.41$    & $-1.39$   & $5.86$   & $0.270$   & 0.013   & 0.11    \\
        $f_R$  & $0.744$   & $-0.839$  & $2.45$   & $0.357$   & 0.021   & 0.12    \\
        $f_c$  & $0.985$   & $-0.990$  & $3.02$   & $0.529$   & 0.048   & 0.22     \\
		\hline\hline
	\end{tabular}
\end{table}

Since the OV mapping is a bijection \cite{Lind1992}, the manifold of EoS models is also 3-parametric (e.g. \cite{OzelPsaltis2009PhRvD}), at least in the density range occupying most of the volume of a NS with $M\gtrsim 1$\Msun. Correlations~(\ref{eq:maxCorrs}) provide a one-to-one mapping between the $(\Mmax,\Rmax,\ratio)$ and $(\rhomax,\Pmax,\ratio)$ parameter spaces. The latter one is suitable for constructing
the universal $P-\rho$ approximation, which can be conveniently split into a high and low density parts. 

\begin{figure*}
    \centering
    \includegraphics[width=\textwidth]{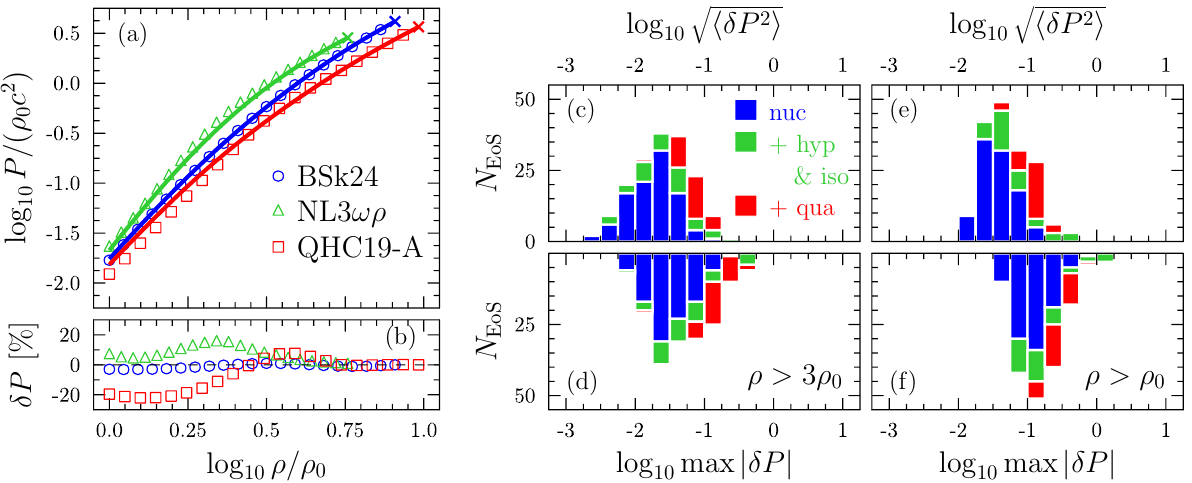}
    \caption{Demonstration of the approximations of $P(\rho)$ dependencies [Eqs.~(\ref{eq:highPrho}), (\ref{eq:lowPrho}), panel (a)] for selected nucleonic (BSk24), hyperonic (NL3$\omega\rho$), and hybrid (QHC19-A) EoSs. Symbols stand for exact values for given models, lines show the approximations with $\rhomax$, $\Pmax$, and $\ratio$ taken from the corresponding EoSs. Crosses mark the $(\rhomax,\Pmax)$ points. Panel (b) displays the relative errors for these EoSs. Histograms show the distributions of relative root mean square error (top) and maximum relative errors (bottom) for Eq.~(\ref{eq:highPrho}) (c,d) and Eq.~(\ref{eq:lowPrho}) (e,f) over the whole EoS collection. }
    \label{fig:demoPrho}
\end{figure*}

For $\rho>3\rho_0$, there exist an $\ratio$-independent fit $P = P_\text{(high)}(\rho; \Pmax,\rhomax)$:
\begin{multline}    
\label{eq:highPrho}
     \frac{P_\text{(high)}}{\Pmax} = \left( \frac{\rho}{\rhomax} \right)^{\gamma(\Pmax,\rhomax)-a_0}\bigg\{ 1 + a_0 \biggl( 1 
     \\ \left. \left. - 
     \frac{\rho}{\rhomax} \right)  
     + \left[b_0 + b_1 \zeta(\Pmax,\rhomax)\right]\left( 1 - \frac{\rho}{\rhomax} \right)^p \right\}^{-1} \ , 
\end{multline}
where $\zeta(\Pmax,\rhomax)$ is defined in Eq.~(\ref{eq:maxCorrs-c}), $\gamma(\Pmax,\rhomax) = \zeta^2 \rhomax c^2/\Pmax$, and 
$a_0 = -0.392$, 
$b_0 = -0.869$,
$b_1 = 1.69$, 
$p = 2.98$. 
Root-mean-square errors of this approximation rarely exceed $10\%$ (see Fig.~\ref{fig:demoPrho}c). This fit reveals that dimensionless $P/\Pmax-\rho/\rhomax$ curves also show somewhat universal behavior. 
\footnote{We presented a similar fit in~\cite{OfShtPir2023}, but here we reduce the number of fitting coefficients.} 

For $\rho<3\rho_0$ the dependence on $r_{1/2}$ cannot be ignored  and we built  an approximation with three parameters, $\rhomax$, $\Pmax$, and $\ratio$, valid for $\rho_0 < \rho < \rhomax$:
\begin{multline}
\label{eq:lowPrho}
    P = P_\text{(high)}\exp\left\{ \left( u_0 + u_1 \ratio + u_2\ratio^2 \right) \right.\\
    \left. \times \left( e^{-v\rho/\rho_0} - e^{-v\rhomax/\rho_0} \right) \right\},
\end{multline}
where $u_0 = -61.2$, $u_1 = 81.9$, $u_2 = -27.2$, $v = 1.33$, and $P_\text{(high)}$ is given by Eq.~(\ref{eq:highPrho}). Quality of this approximation for three particular EoSs is shown in Fig.~\ref{fig:demoPrho}a,b. On average, the rrms error is about $5-10\%$, it is smaller for nucleonic EoSs and larger for hyperonic and hybrid models (Fig.~\ref{fig:demoPrho}e). Although the maximum error (Fig.~\ref{fig:demoPrho}f) could reach $3-5$ times the rms, we show below that the  approximation~(\ref{eq:lowPrho}) together with Eqs.~(\ref{eq:MRfit}) and~(\ref{eq:maxCorrs}) allows one to make valuable inferences on the EoS when applied to observational data.

\section{Inverse OV mapping \& constraints on EoS}
\label{sec:IOV}

\subsection{{Inverse OV mapping}}
\label{sec:IOV:IOV}

Using these approximations one can explicitly perform IOVM. If the $M-R$ relation is known, then  $\Mmax$, $\Rmax$, and $\ratio$ are available. From Eqs.~(\ref{eq:maxCorrs-M}) and (\ref{eq:maxCorrs-R}), one derives $\Pmax$ and $\rhomax$. Substituting these values and $\ratio$ to Eq.~(\ref{eq:lowPrho}), one obtains the $P(\rho)$ curve in the range $\rho_0...\rhomax$. 

In practice, instead of the full $M-R$ curve, one has access to only a few measurements of $M$ and $R$ for a set of observed NSs (see Fig.~\ref{fig:IOVM-MR}). In this case, one can fit the observations with Eq.~(\ref{eq:MRfit}) to infer $\Mmax$, $\Rmax$, and $\ratio$ and consequently $\Pmax$, $\rhomax$ via Eqs.~(\ref{eq:maxCorrs}). In fact, one should rather substitute Eqs.~(\ref{eq:maxCorrs}) into Eq.~(\ref{eq:MRfit}), and use the triade $\rhomax$, $\Pmax$, and $\ratio$ as fitting parameters. 

The data typically strongly constrain $\Rhalf=\ratio\Rmax$ and poorly constrain the properties of MMNS, so we add two theoretical conditions. First, we set $\csmax(\Pmax,\rhomax)<1.05 \times c$ that is the causality limit with the safety factor $1.05$ accounting for the uncertainty of the fit~(\ref{eq:maxCorrs-c}). Second, we impose $\ratio \geqslant 1$ since we did not find an EoS model violating this condition (see Table~\ref{tab:app:zooList} and Fig.~\ref{fig:app:eosStat}c).

We turn now to apply this model to current $M-R$ observations. 

\subsection{{The NS $M$ and $R$ data}}
\label{sec:IOV:obs}

The observations of NS masses and radii fall into two classes. The first one consists of those which give joint information on $M$ and $R$. We use 9 such sources, including the recently reported observations of PSR~J0437$-$4715~\cite{Choudhury2024}. The colors of contours in Fig.~\ref{fig:IOVM-MR} correspond to the different types of sources: 
\begin{enumerate}[(i)]
    \item the NICER objects PSR J0030+0451 \cite{Riley+ApJL2019,Miller2019ApJ0030,Raaijmakers+2019,Vinciguerra+ApJ2024} (grey; data from~\cite{Miller2019ApJ0030}), PSR~J0740+6620 \cite{Riley+ApJL2021,Riley+ApJL2019, Raaijmakers+2021,Salmi2022ApJ,Salmi2024jun} (orange; data from~\cite{Riley+ApJL2021}), and PSR~J0437$-$4715 (blue; data from~\cite{Choudhury2024}, their preferred model).
    
    \item the thermally emitting isolated NS in the Cassiopeia A supernova remnant (CasA, green; data from \cite{Shternin2023MNRAS});
    
    \item low-mass X-ray binaries which exhibit X-ray bursts during super-Eddington accretion, namely 4U~J1702--429 (red) \cite{Nattila+AA2017},  4U~1724--307 (turquoise) and SAX~J1810.8--260 (purple) \cite{Nattila+AA2016};

    \item the $M-R$ data inferred from the GW~170817 signal~\cite{GW170817-EoS} (magenta).
    
\end{enumerate} 

The second class consists of NSs with measured masses only, which limit $\Mmax$ from below. We consider 8 sources:
\begin{enumerate}[(i)]
    
    \item the radio sources PSR J1614--2230 \cite{1614paper2010} and PSR J0348+0432 \cite{0348paper2013} (light blue horizontal strips in Fig.~\ref{fig:IOVM-MR}) as the most robust;

    \item the most massive ($M\gtrsim 2$\Msun) ``spiders''---binary systems containing a NS and a low-mass companion---PSR~J0952--0607, PSR~J1311--3430, PSR~J1810--1744, PSR~J1653--0158 \cite{Kandel2023ApJ}, and PSR~J2215+5135 \cite{Kandel2020ApJ} (light green strips in Fig.~\ref{fig:IOVM-MR}), which are more model-dependent;

    \item an interpretation of the {electromagnetic counterparts of }  GW 170817 
    which yields $\Mmax<2.16\pm0.16\,$\Msun\ \cite{Rezzolla2018ApJ} (thick brown line in Fig.~\ref{fig:IOVM-MR}).
    
\end{enumerate}

Clearly, each data point is obtained under certain model assumptions, and as such, it is model-dependent. Refs.~\cite{Dietrich2020Sci,Ayriyan2021EPJA,Ascenzi2024arXiv} classify the various astrophysical measurements by the level of their model dependence. We are not attempting to perform a similar assessment. Nevertheless, the radio timing mass measurements and the gravitational wave data can be considered the most robust ones. The NICER results on radii depend strongly on the complicated emission geometry. The CasA result is based on the specific carbon atmosphere NS thermal emission model, however being consistent with other measurements it does not impose crucial constraints. The three remaining data sets, detailed below, are the most constraining. The ``spiders'' data strongly constrain $M_{\mathrm{TOV}}$ from below; at the same time, they are based on the complicated model for optical emission of the companion, leaving room for serious modifications. The ``kilonova'' constraints (Fig.~\ref{fig:IOVM-MR}) from Ref.~\cite{Rezzolla2018ApJ} are the strongest ones that limit $M_{\mathrm{TOV}}$ from above, while they rely on a particular interpretation of the post-merger electromagnetic emission. See, for instance, \cite{Blinnikov+P2022} and refs. therein for alternative scenarios. The crucial constraints on NS radii come from the inclusion of the data on bursting sources, mainly from 4U~1702$-$429. We discuss elsewhere the implication of using different combinations of data {and approaches to their interpretation}.

We do not consider other thermally emitting isolated NSs \cite{Potekhin2020MNRAS} since they either yield much more significant uncertainties of $M$ and $R$, or the inferred mass range is beyond the range of validity of our approximations (as for XMMU~J1732$-$344 \cite{Doroshenko2022NatAs} with $M=0.8\pm 0.2\,$\Msun; see also \cite{Alford2023ApJ}). We also do not consider observations of LMXBs in quiescence, which can provide mass and radii information, since the results are currently inferior to other NS radii constraints \cite{Steiner2018MNRAS,Marino2018MNRAS}. Many other NS mass measurements exist in binary systems (see, e.g., \cite{LattimerU2019}), but they are less constraining than those we use due to lower masses. Recently, a binary system of a pulsar and a compact object of mass $2.35^{+0.20}_{-0.18}$ was discovered \cite{Barr2024Sci}. The nature of a compact component is unknown; it can be either a black hole or a NS. In the latter case, this mass measurement is consistent with the data based on the spider pulsars, and we checked that it does not provide an additional significant constraint. Because of that, and since the source can be a black hole, we don't include it in our fit.

\subsection{{Constraining EoS via Bayesian fitting}}
\label{sec:IOV:bayes}

We apply the IOVM procedure described in Sec.~\ref{sec:IOV:IOV} to the selected set of NS observations. We use the Bayesian inference (see e.g. \cite{Annala+Nat2023, Brandes+PRD2023, Brandes2023PhRvDa, Ayriyan2021EPJA, Raaijmakers+2021}) to obtain the model parameters $\Pmax$, $\rhomax$, and $r_{1/2}$, see Appendix~\ref{sec:app:methods} for details. \footnote{The code allowing for the reproduction of our results and underlying data is available at \cite{PaperData}.}

\begin{figure}
    \includegraphics[width=\columnwidth]{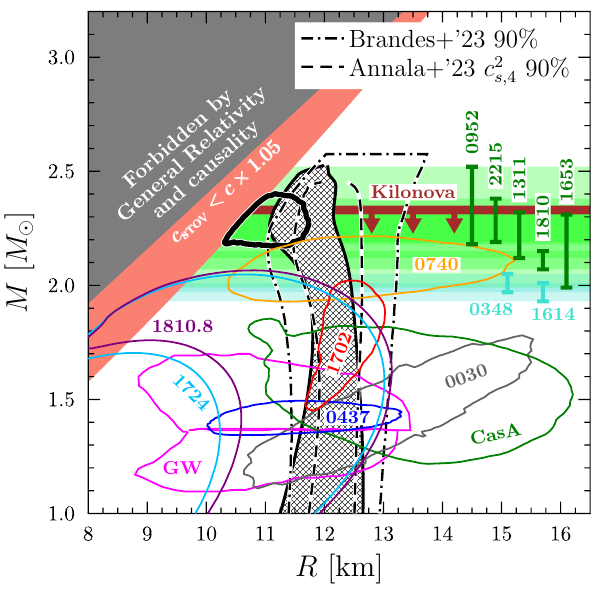}
    \caption{Constraining the $M-R$ curve from NS observations. Solid-colored contours show $90\%$ credible regions for NSs with both $M$ and $R$ measured. Vertical bars and corresponding colored strips display $1\sigma$ mass ranges for NSs in binaries. The thick brown line show the upper limit on $\Mmax$ adopted from the GW~170817 electromagnetic counterpart interpretation \cite{Rezzolla2018ApJ}. The grey-shaded region is prohibited due to the ``maximum compactness'' criterion (see, e.g., \cite{LattPrak2016}). The red-shaded region is excluded due to the causality condition applied to the correlations~(\ref{eq:maxCorrs}). The cross-hatched region is formed by the $90\%$ credible intervals for radii at given NS masses, obtained from fitting the observational data by Eq.~(\ref{eq:MRfit}). Thin dashed and dash-dotted 90\% credible  contours show similar results by Annala et al.~\cite[model ``$c^2_{s,4}$'']{Annala+Nat2023}, and by Brandes et al.~\cite{Brandes+PRD2023}, respectively. The thick black contour shows the 2D-credible region for $\Mmax$ and $\Rmax$ for our results.  Notice a statistical paradox: due to the almost horizontal behavior of the $M-R$ curve near the MMNS point, the most probable radii for fixed $M\sim\Mmax$ are significantly greater than $\Rmax$.
    }
    \label{fig:IOVM-MR}
\end{figure}

\begin{figure}
    \includegraphics[width=\columnwidth]{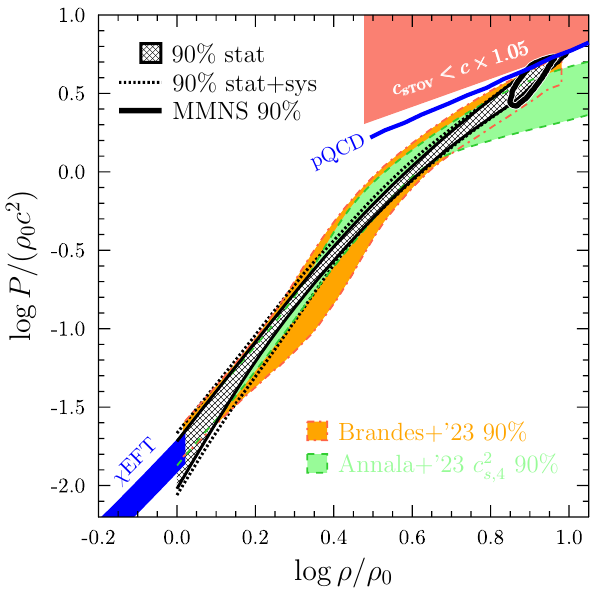}
    \caption{EoS constrained from NS observations. The cross-hatched strip is formed by $90\%$ credible intervals for pressures at given densities, obtained from the proposed IOVM procedure. Dotted lines show the same but with account for fit errors treated as systematic uncertainties. The thick black contour is the $90\%$ 2D-credible area for $\Pmax$ and $\rhomax$. The orange and green bands are similar to the cross-hatched stip but adopted from Brandes et al.~\cite{Brandes+PRD2023} and Annala et al.~\cite[model ``$c^2_{s,4}$'']{Annala+Nat2023}, correspondingly. The red-shaded region is prohibited for $\Pmax$ and $\rhomax$ by the causality condition applied to Eq.~(\ref{eq:maxCorrs-c}). The ``pQCD'' line depicts the upper limit on the EoS which follows from the perturbative QCD matching \cite{KomoltsevKurkelaPRL2022}. The ``$\chi$EFT'' band is the ``chiral corridor'' for the low-density EoS based on the chiral effective field theory \cite{Hebeler+ApJ2013}.}
    \label{fig:IOVM-Prho}
\end{figure}

\begin{figure*}
\centering
\includegraphics[width=0.8\textwidth]{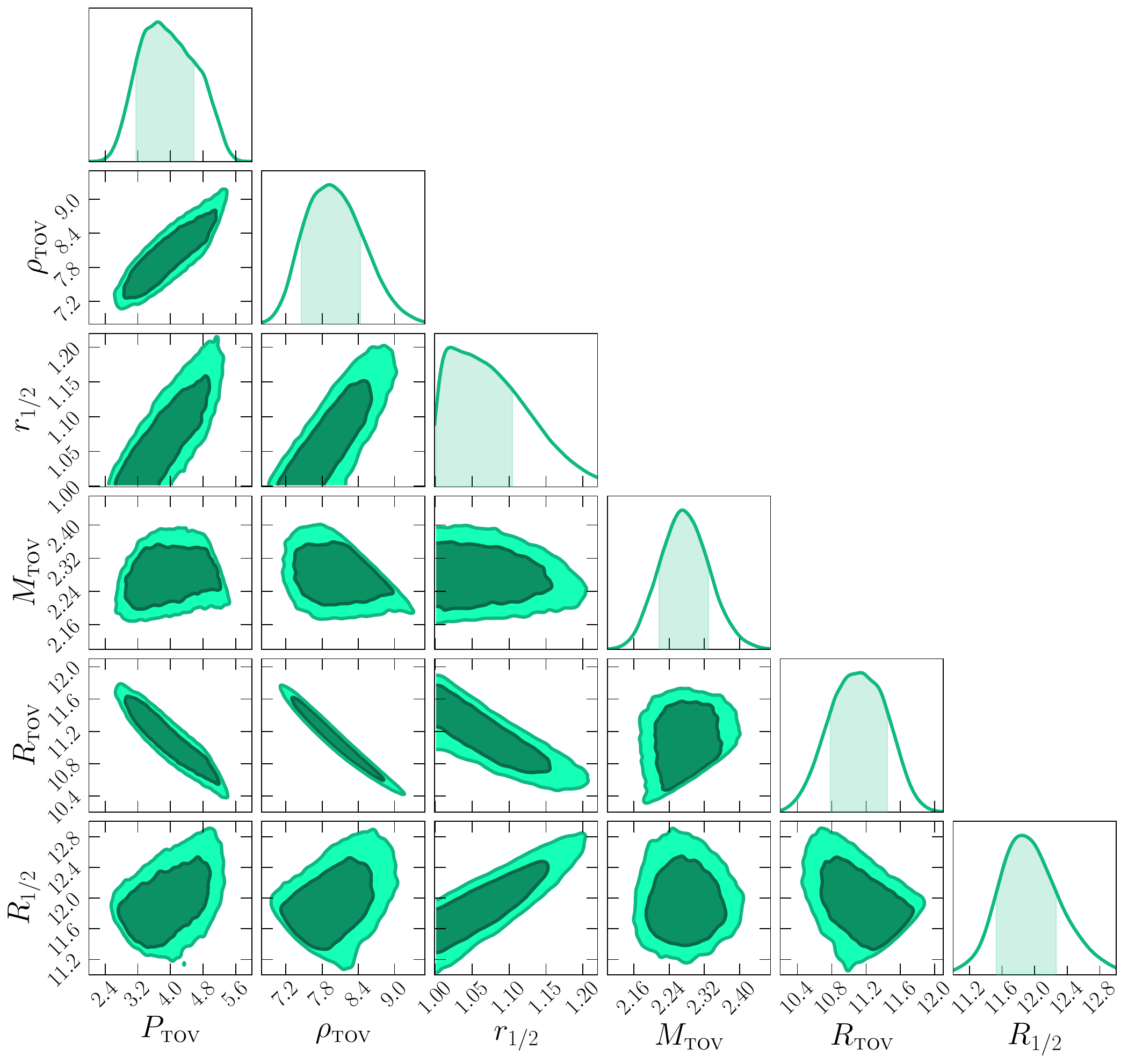}
\caption{Marginal 1D and 2D posterior distributions of model parameters listed in Table~\ref{tab:app:params}. The contours' credibility levels on 2D plots are 68\% and 90\%. The shaded areas of 1D posteriors correspond to 68\% highest posterior density intervals.}
\label{fig:app:tri1}
\end{figure*}

The one-dimensional 68\% credible intervals for the key parameters of the approximations~(\ref{eq:MRfit})---(\ref{eq:lowPrho}) are summarized in Table~\ref{tab:app:params}. The cross-hatched strip in Fig.~\ref{fig:IOVM-MR} shows the 90\% credibility region for the $M-R$ relation while the thick black contour describes the 90\% credible region for $\Mmax$ and $\Rmax$ [obtained with Eqs.~(\ref{eq:MRfit}) and~(\ref{eq:maxCorrs})]. Fig.~\ref{fig:IOVM-Prho} shows the limits on the EoS obtained with Eqs.~(\ref{eq:maxCorrs})---(\ref{eq:lowPrho}). The full set of marginalized 1D and 2D posteriors of the key parameters is shown in Fig.~\ref{fig:app:tri1}.

\begin{table}
\centering
\caption{Fit summary. Uncertainties correspond to 68\% highest posterior density credible intervals.}\label{tab:app:params}
\renewcommand{\arraystretch}{1.4}
    \begin{tabular}{cccccc}
        \hline\hline
  $P_{\mathrm{TOV}}$ & $\rho_{\mathrm{TOV}}$  &  $r_{1/2}$  &  $M_{\mathrm{TOV}}$ &  $R_{\mathrm{TOV}}$ &  $R_{1/2}$\\ 
 $[\rho_0c^2]$ & $[\rho_0]$&&$[M_\odot]$ & [km] &[km] \\
  \hline
  $3.9^{+0.6}_{-0.8}$ &
  $8.0^{+0.4}_{-0.6}$ &
  $1.08_{-0.08}^{+0.02}$ &  
  $2.27^{+0.06}_{-0.06}$ &
  $11.12^{+0.33}_{-0.34}$ &
  $12.06^{+0.21}_{-0.53}$ 
 \\ 
  \hline\hline
    \end{tabular}
\end{table}


\begin{figure*}
\centering
    \includegraphics[width=0.95\textwidth]{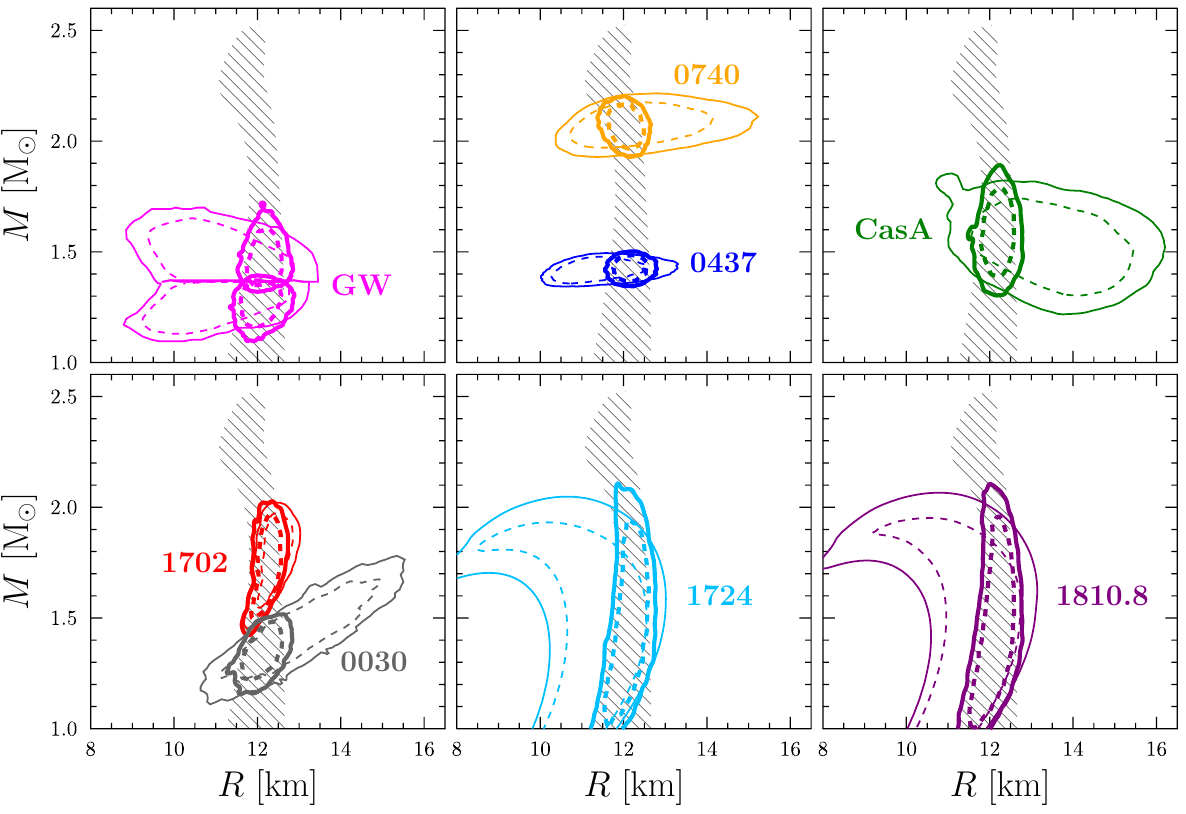}
    \caption{Comparison of initial $M-R$ credible contours (thin) vs posteriors (thick) for the individual sources. Dashed and solid lines stand for 68\% and 90\% credibility levels, correspondingly. The hatched strip is the 90\% credibility area for the $R(M)$ relation (the same as the cross-hatched strip in Fig.~\ref{fig:IOVM-MR}).}
    \label{fig:individuals}
\end{figure*}

The credible area for the EoS matches smoothly the low-density ``EoS corridor'' derived from the chiral effective field theory ($\chi$EFT) \cite{Hebeler+ApJ2013}. Even though our analysis does not include $\chi$EFTs constraints as a prior (the EoS collection contains both $\chi$EFT-consistent and $\chi$EFT-inconsistent models), our results are consistent with the requirement to pass the ``chiral corridor'' \cite{Lattimer2021}. 

The allowed range of EoS is consistent but narrower than those found in recent works \cite{Annala+Nat2023,Brandes+PRD2023} (green and orange bands in Fig.~\ref{fig:IOVM-Prho}). Notice that \cite{Brandes+PRD2023,Annala+Nat2023} use an additional theoretical condition of matching the  perturbative quantum chromodynamics (pQCD) limit. However, this limit \cite{KomoltsevKurkelaPRL2022} coincides numerically with our causality limit (see Fig.~\ref{fig:IOVM-Prho}), so our results satisfy this condition as well.

Another difference between our work and \cite{Annala+Nat2023,Brandes+PRD2023} is that we include the most recent observations of PSR~J0437$-$4715 \cite{Choudhury2024}. An advantage of the semi-analytic IOVM is that various data combinations could be tested at negligible computational cost. We checked that adding or removing PSR~J0437$-$4715 from the observational dataset makes no significant impact neither on the values of fitting parameters nor on their precision.

One has to keep in mind that the universal relations (\ref{eq:MRfit})---(\ref{eq:lowPrho}) are approximate. Their errors could be treated as systematic uncertainties when the IOVM procedure is performed. We accounted for these uncertainties, bootstrapping individual errors of the approximations. Eventually, the $M-R$ area and MMNS properties remain unaffected, and the effect for EoS is shown by the black dotted lines in Fig.~\ref{fig:IOVM-Prho}. Indeed, these systematics slightly widen the EoS credible area at low densities but does not affect the high-density part.

A by-product of simultaneous Bayesian fitting of multiple sources is mutual constraining of their parameters. In Fig.~\ref{fig:individuals} we compare the initial inferences on masses and radii of individual sources (thin lines) with the updated ones according to the Bayesian procedure, given they all have the same EoS in their interiors (thick contours). One can notice the shrinkage of the credible intervals on $R$ for each source (the least pronounced for 4U~J1702$-$429).

\section{Discussion \& Conclusions}
\label{sec:concl}

We presented here the universal approximations describing the $M-R$ curves and $P(\rho)$ dependencies for a wide set of EoSs. This set includes models with various compositions (nucleonic, hyperonic, hybrid) that are based on different approaches to the underlying physics. The key idea is that three parameters are sufficient to describe the EoS to a good accuracy. We show that it is convenient to choose two of those as MMNS properties (say, $\Pmax$ and $\rhomax$) and the third as the ratio $\ratio = \Rhalf/\Rmax$.

These approximations have numerous applications in NS studies. Here we explored the very basic one---explicit implementation 
\footnote{See other explicit implementations in, e.g., \cite{Lind1992,LindInd2012,Soma+2022}.}
of IOVM and inference of the EoS from observations. Our semi-analytic method avoids the time consuming need of repeatedly solving TOV equations when attempting to infer the EoS from  observed NSs masses and radii. The obtained range of EoS is consistent but narrower than those found in recent works \cite{Annala+Nat2023,Brandes+PRD2023}. Some of recent results \cite{Annala+Nat2023} suggest that the true EoS should be hybrid (see, however, \cite{Brandes+PRD2023}). While our approximations show the largest errors for this subclass of models, they are  accurate enough to make our method  useful.  It may artificially smooth out kinks and bends of the real $P-\rho$ curve, but should definitely reproduce its averaged behavior.

Notice that our approximations are fitted to a wide but finite set of theoretical EoS models, so our semi-analytic IOVM may fail if none of these models is close to the true EoS of the NS matter. The agreement of our results with those of \cite{Annala+Nat2023,Brandes+PRD2023}, which are applicable even if the real EoS differs strongly from those that are in our collection, indicates that the true EoS is somewhere in between the existing theoretical models.

Other applications of the universal relations presented here could be, e.g., simultaneous fitting of X-ray spectra of several NSs by atmospheric models using Eq.~(\ref{eq:MRfit}) as the form of the $M-R$ relation; or using Eqs.~(\ref{eq:highPrho}) or~(\ref{eq:lowPrho}) as an analytic form of $\beta$-equilibrium EoS in hydrodynamic studies (somewhat smoothed, however); or, together with existing constraints, Eq.~(\ref{eq:MRfit}) could be used as a simple informed prior for any study that requires an explicit $M-R$ relation for NSs. This is illustrated nicely in Fig.~\ref{fig:individuals}  were one can notice the shrinkage of the credible intervals on $R$ for each source (the least pronounced for 4U~J1702$-$429).

Finally, we speculate that the origin of the universal relations presented here is probably some combination of the smoothing nature of the OV mapping from the EoS to $M-R$ and some general property of low-energy nuclear interactions, which are dominated by just three degrees of freedom.  See \cite{Lindblom2010} for discussion of the sufficient number of EoS parameters, and \cite{OfShtPir2023} for a detailed discussion of possible origins of universal relations similar to Eqs.~(\ref{eq:MRfit})---(\ref{eq:lowPrho}).

{The data for the paper and a link to the \texttt{python} package {\texttt{unicorrns}}, which implements the universal relations (\ref{eq:MRfit})---(\ref{eq:lowPrho}) and allows to sample the $R$ distribution at a given $M$ and the $P$ distribution at a given $\rho$, is available at \cite{PaperData}.}

\begin{acknowledgments}
We are grateful to V.~F.~Suleimanov and J.~N{\"a}ttil{\"a} for providing us the data for the three X-ray bursters, as well as sharing us the posteriors of the paper \cite{Annala+Nat2023}; to L.~Brandes for sharing us the contours from \cite{Brandes+PRD2023}; to E.~E.~Kolomeitsev for providing us the tables for nine KVOR EoSs; to N.~Jokela and M.~J{\"a}rvinen for consulting us on the correct MMNS treatment for JJ EoSs. 
DO and TP were supported by Advanced ERC grant MultiJets. Work of PS was supported by the 
Russian Science Foundation (grant 24-12-00320).
\end{acknowledgments}


%


\begin{widetext}
\appendix

\setcounter{table}{0}
\renewcommand{\thetable}{\Alph{section}\arabic{table}}
\setcounter{figure}{0}
\renewcommand{\thefigure}{\Alph{section}\arabic{figure}}
\section{Collection of the equations of state}
\label{sec:app:EoS}

\begin{figure*}[tttt]
    \includegraphics[width=0.325\textwidth]{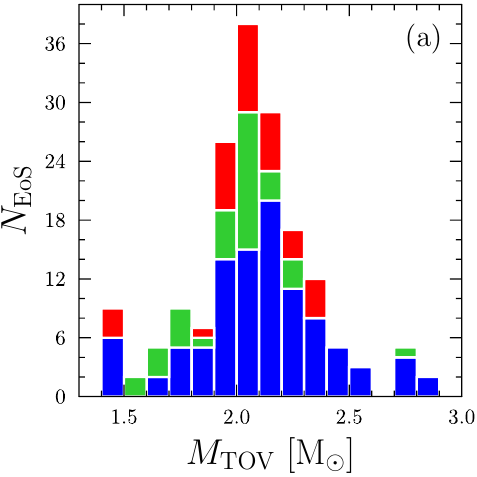}
    \includegraphics[width=0.325\textwidth]{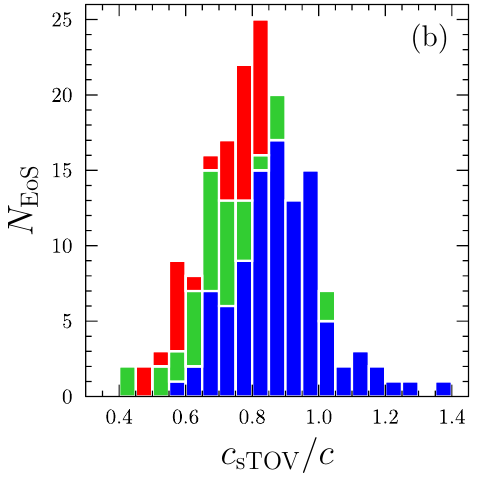}
    \hspace{0.003\textwidth}
    \includegraphics[width=0.325\textwidth]{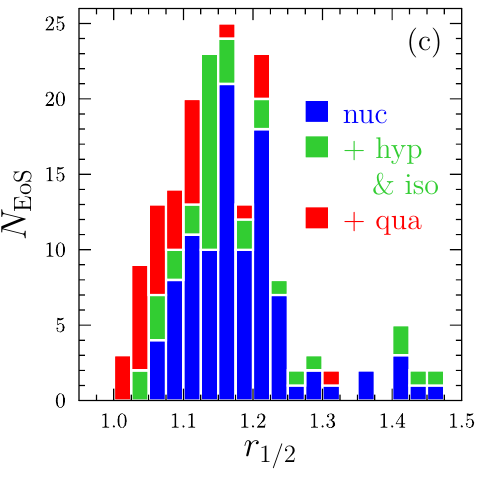}
    \caption{
    Distributions  of (a) NS maximal mass, (b) speed of sound in the center of the star, (c) $r_{1/2} = R_{1/2}/\Rmax$ ratio for the sample of equations of state used.
    }
    \label{fig:app:eosStat}
\end{figure*}

The EoSs for this work were selected as follows. From the CompOSE database \cite{CompOSE2015} (as of November 2022), we took the subset \texttt{Cold Neutron Star EoS}, keeping only the models providing data for both the NS crust and core. Then we added models from \cite{Read+2009} (some time ago, this EoS collection was available by the link given in \cite{OzelFreire2016}; now one can access an archived copy), nine models from \cite{MKV2016}, and several dozen models more which are available to the authors (see \cite{Ofengeim2020} for details). We discarded duplicates and tables ended at $\rho<\rhomax$. Ultimately, we got a sample of 169 models listed in Table~\ref{tab:app:zooList}.

The considered models are based on a variety of approaches to modeling interactions and the microphysics of superdense matter. These include models based on effective energy density functionals, including purely phenomenological (PAL, PAPAL, BGN families), non-relativistic Skyrme-type (SLy, BSk, SkI, etc.) or Gogni (D1M*) and others, numerous relativistic mean field functionals; models derived from microscopic baryon interaction potentials using many-body methods (APR, WFF, BBB); and also some other models. Among the hybrid equations of state, both those in which a first-order phase transition occurs between hadronic and quark matter (for example, CMF, VQCD) and those in which there is a quark-hadron crossover (QHC) between these phases are present.

We name models according to the CompOSE tradition, following the principle \textit{original paper acronym(model name)}. For each EoS, tables and detailed description (choice of microphysical parameters, stitching the core and the crust, etc.) can be found in the references given in the ``source'' column. In these sources, one can find a decoding of the acronym of the original work in which this particular model proposed (typically not coinciding with the source). If neither the source nor the references given there describe a crust model, the BSk24 crust \cite{BSk2018} was stitched to the core under the condition of pressure continuity. Model names in Table~\ref{tab:app:zooList} may be reasonably abbreviated with respect to those used in the references.

Additionally, for each EoS, the mass $\Mmax$, its radius $\Rmax$, pressure $\Pmax$, density $\rhomax$, and the speed of sound $\csmax$ at the center of such MMNS, as well as the radius of the half-maximum-mass NS $\Rhalf$ are indicated in the Table~\ref{tab:app:zooList}. In Fig.~\ref{fig:app:eosStat} we show distributions of this sample of models over $\Mmax$ (panel a), over $\csmax$ (b), and over the parameter $\ratio = \Rhalf/\Rmax$ (c). The relative root mean square (rrms) and maximum relative fitting errors of the $M-R$ [Eq.~(\ref{eq:MRfit})] and $P(\rho)$ [Eqs.~(\ref{eq:highPrho}) and (\ref{eq:lowPrho})] fits are also provided in the table~\ref{tab:app:zooList}. For the latter fit, we give the errors in two density ranges $\rho\in (3\rho_0...\rhomax)$ and $\rho\in (1\rho_0...\rhomax)$ called ``all $\rho$''.

\setlength{\tabcolsep}{0.0075\textwidth}
\newcommand{\marka}{$^{\text{a}}$}
\newcommand{\markb}{$^{\text{b}}$}
\newcommand{\markc}{$^{\text{c}}$}
\renewcommand{\arraystretch}{1.4}
\begin{longtable}{l|c|ccc|cc|ccc|cccc}
    \caption{\label{tab:app:zooList} Neutron star equations of state }\\
    \hline\hline
    Name & Source & $\Mmax$ & $\Rmax$ & $\Rhalf$ & \multicolumn{2}{c|}{Fit $R(M)$ [\%]} & $\rhomax$  & $\Pmax$        & $\csmax$ & \multicolumn{4}{c}{Fit $P(\rho)$ [\%]}\vspace{-1ex}\\
         &        &         &         &          &             &                        &            &                &          & \multicolumn{2}{c}{$\rho>3\rho_0$} & \multicolumn{2}{c}{all $\rho$}\vspace{-1ex}\\
         &        & [\Msun] & [km]    & [km]     & rrms        & max                    & [$\rho_0$] & [$\rho_0 c^2$] & [c]      & rrms            & max              & rrms            & max           \\
    \hline \endfirsthead 
    \multicolumn{12}{l}{\textit{Continued Table \ref{tab:app:zooList}}}\\
    \hline
    Name & Source & $\Mmax$ & $\Rmax$ & $\Rhalf$ & \multicolumn{2}{c|}{Fit $R(M)$ [\%]} & $\rhomax$  & $\Pmax$        & $\csmax$ & \multicolumn{4}{c}{Fit $P(\rho)$ [\%]}                               \vspace{-1ex}\\
         &        &         &         &          &             &                        &            &                &          & \multicolumn{2}{c}{$\rho>3\rho_0$} & \multicolumn{2}{c}{all $\rho$}  \vspace{-1ex}\\
         &        & [\Msun] & [km]    & [km]     & rrms        & max                    & [$\rho_0$] & [$\rho_0 c^2$] & [c]      & rrms            & max              & rrms            & max           \\
    \hline \endhead 
    \hline\multicolumn{12}{l}{\textit{To be continued}}\\
    \endfoot
    \hline\hline 
    \multicolumn{14}{c}{\parbox{0.98\textwidth}{\vspace*{1ex}
\marka accessible via the archive link \texttt{https://web.archive.org/web/20200830055909/http://xtreme.as.arizona.edu/neutronstars/}. \\
\markb CompStar Online Supernova Equations of State database, \texttt{https://compose.obspm.fr}. \\
\markc See also \cite{JarvinenEPJC2022}.
}}
    \endlastfoot
    \multicolumn{14}{c}{Nucleonic models}\\
    APR(APR1) & \cite{Read+2009,OzelFreire2016}\marka & 1.68 & 8.28 & 9.70 & 0.89 & 1.1 & 14.9 & 7.42 & 0.839 & 8.9 & 13 & 9.1 & 17 \\
APR(APR2) & \cite{Read+2009,OzelFreire2016}\marka & 1.81 & 8.72 & 10.51 & 0.50 & 0.62 & 13.4 & 7.63 & 1.03 & 5.9 & 10 & 7.4 & 31 \\
APR(APR3) & \cite{Read+2009,OzelFreire2016}\marka & 2.39 & 10.76 & 12.02 & 0.32 & 0.54 & 8.35 & 5.78 & 1.13 & 0.90 & 1.5 & 4.1 & 25 \\
ZBL(BPAL12) & \cite{Read+2009,OzelFreire2016}\marka & 1.45 & 9.01 & 12.69 & 1.1 & 1.3 & 14.2 & 4.29 & 0.704 & 2.1 & 5.1 & 2.5 & 6.2 \\
EOHJBO(ENG) & \cite{Read+2009,OzelFreire2016}\marka & 2.24 & 10.41 & 11.97 & 0.42 & 0.87 & 9.19 & 5.68 & 1.00 & 3.8 & 6.6 & 4.3 & 7.9 \\
FP(FPS) & \cite{Read+2009,OzelFreire2016}\marka & 1.80 & 9.28 & 11.18 & 0.20 & 0.29 & 12.1 & 5.40 & 0.870 & 3.0 & 6.0 & 3.2 & 6.4 \\
PAL(PAL6) & \cite{Read+2009,OzelFreire2016}\marka & 1.48 & 9.27 & 13.07 & 1.0 & 1.2 & 13.5 & 3.93 & 0.692 & 1.2 & 2.7 & 2.0 & 11 \\
WFF(WFF1) & \cite{Read+2009,OzelFreire2016}\marka & 2.13 & 9.41 & 10.35 & 0.68 & 0.96 & 10.8 & 8.04 & 1.18 & 5.9 & 11 & 12 & 72 \\
WFF(WFF2) & \cite{Read+2009,OzelFreire2016}\marka & 2.20 & 9.81 & 11.14 & 0.60 & 1.0 & 10.1 & 7.37 & 1.15 & 3.2 & 5.1 & 7.0 & 44 \\
WFF(WFF3) & \cite{Read+2009,OzelFreire2016}\marka & 1.84 & 9.50 & 11.06 & 0.24 & 0.36 & 11.3 & 4.86 & 0.836 & 3.0 & 5.3 & 5.6 & 20 \\
MPA(MPA1) & \cite{Read+2009,OzelFreire2016}\marka & 2.46 & 11.32 & 12.38 & 0.40 & 0.66 & 7.55 & 4.56 & 0.991 & 2.7 & 3.9 & 4.2 & 8.8 \\
MS(MS1) & \cite{Read+2009,OzelFreire2016}\marka & 2.77 & 13.33 & 14.83 & 0.71 & 1.1 & 5.53 & 2.80 & 0.892 & 2.6 & 4.9 & 6.8 & 33 \\
MS(MS1b) & \cite{Read+2009,OzelFreire2016}\marka & 2.78 & 13.29 & 14.52 & 0.44 & 0.74 & 5.51 & 2.80 & 0.894 & 3.0 & 6.2 & 8.2 & 31 \\
MS(MS2) & \cite{Read+2009,OzelFreire2016}\marka & 1.81 & 11.68 & 14.24 & 0.47 & 0.69 & 7.79 & 1.89 & 0.582 & 0.76 & 2.4 & 2.5 & 5.2 \\
ABHT(QMC-RMF1) & \cite{CompOSE2015}\markb & 1.95 & 10.25 & 12.02 & 0.37 & 0.47 & 9.82 & 3.96 & 0.806 & 1.5 & 2.5 & 7.0 & 24 \\
ABHT(QMC-RMF2) & \cite{CompOSE2015}\markb & 2.04 & 10.49 & 12.07 & 0.26 & 0.40 & 9.26 & 3.93 & 0.828 & 1.5 & 2.5 & 7.2 & 27 \\
ABHT(QMC-RMF3) & \cite{CompOSE2015}\markb & 2.14 & 10.71 & 12.32 & 0.66 & 0.92 & 8.81 & 4.11 & 0.862 & 2.2 & 3.3 & 6.5 & 17 \\
ABHT(QMC-RMF4) & \cite{CompOSE2015}\markb & 2.21 & 11.03 & 12.22 & 0.13 & 0.25 & 8.15 & 3.69 & 0.852 & 1.8 & 2.6 & 6.0 & 30 \\
APR(APR) & \cite{CompOSE2015}\markb & 2.19 & 9.93 & 11.37 & 0.53 & 0.75 & 9.93 & 6.87 & 1.15 & 2.0 & 3.3 & 6.6 & 42 \\
BG(BGN1) & \cite{HPY2007} & 2.17 & 10.91 & 13.13 & 0.14 & 0.27 & 8.66 & 4.24 & 0.939 & 2.0 & 2.7 & 2.5 & 8.0 \\
BG(BGN2) & \cite{HPY2007} & 2.48 & 11.74 & 13.51 & 0.11 & 0.22 & 7.21 & 4.23 & 1.04 & 2.3 & 3.1 & 4.5 & 16 \\
BBB(BBB1) & \cite{HPY2007} & 1.79 & 9.67 & 11.23 & 0.32 & 0.42 & 11.0 & 4.05 & 0.765 & 3.0 & 6.1 & 8.7 & 29 \\
BBB(BHF-BBB2) & \cite{CompOSE2015}\markb & 1.92 & 9.50 & 11.37 & 0.56 & 0.92 & 11.4 & 5.75 & 0.915 & 2.6 & 3.7 & 3.9 & 18 \\
BL(chiral) & \cite{CompOSE2015}\markb & 2.08 & 10.28 & 12.56 & 0.41 & 0.54 & 9.78 & 5.13 & 0.985 & 1.2 & 1.8 & 2.2 & 11 \\
GDTB(DDH$\delta$) & \cite{CompOSE2015}\markb & 2.15 & 11.23 & 12.49 & 0.45 & 0.65 & 7.92 & 3.13 & 0.790 & 0.73 & 1.9 & 2.1 & 5.5 \\
GKYG(APR-I) & \cite{Gusakov+2005} & 1.92 & 10.31 & 12.42 & 0.31 & 0.41 & 9.85 & 3.86 & 0.801 & 0.34 & 0.91 & 3.2 & 13 \\
GKYG(APR-II) & \cite{Gusakov+2005} & 1.92 & 10.27 & 12.34 & 0.29 & 0.39 & 9.91 & 3.89 & 0.802 & 0.43 & 0.85 & 3.3 & 13 \\
GKYG(APR-III) & \cite{Gusakov+2005} & 1.93 & 10.38 & 12.57 & 0.35 & 0.45 & 9.75 & 3.80 & 0.799 & 0.27 & 1.1 & 3.0 & 11 \\
GM(GM1) & \cite{CompOSE2015}\markb & 2.38 & 12.16 & 14.23 & 0.39 & 0.63 & 6.91 & 3.07 & 0.848 & 0.90 & 1.8 & 1.6 & 6.4 \\
GMSR(BSk14) & \cite{CompOSE2015}\markb & 1.93 & 9.29 & 11.50 & 0.059 & 0.076 & 11.9 & 7.10 & 1.13 & 5.2 & 14 & 11 & 31 \\
GMSR(H7) & \cite{CompOSE2015}\markb & 2.48 & 11.23 & 12.61 & 1.4 & 2.6 & 7.64 & 5.64 & 1.38 & 10 & 16 & 10 & 17 \\
GMSR(LNS5) & \cite{CompOSE2015}\markb & 1.97 & 9.92 & 11.65 & 0.50 & 0.75 & 10.4 & 4.84 & 0.890 & 1.9 & 3.2 & 2.2 & 3.6 \\
GMSR(SLy5) & \cite{CompOSE2015}\markb & 2.10 & 10.06 & 11.93 & 0.48 & 0.74 & 10.0 & 5.67 & 1.00 & 0.90 & 1.1 & 2.4 & 13 \\
GPPVA(DD2) & \cite{CompOSE2015}\markb & 2.42 & 11.88 & 13.13 & 0.26 & 0.46 & 6.99 & 3.30 & 0.866 & 2.0 & 3.1 & 2.4 & 5.5 \\
GPPVA(DDME2) & \cite{CompOSE2015}\markb & 2.48 & 12.07 & 13.16 & 0.23 & 0.43 & 6.71 & 3.24 & 0.873 & 2.7 & 4.7 & 3.4 & 7.5 \\
GPPVA(FSU2) & \cite{CompOSE2015}\markb & 2.07 & 12.06 & 14.10 & 0.98 & 1.3 & 7.11 & 2.08 & 0.632 & 2.3 & 3.2 & 2.9 & 6.3 \\
GPPVA(FSU2H) & \cite{CompOSE2015}\markb & 2.38 & 12.36 & 13.16 & 0.38 & 0.67 & 6.36 & 2.33 & 0.703 & 6.1 & 11 & 7.1 & 11 \\
GPPVA(FSU2R) & \cite{CompOSE2015}\markb & 2.05 & 11.65 & 12.87 & 0.40 & 0.64 & 7.35 & 2.18 & 0.632 & 4.2 & 6.1 & 4.7 & 15 \\
GPPVA(NL3$\omega\rho$L55) & \cite{CompOSE2015}\markb & 2.75 & 13.00 & 13.71 & 0.23 & 0.46 & 5.66 & 2.92 & 0.894 & 4.6 & 9.6 & 7.4 & 13 \\
GPPVA(TM1e) & \cite{CompOSE2015}\markb & 2.12 & 11.84 & 13.07 & 0.37 & 0.61 & 7.12 & 2.24 & 0.654 & 3.9 & 5.9 & 4.9 & 19 \\
GPPVA(TW) & \cite{CompOSE2015}\markb & 2.08 & 10.64 & 12.48 & 0.60 & 0.81 & 9.05 & 3.95 & 0.839 & 1.3 & 2.3 & 3.9 & 10 \\
KKPY(APR-IV) & \cite{Kaurov+2014} & 2.16 & 10.86 & 12.81 & 0.23 & 0.38 & 8.68 & 4.12 & 0.904 & 0.72 & 1.1 & 1.0 & 3.4 \\
PCGS(PCSB0) & \cite{CompOSE2015}\markb & 2.53 & 12.13 & 13.26 & 0.54 & 0.81 & 6.63 & 3.35 & 0.888 & 3.5 & 5.3 & 4.1 & 9.5 \\
PFCPG(BSk19) & \cite{BSk2013} & 1.86 & 9.10 & 11.08 & 0.35 & 0.47 & 12.4 & 6.76 & 1.01 & 4.3 & 8.9 & 5.5 & 14 \\
PFCPG(BSk20) & \cite{BSk2013} & 2.16 & 10.17 & 11.76 & 0.16 & 0.33 & 9.61 & 5.79 & 1.06 & 0.98 & 2.0 & 1.9 & 12 \\
PFCPG(BSk21) & \cite{BSk2013} & 2.27 & 11.04 & 12.52 & 0.23 & 0.34 & 8.17 & 4.24 & 0.963 & 0.96 & 1.5 & 1.3 & 4.1 \\
PCPFDDG(BSk22) & \cite{BSk2018} & 2.26 & 11.20 & 13.03 & 0.19 & 0.29 & 8.07 & 4.02 & 0.951 & 2.0 & 2.9 & 2.4 & 7.2 \\
PCPFDDG(BSk24) & \cite{BSk2018} & 2.28 & 11.07 & 12.51 & 0.12 & 0.19 & 8.10 & 4.15 & 0.955 & 0.74 & 1.1 & 1.2 & 3.1 \\
PCPFDDG(BSk25) & \cite{BSk2018} & 2.22 & 11.04 & 12.25 & 0.20 & 0.28 & 8.11 & 3.76 & 0.896 & 1.4 & 2.8 & 4.2 & 17 \\
PCPFDDG(BSk26) & \cite{BSk2018} & 2.17 & 10.19 & 11.79 & 0.21 & 0.38 & 9.57 & 5.76 & 1.06 & 0.86 & 1.7 & 1.9 & 12 \\
RG(KDE0v) & \cite{CompOSE2015}\markb & 1.96 & 9.66 & 11.71 & 0.42 & 0.54 & 11.0 & 5.82 & 0.984 & 2.0 & 3.9 & 2.2 & 6.5 \\
RG(KDE0v1) & \cite{CompOSE2015}\markb & 1.97 & 9.79 & 11.96 & 0.52 & 0.66 & 10.8 & 5.53 & 0.965 & 1.4 & 2.8 & 1.6 & 6.3 \\
RG(Rs) & \cite{CompOSE2015}\markb & 2.12 & 10.76 & 13.11 & 0.13 & 0.19 & 8.96 & 4.23 & 0.922 & 2.4 & 3.3 & 4.6 & 20 \\
RG(Sk255) & \cite{CompOSE2015}\markb & 2.14 & 10.85 & 13.44 & 0.46 & 0.63 & 8.88 & 4.32 & 0.937 & 2.3 & 3.0 & 2.6 & 10 \\
RG(Sk272) & \cite{CompOSE2015}\markb & 2.23 & 11.09 & 13.51 & 0.41 & 0.57 & 8.40 & 4.30 & 0.965 & 2.3 & 3.0 & 2.8 & 11 \\
RG(Ska) & \cite{CompOSE2015}\markb & 2.21 & 10.89 & 13.03 & 0.20 & 0.35 & 8.63 & 4.46 & 0.970 & 1.8 & 2.5 & 3.4 & 16 \\
RG(Skb) & \cite{CompOSE2015}\markb & 2.19 & 10.61 & 12.15 & 0.40 & 0.59 & 8.85 & 4.64 & 0.974 & 1.0 & 1.5 & 5.5 & 25 \\
RG(SkI2) & \cite{CompOSE2015}\markb & 2.16 & 11.12 & 13.64 & 0.27 & 0.41 & 8.45 & 3.89 & 0.914 & 3.5 & 4.8 & 5.0 & 18 \\
RG(SkI3) & \cite{CompOSE2015}\markb & 2.24 & 11.31 & 13.65 & 0.22 & 0.33 & 8.08 & 3.89 & 0.937 & 3.4 & 4.5 & 4.5 & 15 \\
RG(SkI4) & \cite{CompOSE2015}\markb & 2.17 & 10.67 & 12.39 & 0.16 & 0.24 & 8.86 & 4.47 & 0.949 & 0.86 & 1.3 & 2.3 & 11 \\
RG(SkI5) & \cite{CompOSE2015}\markb & 2.24 & 11.47 & 14.22 & 0.36 & 0.53 & 7.96 & 3.76 & 0.932 & 4.7 & 6.3 & 6.0 & 20 \\
RG(SkI6) & \cite{CompOSE2015}\markb & 2.19 & 10.76 & 12.51 & 0.12 & 0.18 & 8.71 & 4.42 & 0.951 & 0.94 & 1.4 & 2.1 & 10 \\
RG(SkMp) & \cite{CompOSE2015}\markb & 2.11 & 10.53 & 12.64 & 0.10 & 0.19 & 9.26 & 4.58 & 0.946 & 1.5 & 2.2 & 4.2 & 20 \\
RG(SkOp) & \cite{CompOSE2015}\markb & 1.97 & 10.13 & 12.47 & 0.31 & 0.43 & 10.2 & 4.72 & 0.906 & 0.91 & 1.4 & 2.0 & 10 \\
RG(SLy2) & \cite{CompOSE2015}\markb & 2.05 & 10.05 & 11.95 & 0.31 & 0.46 & 10.1 & 5.34 & 0.977 & 0.90 & 1.8 & 1.2 & 6.4 \\
RG(SLy4) & \cite{CompOSE2015}\markb & 2.05 & 9.99 & 11.87 & 0.30 & 0.44 & 10.2 & 5.46 & 0.985 & 1.1 & 2.1 & 1.4 & 6.9 \\
RG(SLy9) & \cite{CompOSE2015}\markb & 2.16 & 10.63 & 12.57 & 0.23 & 0.37 & 9.01 & 4.59 & 0.952 & 0.69 & 1.1 & 1.1 & 5.8 \\
RG(SLy230a) & \cite{CompOSE2015}\markb & 2.10 & 10.25 & 11.91 & 0.31 & 0.48 & 9.60 & 4.95 & 0.948 & 1.1 & 1.8 & 2.2 & 5.6 \\
VGBCMR(D1M*) & \cite{CompOSE2015}\markb & 2.00 & 10.21 & 11.77 & 0.22 & 0.37 & 9.72 & 4.28 & 0.855 & 1.6 & 2.7 & 4.4 & 15 \\
XMLSLZ(DD-LZ1) & \cite{CompOSE2015}\markb & 2.56 & 12.28 & 13.05 & 0.14 & 0.28 & 6.38 & 3.12 & 0.875 & 3.7 & 7.5 & 5.9 & 16 \\
XMLSLZ(DDME-X) & \cite{CompOSE2015}\markb & 2.56 & 12.35 & 13.28 & 0.15 & 0.33 & 6.36 & 3.10 & 0.876 & 3.1 & 6.2 & 4.1 & 7.1 \\
XMLSLZ(GM1) & \cite{CompOSE2015}\markb & 2.36 & 11.94 & 13.73 & 0.21 & 0.43 & 7.08 & 3.17 & 0.851 & 0.52 & 1.2 & 4.0 & 20 \\
XMLSLZ(MTVTC) & \cite{CompOSE2015}\markb & 2.02 & 10.90 & 13.32 & 0.14 & 0.26 & 8.87 & 3.46 & 0.803 & 2.0 & 2.9 & 2.5 & 9.0 \\
XMLSLZ(NL3) & \cite{CompOSE2015}\markb & 2.77 & 13.28 & 14.57 & 0.64 & 0.95 & 5.53 & 2.81 & 0.890 & 3.2 & 6.3 & 7.2 & 33 \\
XMLSLZ(PK1) & \cite{CompOSE2015}\markb & 2.31 & 12.66 & 14.40 & 0.82 & 1.2 & 6.33 & 2.14 & 0.682 & 3.1 & 4.4 & 3.4 & 12 \\
XMLSLZ(PKDD) & \cite{CompOSE2015}\markb & 2.33 & 11.77 & 13.69 & 0.61 & 0.85 & 7.33 & 3.30 & 0.852 & 0.31 & 1.0 & 2.5 & 7.5 \\
XMLSLZ(TM1) & \cite{CompOSE2015}\markb & 2.18 & 12.37 & 14.37 & 0.72 & 1.1 & 6.72 & 2.11 & 0.659 & 1.8 & 2.4 & 1.9 & 7.9 \\
XMLSLZ(TW99) & \cite{CompOSE2015}\markb & 2.08 & 10.62 & 12.37 & 0.49 & 0.68 & 9.06 & 3.96 & 0.839 & 1.3 & 2.1 & 3.4 & 8.8 \\
YHSHP(PAL1\_120) & \cite{Yakovlev+2011} & 1.47 & 9.18 & 13.08 & 1.3 & 1.4 & 13.8 & 4.09 & 0.699 & 1.5 & 3.7 & 2.0 & 8.0 \\
YHSHP(PAL1\_180) & \cite{Yakovlev+2011} & 1.74 & 9.93 & 13.06 & 0.60 & 0.71 & 11.2 & 3.97 & 0.771 & 1.1 & 1.7 & 1.8 & 8.5 \\
YHSHP(PAL1\_240) & \cite{Yakovlev+2011} & 1.95 & 10.60 & 13.18 & 0.28 & 0.38 & 9.50 & 3.70 & 0.811 & 1.5 & 1.9 & 2.0 & 6.1 \\
YHSHP(PAL2\_120) & \cite{Yakovlev+2011} & 1.48 & 9.72 & 14.32 & 0.32 & 0.49 & 12.6 & 3.47 & 0.677 & 2.6 & 3.8 & 2.7 & 3.9 \\
YHSHP(PAL2\_180) & \cite{Yakovlev+2011} & 1.75 & 10.35 & 14.11 & 0.18 & 0.33 & 10.5 & 3.52 & 0.753 & 3.1 & 4.1 & 2.9 & 6.0 \\
YHSHP(PAL2\_240) & \cite{Yakovlev+2011} & 1.97 & 10.97 & 14.06 & 0.14 & 0.27 & 9.03 & 3.35 & 0.795 & 3.2 & 4.3 & 3.1 & 4.3 \\
YHSHP(PAL2\_400\_0) & \cite{Yakovlev+2011} & 2.85 & 12.69 & 14.40 & 0.19 & 0.41 & 6.00 & 4.49 & 1.23 & 3.1 & 4.2 & 10 & 42 \\
YHSHP(PAL3\_120) & \cite{Yakovlev+2011} & 1.42 & 8.45 & 11.56 & 1.4 & 1.6 & 15.9 & 5.14 & 0.729 & 4.8 & 9.0 & 4.8 & 9.0 \\
YHSHP(PAL3\_180) & \cite{Yakovlev+2011} & 1.69 & 9.36 & 11.89 & 0.55 & 0.63 & 12.3 & 4.66 & 0.799 & 1.8 & 3.6 & 2.5 & 11 \\
YHSHP(PAL3\_240) & \cite{Yakovlev+2011} & 1.91 & 10.12 & 12.22 & 0.16 & 0.26 & 10.2 & 4.18 & 0.828 & 0.50 & 0.99 & 2.4 & 12 \\
YHSHP(PAL3\_260\_01) & \cite{Yakovlev+2011} & 2.06 & 10.41 & 12.32 & 0.12 & 0.24 & 9.47 & 4.46 & 0.907 & 0.52 & 0.81 & 1.2 & 6.2 \\
YHSHP(PAL3\_300) & \cite{Yakovlev+2011} & 1.83 & 10.51 & 12.43 & 0.16 & 0.23 & 9.47 & 3.02 & 0.709 & 0.40 & 0.66 & 4.3 & 24 \\
YHSHP(PAL3\_300\_0) & \cite{Yakovlev+2011} & 2.34 & 11.08 & 12.68 & 0.11 & 0.19 & 8.07 & 4.69 & 1.04 & 1.2 & 1.7 & 1.8 & 4.8 \\
YHSHP(PAL3\_400) & \cite{Yakovlev+2011} & 2.83 & 12.32 & 13.31 & 0.26 & 0.70 & 6.20 & 4.92 & 1.26 & 1.1 & 1.7 & 6.3 & 18 \\
YHSHP(PAPAL\_120) & \cite{Yakovlev+2011} & 1.44 & 8.74 & 12.24 & 1.5 & 1.7 & 15.0 & 4.70 & 0.717 & 3.3 & 6.8 & 3.5 & 6.8 \\
YHSHP(PAPAL\_180) & \cite{Yakovlev+2011} & 1.71 & 9.59 & 12.41 & 0.67 & 0.77 & 11.9 & 4.37 & 0.786 & 1.0 & 2.1 & 1.8 & 8.6 \\
YHSHP(PAPAL\_240) & \cite{Yakovlev+2011} & 1.93 & 10.32 & 12.64 & 0.27 & 0.36 & 9.92 & 3.98 & 0.822 & 0.61 & 0.95 & 1.9 & 8.3 \\
MKV(KVOR) & \cite{MKV2016} & 2.02 & 10.72 & 12.92 & 0.40 & 0.55 & 9.10 & 3.68 & 0.813 & 0.58 & 1.9 & 2.2 & 5.5 \\
MKV(KVORcut02) & \cite{MKV2016} & 2.27 & 12.61 & 13.74 & 0.32 & 0.47 & 6.23 & 1.99 & 0.692 & 1.8 & 5.7 & 9.1 & 22 \\
MKV(KVORcut03) & \cite{MKV2016} & 2.17 & 11.58 & 12.97 & 0.60 & 0.77 & 7.49 & 2.74 & 0.741 & 2.9 & 5.0 & 5.4 & 15 \\
MKV(KVORcut04) & \cite{MKV2016} & 2.09 & 10.97 & 12.90 & 1.2 & 1.6 & 8.54 & 3.42 & 0.780 & 2.8 & 6.4 & 5.1 & 10 \\
MKV(MKVOR) & \cite{MKV2016} & 2.34 & 11.33 & 12.07 & 0.37 & 0.57 & 7.52 & 3.54 & 0.846 & 5.3 & 8.0 & 10 & 28 \\
    \hline
    \multicolumn{14}{c}{Models with hyperons \& $\Delta$-isobars}\\
    G(GNH3) & \cite{Read+2009,OzelFreire2016}\marka & 1.96 & 11.39 & 14.62 & 0.49 & 0.81 & 8.49 & 2.85 & 0.747 & 4.2 & 6.9 & 4.7 & 14 \\
LNO(H1) & \cite{Read+2009,OzelFreire2016}\marka & 1.55 & 10.97 & 13.88 & 1.5 & 1.9 & 9.16 & 1.95 & 0.541 & 3.4 & 5.4 & 4.3 & 17 \\
LNO(H2) & \cite{Read+2009,OzelFreire2016}\marka & 1.67 & 11.51 & 13.87 & 1.1 & 1.5 & 8.02 & 1.71 & 0.564 & 2.8 & 4.4 & 3.5 & 9.8 \\
LNO(H3) & \cite{Read+2009,OzelFreire2016}\marka & 1.79 & 11.85 & 14.04 & 0.52 & 0.74 & 7.52 & 1.68 & 0.518 & 1.6 & 2.6 & 3.2 & 9.3 \\
LNO(H4) & \cite{Read+2009,OzelFreire2016}\marka & 2.03 & 11.76 & 13.69 & 0.77 & 1.1 & 7.51 & 2.30 & 0.653 & 1.8 & 3.3 & 3.7 & 11 \\
LNO(H5) & \cite{Read+2009,OzelFreire2016}\marka & 1.73 & 11.31 & 13.38 & 1.3 & 1.8 & 8.19 & 1.92 & 0.596 & 2.9 & 4.4 & 3.7 & 9.0 \\
LNO(H7) & \cite{Read+2009,OzelFreire2016}\marka & 1.68 & 10.85 & 13.42 & 1.2 & 1.7 & 9.18 & 2.32 & 0.630 & 3.4 & 5.1 & 3.5 & 8.8 \\
RSGMT(QMC700) & \cite{OzelFreire2016}\marka & 1.98 & 12.36 & 12.70 & 2.4 & 3.5 & 5.99 & 1.17 & 0.404 & 26 & 42 & 23 & 42 \\
BG(BGN1H1) & \cite{Read+2009,OzelFreire2016}\marka & 1.63 & 9.33 & 13.17 & 7.8 & 12 & 13.6 & 5.48 & 0.876 & 10 & 48 & 26 & 80 \\
BG(BGN1H2) & \cite{HPY2007} & 1.57 & 9.04 & 13.00 & 9.8 & 14 & 14.6 & 5.95 & 0.897 & 12 & 53 & 32 & 100 \\
BG(BGN2H1) & \cite{HPY2007} & 1.82 & 9.57 & 13.43 & 6.3 & 11 & 12.4 & 6.36 & 1.01 & 6.8 & 27 & 34 & 120 \\
BG(BGN2H2) & \cite{HPY2007} & 1.74 & 9.12 & 13.42 & 7.0 & 11 & 13.9 & 7.45 & 1.05 & 10 & 41 & 45 & 160 \\
DNS(CMF) & \cite{CompOSE2015}\markb & 2.07 & 11.88 & 13.51 & 0.32 & 0.44 & 7.19 & 2.19 & 0.686 & 0.96 & 1.9 & 4.6 & 18 \\
DS(CMF)-1 & \cite{CompOSE2015}\markb & 2.07 & 11.88 & 13.51 & 0.32 & 0.44 & 7.20 & 2.19 & 0.686 & 0.96 & 2.0 & 4.6 & 18 \\
DS(CMF)-2 & \cite{CompOSE2015}\markb & 2.13 & 12.00 & 13.60 & 0.39 & 0.57 & 7.05 & 2.25 & 0.692 & 0.33 & 1.1 & 5.6 & 20 \\
DS(CMF)-3 & \cite{CompOSE2015}\markb & 2.00 & 11.54 & 13.17 & 0.17 & 0.21 & 7.65 & 2.32 & 0.685 & 1.1 & 2.5 & 2.0 & 7.5 \\
DS(CMF)-4 & \cite{CompOSE2015}\markb & 2.05 & 11.59 & 13.24 & 0.21 & 0.31 & 7.59 & 2.41 & 0.689 & 0.37 & 0.96 & 2.3 & 8.5 \\
DS(CMF)-5 & \cite{CompOSE2015}\markb & 2.07 & 11.48 & 13.20 & 0.36 & 0.45 & 7.74 & 2.65 & 0.753 & 1.6 & 2.5 & 2.3 & 8.3 \\
DS(CMF)-6 & \cite{CompOSE2015}\markb & 2.11 & 11.57 & 13.27 & 0.33 & 0.47 & 7.62 & 2.65 & 0.744 & 0.79 & 1.1 & 2.6 & 9.4 \\
DS(CMF)-7 & \cite{CompOSE2015}\markb & 2.07 & 11.49 & 13.20 & 0.35 & 0.43 & 7.73 & 2.64 & 0.752 & 1.6 & 2.5 & 2.3 & 8.3 \\
DS(CMF)-8 & \cite{CompOSE2015}\markb & 2.09 & 11.58 & 13.26 & 0.39 & 0.50 & 7.60 & 2.58 & 0.744 & 1.5 & 2.3 & 2.8 & 9.0 \\
GHK(GM1A) & \cite{Ofengeim+2019} & 1.99 & 11.94 & 13.63 & 1.2 & 2.0 & 7.18 & 2.02 & 0.686 & 4.4 & 6.3 & 6.7 & 15 \\
GHK(TM1C) & \cite{Ofengeim+2019} & 2.05 & 12.47 & 14.41 & 0.25 & 0.39 & 6.58 & 1.77 & 0.644 & 1.8 & 2.7 & 3.3 & 8.2 \\
GHK(GM1'B) & \cite{Gusakov+2014} & 2.01 & 11.46 & 14.00 & 0.76 & 1.4 & 8.13 & 2.78 & 0.748 & 4.4 & 6.6 & 4.4 & 6.6 \\
OGHF(FSU2H) & \cite{Ofengeim+2019} & 1.99 & 11.98 & 13.03 & 0.96 & 1.5 & 6.90 & 1.83 & 0.637 & 2.0 & 5.6 & 10 & 29 \\
OGHF(NL3$\omega\rho$) & \cite{Ofengeim+2019} & 2.71 & 12.94 & 13.69 & 0.090 & 0.22 & 5.73 & 2.85 & 0.889 & 3.5 & 8.3 & 7.5 & 15 \\
OPGR(GM1Y4) & \cite{CompOSE2015}\markb & 1.79 & 13.01 & 13.52 & 1.2 & 1.6 & 5.23 & 0.823 & 0.424 & 6.7 & 17 & 33 & 76 \\
OPGR(GM1Y5) & \cite{CompOSE2015}\markb & 2.12 & 12.30 & 13.64 & 0.45 & 0.91 & 6.61 & 1.91 & 0.667 & 2.2 & 6.6 & 8.3 & 24 \\
OPGR(GM1Y6) & \cite{CompOSE2015}\markb & 2.29 & 12.12 & 13.69 & 0.34 & 0.49 & 6.84 & 2.64 & 0.808 & 2.3 & 3.5 & 5.5 & 22 \\
R(DD2Y$\Delta$)1.1-1.1 & \cite{CompOSE2015}\markb & 2.04 & 11.21 & 13.11 & 0.26 & 0.41 & 8.22 & 2.89 & 0.741 & 0.43 & 1.4 & 3.6 & 17 \\
R(DD2Y$\Delta$)1.2-1.1 & \cite{CompOSE2015}\markb & 2.05 & 10.98 & 12.41 & 0.99 & 1.2 & 8.39 & 3.05 & 0.731 & 4.0 & 6.8 & 9.8 & 32 \\
R(DD2Y$\Delta$)1.2-1.3 & \cite{CompOSE2015}\markb & 2.03 & 11.44 & 13.14 & 1.2 & 1.7 & 7.89 & 2.63 & 0.739 & 2.7 & 4.0 & 5.9 & 18 \\
MKV(KVORHcut03$\phi$) & \cite{MKV2016} & 1.98 & 11.76 & 12.98 & 0.79 & 1.1 & 7.17 & 1.93 & 0.630 & 4.4 & 13 & 6.9 & 14 \\
MKV(KVORHcut03H$\phi\sigma$) & \cite{MKV2016} & 2.08 & 11.94 & 12.97 & 1.4 & 2.0 & 6.85 & 1.93 & 0.621 & 8.6 & 16 & 8.8 & 16 \\
MKV(MKVORH$\phi$) & \cite{MKV2016} & 2.23 & 11.23 & 12.03 & 0.13 & 0.27 & 7.73 & 3.23 & 0.790 & 4.4 & 7.7 & 10 & 29 \\
MKV(MKVORH$\phi\sigma$) & \cite{MKV2016} & 2.30 & 11.43 & 12.05 & 0.60 & 0.89 & 7.35 & 3.09 & 0.782 & 7.2 & 13 & 12 & 30 \\
    \hline
    \multicolumn{14}{c}{Hybrid models}\\
    ABPR(ALF1) & \cite{Read+2009,OzelFreire2016}\marka & 1.49 & 9.21 & 10.35 & 1.4 & 1.6 & 11.9 & 2.95 & 0.565 & 9.0 & 22 & 17 & 51 \\
ABPR(ALF2) & \cite{Read+2009,OzelFreire2016}\marka & 2.09 & 11.96 & 12.86 & 0.58 & 0.87 & 6.94 & 1.95 & 0.553 & 5.7 & 7.4 & 15 & 45 \\
ABPR(ALF3) & \cite{Read+2009,OzelFreire2016}\marka & 1.47 & 9.50 & 11.43 & 1.7 & 2.0 & 11.6 & 2.75 & 0.565 & 5.5 & 12 & 11 & 37 \\
ABPR(ALF4) & \cite{Read+2009,OzelFreire2016}\marka & 1.94 & 10.88 & 11.58 & 1.6 & 2.3 & 8.28 & 2.33 & 0.506 & 14 & 21 & 17 & 35 \\
PCL(PCL2) & \cite{Read+2009,OzelFreire2016}\marka & 1.48 & 10.18 & 13.44 & 2.1 & 2.6 & 10.8 & 2.54 & 0.599 & 4.6 & 7.6 & 5.5 & 21 \\
BFH(QHC19-A) & \cite{CompOSE2015}\markb & 1.93 & 10.23 & 11.49 & 0.32 & 0.59 & 9.63 & 3.66 & 0.781 & 3.3 & 7.8 & 8.0 & 22 \\
BFH(QHC19-B) & \cite{CompOSE2015}\markb & 2.07 & 10.55 & 11.51 & 0.54 & 0.76 & 8.82 & 3.59 & 0.799 & 4.9 & 9.9 & 10 & 23 \\
BFH(QHC19-C) & \cite{CompOSE2015}\markb & 2.18 & 10.74 & 11.53 & 1.1 & 1.5 & 8.37 & 3.70 & 0.815 & 7.7 & 14 & 13 & 25 \\
BFH(QHC19-D) & \cite{CompOSE2015}\markb & 2.28 & 10.86 & 11.54 & 1.5 & 2.1 & 8.12 & 3.92 & 0.833 & 11 & 18 & 15 & 26 \\
BFH(QHC18) & \cite{CompOSE2015}\markb & 2.04 & 10.38 & 11.62 & 1.5 & 2.0 & 9.22 & 3.88 & 0.804 & 6.9 & 16 & 13 & 31 \\
DS(CMF)-1 & \cite{CompOSE2015}\markb & 1.97 & 11.19 & 13.49 & 3.3 & 5.8 & 8.54 & 2.98 & 0.784 & 9.7 & 19 & 12 & 19 \\
DS(CMF)-2 & \cite{CompOSE2015}\markb & 1.96 & 11.14 & 13.55 & 4.1 & 7.2 & 8.70 & 3.08 & 0.787 & 10 & 21 & 14 & 24 \\
DS(CMF)-3 & \cite{CompOSE2015}\markb & 1.99 & 11.24 & 13.17 & 1.5 & 2.8 & 8.21 & 2.78 & 0.777 & 6.6 & 14 & 6.2 & 14 \\
DS(CMF)-4 & \cite{CompOSE2015}\markb & 1.98 & 11.24 & 13.23 & 2.0 & 3.8 & 8.27 & 2.82 & 0.778 & 8.2 & 16 & 8.0 & 16 \\
DS(CMF)-5 & \cite{CompOSE2015}\markb & 2.02 & 11.88 & 13.19 & 0.66 & 1.1 & 6.92 & 1.93 & 0.727 & 5.3 & 9.5 & 4.8 & 11 \\
DS(CMF)-6 & \cite{CompOSE2015}\markb & 2.01 & 11.99 & 13.25 & 0.54 & 1.2 & 6.76 & 1.84 & 0.728 & 7.2 & 13 & 6.4 & 13 \\
DS(CMF)-7 & \cite{CompOSE2015}\markb & 2.02 & 11.89 & 13.19 & 0.66 & 1.1 & 6.92 & 1.92 & 0.727 & 5.5 & 10 & 4.9 & 11 \\
DS(CMF)-8 & \cite{CompOSE2015}\markb & 2.02 & 11.89 & 13.25 & 0.30 & 1.3 & 6.92 & 1.92 & 0.727 & 6.6 & 13 & 5.8 & 13 \\
JJ(VQCD)intermediate & \cite{CompOSE2015}\markb\markc & 2.15 & 11.80 & 12.15 & 0.97 & 1.9 & 6.12 & 1.92 & 0.689 & 1.1 & 1.6 & 16 & 49 \\
JJ(VQCD)soft & \cite{CompOSE2015}\markb\markc & 2.03 & 11.85 & 12.06 & 1.7 & 2.8 & 5.48 & 1.48 & 0.648 & 3.7 & 6.4 & 19 & 51 \\
JJ(VQCD)stiff & \cite{CompOSE2015}\markb\markc & 2.33 & 11.86 & 12.32 & 0.19 & 0.30 & 6.81 & 2.68 & 0.763 & 5.4 & 12 & 14 & 44 \\
KBH(QHC21\_A) & \cite{CompOSE2015}\markb & 2.18 & 11.24 & 12.27 & 0.20 & 0.33 & 7.83 & 3.15 & 0.798 & 3.5 & 12 & 7.9 & 21 \\
KBH(QHC21\_AT) & \cite{CompOSE2015}\markb & 2.13 & 10.87 & 11.64 & 0.26 & 0.44 & 8.25 & 3.35 & 0.798 & 4.3 & 12 & 10 & 26 \\
KBH(QHC21\_B) & \cite{CompOSE2015}\markb & 2.25 & 11.41 & 12.28 & 0.37 & 0.59 & 7.49 & 3.08 & 0.803 & 5.0 & 15 & 9.5 & 24 \\
KBH(QHC21\_BT) & \cite{CompOSE2015}\markb & 2.20 & 11.07 & 11.65 & 0.14 & 0.23 & 7.86 & 3.25 & 0.803 & 5.7 & 15 & 12 & 28 \\
KBH(QHC21\_C) & \cite{CompOSE2015}\markb & 2.31 & 11.56 & 12.28 & 0.60 & 0.88 & 7.22 & 3.03 & 0.808 & 6.5 & 18 & 11 & 26 \\
KBH(QHC21\_CT) & \cite{CompOSE2015}\markb & 2.26 & 11.13 & 11.60 & 0.54 & 0.77 & 7.68 & 3.31 & 0.812 & 8.1 & 17 & 14 & 30 \\
KBH(QHC21\_D) & \cite{CompOSE2015}\markb & 2.36 & 11.69 & 12.27 & 0.82 & 1.2 & 6.99 & 3.01 & 0.813 & 8.2 & 20 & 13 & 28 \\
KBH(QHC21\_DT) & \cite{CompOSE2015}\markb & 2.32 & 11.28 & 11.62 & 0.70 & 0.99 & 7.41 & 3.27 & 0.817 & 9.7 & 20 & 15 & 32 \\
OOS(DD2\_FRG)(2) & \cite{CompOSE2015}\markb & 2.05 & 12.53 & 13.08 & 1.4 & 1.9 & 5.92 & 1.36 & 0.551 & 10 & 31 & 15 & 36 \\
OOS(DD2\_FRG)(2+1) & \cite{CompOSE2015}\markb & 1.84 & 12.75 & 13.03 & 2.1 & 2.7 & 5.33 & 0.897 & 0.466 & 11 & 31 & 26 & 56 \\
OOS(DD2-FRG)vec(2) & \cite{CompOSE2015}\markb & 2.12 & 12.69 & 13.09 & 1.9 & 2.8 & 5.67 & 1.32 & 0.559 & 17 & 39 & 17 & 38 \\
OOS(DD2-FRG)vec(2+1) & \cite{CompOSE2015}\markb & 1.93 & 12.98 & 13.05 & 2.8 & 3.8 & 4.99 & 0.844 & 0.467 & 21 & 53 & 31 & 61 
\end{longtable}

\setcounter{table}{0}
\setcounter{figure}{0}
\section{Methods}
\label{sec:app:methods}
We infer constraints on the model parameters using the Bayesian approach, e.g., \cite{Annala+Nat2023, Brandes+PRD2023, Brandes2023PhRvDa, Ayriyan2021EPJA, Raaijmakers+2021, Brandes+2024mar, Fan+PRD2024}. The construction is standard, and we briefly describe the key points here.
We are interested in the posterior probability $ \mathcal{P}(\bm{\theta}|\mathrm{data};\,\mathcal{M}) $ of the model parameters $\theta$ given the observed data and the model $
\mathcal{M}$ which in our case is the $M-R$ curve parameterization described in the main text. According to the Bayes rule, the posterior probability of the parameters given data can be expressed as
\begin{equation}\label{eq:Bayes_general}
    \mathcal{P}(\bm{\theta}|\mathrm{data};\,\mathcal{M}) =\frac{\mathcal{L}(\mathrm{data}|\bm{\theta};\ \mathcal{M})\pi(\bm{\theta}|\mathcal{M})}{p(\mathrm{data}|\mathcal{M})},
\end{equation}
where 
vector of model parameters
assuming $N$ pairs of  mass and radii measurements is 
\begin{equation}
    \bm{\theta}= \left\{M_i,\,R_i\right\}_{i=1}^N\bigcup\bm{\theta}_{\mathrm{EoS}},\quad \bm{\theta}_{\mathrm{EoS}}=(\Pmax,\,\rhomax,\, r_{1/2}).
\end{equation}
The inference on individual $\{M_i,\,R_i\}$ is updated given the structural information contained in the model $\mathcal{M}$ therefore, they are treated as the model parameters as well as the EoS parameters $\bm{\theta}_{\mathrm{EoS}}$. In Eq.~(\ref{eq:Bayes_general}), $\mathcal{L}(\mathrm{data}|\bm{\theta};\ \mathcal{M})$ is the observed data likelihood, $\pi(\bm{\theta}|\mathcal{M})$ is the prior probability density of $\bm{\theta}$ (given the model) and $p(\mathrm{data}|\mathcal{M})$ is a normalization factor (or the evidence -- the prior probability of data given the model). The likelihood of the data for each source depends only on its mass and radius. Thus one can express $\mathcal{L}(\mathrm{data}|\bm{\theta};\ \mathcal{M}) = \prod_{i=1}^N \mathcal{L}_i(\mathrm{data_i}|M_i,\,R_i)$. Then Eq.~(\ref{eq:Bayes_general}) becomes
\begin{equation}\label{eq:Post_fin}
    \mathcal{P}(\bm{\theta}|\mathrm{data};\,\mathcal{M})=\frac{\pi(\bm{\theta}_{\mathrm{EoS}}| \mathcal{M})}{p(\mathrm{data}|\mathcal{M})} 
    \prod_i \mathcal{L}_i(\mathrm{data}_i|M_i,\,R_i)  \mathcal{P}(M_i,\,R_i|\bm{\theta}_{\mathrm{EoS}},\, \mathcal{M}) ,
\end{equation}
where $\mathcal{P}(M_i,\,R_i|\bm{\theta}_{\mathrm{EoS}},\, \mathcal{M})$ is the probability of a particular source having mass $M_i$ and radius $R_i$ given the EoS parameter vector $\bm{\theta}_{\mathrm{EoS}}$, and $\pi(\bm{\theta}_{\mathrm{EoS}}| \mathcal{M})$ is the prior probability of the EoS parameters. Both terms under the product sign require specific consideration.  
The likelihood for the individual source can be constructed from scratch based on the data generation statistical model, basically repeating the analysis of the original works on the source, which is usually unfeasible in practice. Alternatively, the likelihood needs to be deduced from the published analysis results which are usually available in the form of the posterior probabilities $\mathcal{P}(M_i,R_i|\mathrm{data}_i)$ if Bayesian approach is used or in the form of confidence regions or confidence distributions \cite{Schweder_Hjort_2016} if the frequentist approach is adopted. In rare cases, the likelihood function itself is reported, like in the case of the \textsc{toast} likelihood for the GW170817 \cite{Hernandez2020MNRAS}, see below. When Bayesian approach was used for the $i$th source in the reference study, the likelihood can be expressed
via the Bayes rule
\begin{equation}\label{eq:Likei}
\mathcal{L}_i(\mathrm{data_i}|M_i,\,R_i)=\frac{p(\mathrm{data}_i)\mathcal{P}(M_i,\,R_i|\mathrm{data})}{\pi_{\mathrm{ref}}(M_i,R_i)}. 
\end{equation}
Therefore, either the prior $\pi_{\mathrm{ref}}(M_i,\,R_i)$ of the reference study needs to be explicitly specified, or the likelihood function can be approximated by the posterior probability itself provided the prior is sufficiently 
flat in the region of support of the final posterior probability (so it can be treated as a constant during marginalization). 


The conditional probability $\mathcal{P}(M_i,\,R_i|\bm{\theta}_{\mathrm{EoS}})$ is also a subtle term since it involves conditioning over a null set. Depending on a choice of a parameterization, the relationship between $M_i,\, R_i$ and $\bm{\theta}_{\mathrm{Eos}}$ can be expressed in different ways, leading, in principle, to different inferences. For instance, specification of the model as the relation $R=R_i(M_i,\,\bm{\theta}_{\mathrm{EoS}})$ supplemented with prior distributions on $\bm{\theta}_{\mathrm{EoS}}$ and 
$M_i$ induces a specific prior distribution on $R_i$ which is not expected to be uniform or in any other ways non-informative alike. Adopting a different parameterization of the $M-R$ curve would result in different $M_i,\,R_i$ prior distribution given the structural model $\mathcal{M}$ and, consequently, in different posterior distributions.
This difficulty is a manifestation of the 
Borel-Kolmogorov paradox, e.g., \cite{Bungert2020arXiv}. One of the strategies to overcome it is in some way to combine, or `meld' the induced priors with the prior distributions on $M_i,\, R_i$ specified before the one is informed about the model $\mathcal{M}$ \cite{Poole2000} putting in a sense a selection of a specific parametrization in Bayesian context. 
In the astrophysical context, the dependence of the inference on low-dimensional relations embedded in higher-dimension space on the parameterization of this space and a related problem of choice of priors on the parameters is discussed, e.g., in
\cite{Steiner2018arXiv,Pihajoki2017MNRAS}.

Having said this we nevertheless follow the previous studies and formulate the $M-R$ relationship  in a form $R=R_i(M_i,\,\bm{\theta}_{\mathrm{EoS}})$ with uniform prior on $M_i$ in the range 
$M_{\mathrm{min}}...\Mmax(\bm{\theta}_\mathrm{EoS})$, 
where $M_{\mathrm{min}}=1M_\odot$ in our case. In a sense, a specific choice of the parametrization can be regarded as a part of the structure model. It is believed that it is the data likelihood that mainly determines the posteriors; therefore, variations in a choice of the parametrization would not produce much harm. An alternative would be to select some physics-motivated prior on $M_i$, e.g., equate it to the observed NS mass distribution. Given the observed NS distribution and assuming that the structural relationship $\mathcal{M}$ is the correct one, we then 
\textit{a priori} expect a specific NS radii distribution. However, the sample of the NSs with measured masses as well as the sample of NSs for which $M-R$ measurements exist are not fair samples from the NS population due to unquantified selection effects.
In the absence of a better solution, the conditional probability for mass and radius of $i$th source given EoS parameters is then
\begin{equation}\label{eq:Peos}
    \mathcal{P}(M_i,\,R_i|\bm{\theta}_{\mathrm{EoS}})=
    \mathcal{P}(R_i| M_i,\,\bm{\theta}_{\mathrm{EoS}})\mathcal{P}(M_i|\bm{\theta}_{\mathrm{EoS}})=
    \delta\left(R_i-R(M_i,\bm{\theta}_{\mathrm{EoS}})\right)\frac{\Theta(\Mmax(\bm{\theta}_{\mathrm{EoS}})-M_i)\Theta(M_i-M_{\mathrm{min}})}{\Mmax(\bm{\theta}_{\mathrm{EoS}})-M_{\mathrm{min}}},
\end{equation}
with $\Theta(x)$ for the Heaviside step function.
Inserting Eqs.~(\ref{eq:Likei}) and (\ref{eq:Peos}) in Eq.~(\ref{eq:Post_fin}) we can infer the posterior probability densities on the parameters $\bm{\theta}$.
For NICER and CasA data, the samples from posterior distributions on $M_i$ and $R_i$ are available \cite{Riley+ApJL2021,Miller2019ApJ0030,Choudhury2024, Shternin2023MNRAS} and we use the kernel density estimation (KDE) for  $\mathcal{P}(M_i,R_i|\mathrm{data}_i)$ in Eq.~(\ref{eq:Likei}). For the GW data we also use 4D KDE of the $M_{\mathrm{GW}1},\,R_{\mathrm{GW}1},\,M_{\mathrm{GW}2},\,R_{\mathrm{GW}2}$
samples from Ref.~\cite{Raaijmakers+2021}. In the latter case it is also possible to use directly the marginalized (overall nuisance parameters of GW wavefront modeling) likelihood $L_{\mathrm{GW}}(\mathrm{data}|M_{\mathrm{GW}1},\,R_{\mathrm{GW}1},\,M_{\mathrm{GW}2},\,R_{\mathrm{GW}2})$ provided by Ref.~\cite{Hernandez2020MNRAS}. We checked, and both approaches give similar results.

When only the mass $M_i$ of a given source is measured, the sole information one gets is that $\Mmax>M_{i}$. Then Eq.~(\ref{eq:Peos}) can be marginalized over $R_i$ giving
\begin{equation}\label{eq:PMeos}
    \mathcal{P}(M_i|\bm{\theta}_{\mathrm{EoS}})=\mathcal{P}(M_i|\Mmax)=\frac{\Theta(\Mmax(\bm{\theta}_{\mathrm{EoS}})-M_i)\Theta(M_i-M_{\mathrm{min}})}{\Mmax(\bm{\theta}_{\mathrm{EoS}})-M_{\mathrm{min}}}.
\end{equation}
The likelihood of such observation, marginalized over $M_i$, is then
\begin{equation}
    \mathcal{L}(\mathrm{data}_i|\bm{\theta}_{\mathrm{EoS}})=\mathcal{L}(\mathrm{data_i}|\Mmax)=\int\mathrm{d} M_i\, \mathcal{L}(\mathrm{data}_i|M_i)\mathcal{P}(M_i|\Mmax)\propto \mathrm{CDF}_i(\Mmax)
\end{equation}
if we approximate likelihood $\mathcal{L}(\mathrm{data}_i|M_i)$ as the posterior distribution of $M_i$ as above. Here $\mathrm{CDF}_i(\Mmax)$ is the cumulative posterior distribution for $M_i$, i.e., the posterior probability that $M_{i}<\Mmax$ given the data. When the inference is summarized in the form of a frequentists confidence interval, we assume the normal (Gaussian) likelihood \cite{Schweder_Hjort_2016}, which leads to the same expression with $\mathrm{CDF}_i(\Mmax)$ being the normal distribution cumulative distribution function (namely, the error function).
Likewise, for the kilonova constraint, where $\Mmax$ is limited from above, the complementary cumulative distribution function is used
\begin{equation}
    \mathcal{P}_{\mathrm{kilonova}}(\mathrm{data}|\Mmax)\propto 1-\mathrm{CDF}_{\mathrm{kilonova}}(\Mmax).
\end{equation}



Apart from the causality condition (see Sec.~III in the main text) and priors on $M_i$ described above, we use the following weak uniform non-informative priors on the rest of model parameters: $1<r_{1/2}<2$, $0.1<\Pmax/\rho_0<100$, and $0.1<\rhomax/\rho_0<100$. Additionally, we set the conservative limit $\Mmax < 3\,$\Msun. To sample from posteriors and obtain credible regions for the model parameters we employ affine-invariant Markov Chain Monte Carlo (MCMC) sampler \textsc{emcee} \cite{Foreman-Mackey2013}. Given the number of model parameters (12), 32 walkers were used. The convergence of the MCMC chain was controlled by the autocorrelation time following the \textsc{emcee} manual recommendations.\footnote{\texttt{https://emcee.readthedocs.io/en/stable/tutorials/autocorr/}} The chain is considered to be converged when the autocorrelation time estimate is stable and the number of MCMC steps is more than 50 times the autocorrelation time. A typical number of steps in the converged chain was about 20000, and the first half of the chain was burned in. All credible intervals and regions given in the tables and shown in the plots correspond to the highest posterior density intervals while the point estimates correspond to the best-fit (the mode of the multidimensional posterior) estimated by the separate  maximization procedure.

The model parameters are summarized in Table~\ref{tab:app:params} and marginalized 1D and 2D posterior distributions are presented in Fig.~\ref{fig:app:tri1}. We show the fit parameters $P_{\mathrm{TOV}}\,\rho_{\mathrm{TOV}},\,r_{1/2}$ as well as the derived parameters $M_{\mathrm{TOV}},\,R_{\mathrm{TOV}},\, R_{1/2}$. The uncertainties in Table~\ref{tab:app:params} correspond to 68\% highest posterior density intervals.
We also provide the $68\%$ credible intervals for masses and radii of the individual sources updated during the Bayesian fitting procedure (Table~\ref{tab:app:indMR}), see Sec.~\ref{sec:concl} and Fig.~\ref{fig:individuals}. 

\begin{table*}
    \centering
    \caption{Masses and radii of individual sources updated according to the Bayesian fit. Uncertainties correspond to 68\% highest posterior density credible intervals.}
    \renewcommand{\arraystretch}{1.5}
    \label{tab:app:indMR}
    \begin{tabular}{cccccccccc}
        \hline
		 & GW170817-1 & GW170817-2 & 0740 &  0030 &0437& CasA    & 1702 & 1724 & 1810 \\ 
		\hline
    $M$ [$M_\odot$]&
    $1.48^{+0.06}_{-0.11}$ &
    $1.26^{+0.09}_{-0.05}$ &
    $2.07^{+0.06}_{-0.07}$ &
    
    $1.37^{+0.05}_{-0.11}$ &
    $1.43^{+0.04}_{-0.04}$&
    $1.60^{+0.11}_{-0.14}$ &
    $1.75^{+0.16}_{-0.15}$ &
     $1.42^{+0.32}_{-0.28}$ &
    $1.42^{+0.30}_{-0.29}$ \\
    $R$ [km]& 
    $12.21^{+0.24}_{-0.32}$ &
     $12.12^{+0.26}_{-0.39}$ &
    $12.06^{+0.30}_{-0.27}$ &
   
     $12.18^{+0.25}_{-0.39}$ &
     $12.20^{+0.23}_{-0.33}$&
      $12.24^{+0.22}_{-0.29}$ & 
    $12.25^{+0.22}_{-0.24}$ &
     $12.20^{+0.27}_{-0.37}$ & 
    $12.20^{+0.28}_{-0.35}$ \\
\hline
     \end{tabular}
\end{table*}

\end{widetext}

\begin{thebibliography}{82}%
\makeatletter
\providecommand \@ifxundefined [1]{%
 \@ifx{#1\undefined}
}%
\providecommand \@ifnum [1]{%
 \ifnum #1\expandafter \@firstoftwo
 \else \expandafter \@secondoftwo
 \fi
}%
\providecommand \@ifx [1]{%
 \ifx #1\expandafter \@firstoftwo
 \else \expandafter \@secondoftwo
 \fi
}%
\providecommand \natexlab [1]{#1}%
\providecommand \enquote  [1]{``#1''}%
\providecommand \bibnamefont  [1]{#1}%
\providecommand \bibfnamefont [1]{#1}%
\providecommand \citenamefont [1]{#1}%
\providecommand \href@noop [0]{\@secondoftwo}%
\providecommand \href [0]{\begingroup \@sanitize@url \@href}%
\providecommand \@href[1]{\@@startlink{#1}\@@href}%
\providecommand \@@href[1]{\endgroup#1\@@endlink}%
\providecommand \@sanitize@url [0]{\catcode `\\12\catcode `\$12\catcode `\&12\catcode `\#12\catcode `\^12\catcode `\_12\catcode `\%12\relax}%
\providecommand \@@startlink[1]{}%
\providecommand \@@endlink[0]{}%
\providecommand \url  [0]{\begingroup\@sanitize@url \@url }%
\providecommand \@url [1]{\endgroup\@href {#1}{\urlprefix }}%
\providecommand \urlprefix  [0]{URL }%
\providecommand \Eprint [0]{\href }%
\providecommand \doibase [0]{https://doi.org/}%
\providecommand \selectlanguage [0]{\@gobble}%
\providecommand \bibinfo  [0]{\@secondoftwo}%
\providecommand \bibfield  [0]{\@secondoftwo}%
\providecommand \translation [1]{[#1]}%
\providecommand \BibitemOpen [0]{}%
\providecommand \bibitemStop [0]{}%
\providecommand \bibitemNoStop [0]{.\EOS\space}%
\providecommand \EOS [0]{\spacefactor3000\relax}%
\providecommand \BibitemShut  [1]{\csname bibitem#1\endcsname}%
\let\auto@bib@innerbib\@empty
\bibitem [{\citenamefont {{Degenaar}}\ and\ \citenamefont {{Suleimanov}}(2018)}]{DegenaarSuleimanov2018}%
  \BibitemOpen
  \bibfield  {author} {\bibinfo {author} {\bibfnamefont {N.}~\bibnamefont {{Degenaar}}}\ and\ \bibinfo {author} {\bibfnamefont {V.~F.}\ \bibnamefont {{Suleimanov}}},\ }\bibfield  {title} {\bibinfo {title} {{Testing the Equation of State with Electromagnetic Observations}},\ }in\ \href {https://doi.org/10.1007/978-3-319-97616-7_5} {\emph {\bibinfo {booktitle} {Astrophysics and Space Science Library}}},\ \bibinfo {series} {Astrophysics and Space Science Library}, Vol.\ \bibinfo {volume} {457},\ \bibinfo {editor} {edited by\ \bibinfo {editor} {\bibfnamefont {L.}~\bibnamefont {{Rezzolla}}}, \bibinfo {editor} {\bibfnamefont {P.}~\bibnamefont {{Pizzochero}}}, \bibinfo {editor} {\bibfnamefont {D.~I.}\ \bibnamefont {{Jones}}}, \bibinfo {editor} {\bibfnamefont {N.}~\bibnamefont {{Rea}}}, \bibinfo {editor} {\bibfnamefont {I.}~\bibnamefont {{Vida{\~n}a}}},\ and\ \bibinfo {editor} {\bibnamefont {et~al.}}}\ (\bibinfo {year} {2018})\ p.\ \bibinfo {pages} {185}\BibitemShut {NoStop}%
\bibitem [{\citenamefont {{Tolman}}(1939)}]{Tolman1939}%
  \BibitemOpen
  \bibfield  {author} {\bibinfo {author} {\bibfnamefont {R.~C.}\ \bibnamefont {{Tolman}}},\ }\bibfield  {title} {\bibinfo {title} {{Static Solutions of Einstein's Field Equations for Spheres of Fluid}},\ }\href {https://doi.org/10.1103/PhysRev.55.364} {\bibfield  {journal} {\bibinfo  {journal} {Physical Review}\ }\textbf {\bibinfo {volume} {55}},\ \bibinfo {pages} {364} (\bibinfo {year} {1939})}\BibitemShut {NoStop}%
\bibitem [{\citenamefont {{Oppenheimer}}\ and\ \citenamefont {{Volkoff}}(1939)}]{OppVol1939}%
  \BibitemOpen
  \bibfield  {author} {\bibinfo {author} {\bibfnamefont {J.~R.}\ \bibnamefont {{Oppenheimer}}}\ and\ \bibinfo {author} {\bibfnamefont {G.~M.}\ \bibnamefont {{Volkoff}}},\ }\bibfield  {title} {\bibinfo {title} {{On Massive Neutron Cores}},\ }\href {https://doi.org/10.1103/PhysRev.55.374} {\bibfield  {journal} {\bibinfo  {journal} {Physical Review}\ }\textbf {\bibinfo {volume} {55}},\ \bibinfo {pages} {374} (\bibinfo {year} {1939})}\BibitemShut {NoStop}%
\bibitem [{\citenamefont {{Lindblom}}(1992)}]{Lind1992}%
  \BibitemOpen
  \bibfield  {author} {\bibinfo {author} {\bibfnamefont {L.}~\bibnamefont {{Lindblom}}},\ }\bibfield  {title} {\bibinfo {title} {{Determining the Nuclear Equation of State from Neutron-Star Masses and Radii}},\ }\href {https://doi.org/10.1086/171882} {\bibfield  {journal} {\bibinfo  {journal} {\apj}\ }\textbf {\bibinfo {volume} {398}},\ \bibinfo {pages} {569} (\bibinfo {year} {1992})}\BibitemShut {NoStop}%
\bibitem [{\citenamefont {{Fiorella Burgio}}\ and\ \citenamefont {{Fantina}}(2018)}]{BurgioFantina2018}%
  \BibitemOpen
  \bibfield  {author} {\bibinfo {author} {\bibfnamefont {G.}~\bibnamefont {{Fiorella Burgio}}}\ and\ \bibinfo {author} {\bibfnamefont {A.~F.}\ \bibnamefont {{Fantina}}},\ }\bibfield  {title} {\bibinfo {title} {{Nuclear Equation of State for Compact Stars and Supernovae}},\ }in\ \href {https://doi.org/10.1007/978-3-319-97616-7_6} {\emph {\bibinfo {booktitle} {The Physics and Astrophysics of Neutron Stars}}},\ \bibinfo {series} {Astrophysics and Space Science Library}, Vol.\ \bibinfo {volume} {457},\ \bibinfo {editor} {edited by\ \bibinfo {editor} {\bibfnamefont {L.}~\bibnamefont {{Rezzolla}}}, \bibinfo {editor} {\bibfnamefont {P.}~\bibnamefont {{Pizzochero}}}, \bibinfo {editor} {\bibfnamefont {D.~I.}\ \bibnamefont {{Jones}}}, \bibinfo {editor} {\bibfnamefont {N.}~\bibnamefont {{Rea}}},\ and\ \bibinfo {editor} {\bibfnamefont {I.}~\bibnamefont {{Vida{\~n}a}}}}\ (\bibinfo {year} {2018})\ p.\ \bibinfo {pages} {255},\ \Eprint {https://arxiv.org/abs/1804.03020} {arXiv:1804.03020 [nucl-th]} \BibitemShut {NoStop}%
\bibitem [{\citenamefont {{Andersson}}\ and\ \citenamefont {{Kokkotas}}(1998)}]{AnderssonKokkotasMNRAS1998}%
  \BibitemOpen
  \bibfield  {author} {\bibinfo {author} {\bibfnamefont {N.}~\bibnamefont {{Andersson}}}\ and\ \bibinfo {author} {\bibfnamefont {K.~D.}\ \bibnamefont {{Kokkotas}}},\ }\bibfield  {title} {\bibinfo {title} {{Towards gravitational wave asteroseismology}},\ }\href {https://doi.org/10.1046/j.1365-8711.1998.01840.x} {\bibfield  {journal} {\bibinfo  {journal} {\mnras}\ }\textbf {\bibinfo {volume} {299}},\ \bibinfo {pages} {1059} (\bibinfo {year} {1998})},\ \Eprint {https://arxiv.org/abs/gr-qc/9711088} {arXiv:gr-qc/9711088 [gr-qc]} \BibitemShut {NoStop}%
\bibitem [{\citenamefont {{Lattimer}}\ and\ \citenamefont {{Prakash}}(2001)}]{LattPrakApJ01}%
  \BibitemOpen
  \bibfield  {author} {\bibinfo {author} {\bibfnamefont {J.~M.}\ \bibnamefont {{Lattimer}}}\ and\ \bibinfo {author} {\bibfnamefont {M.}~\bibnamefont {{Prakash}}},\ }\bibfield  {title} {\bibinfo {title} {{Neutron Star Structure and the Equation of State}},\ }\href {https://doi.org/10.1086/319702} {\bibfield  {journal} {\bibinfo  {journal} {\apj}\ }\textbf {\bibinfo {volume} {550}},\ \bibinfo {pages} {426} (\bibinfo {year} {2001})},\ \Eprint {https://arxiv.org/abs/astro-ph/0002232} {arXiv:astro-ph/0002232 [astro-ph]} \BibitemShut {NoStop}%
\bibitem [{\citenamefont {{Bejger}}\ and\ \citenamefont {{Haensel}}(2002)}]{BejerHaenselAA2002}%
  \BibitemOpen
  \bibfield  {author} {\bibinfo {author} {\bibfnamefont {M.}~\bibnamefont {{Bejger}}}\ and\ \bibinfo {author} {\bibfnamefont {P.}~\bibnamefont {{Haensel}}},\ }\bibfield  {title} {\bibinfo {title} {{Moments of inertia for neutron and strange stars: Limits derived for the Crab pulsar}},\ }\href {https://doi.org/10.1051/0004-6361:20021241} {\bibfield  {journal} {\bibinfo  {journal} {\aap}\ }\textbf {\bibinfo {volume} {396}},\ \bibinfo {pages} {917} (\bibinfo {year} {2002})},\ \Eprint {https://arxiv.org/abs/astro-ph/0209151} {arXiv:astro-ph/0209151 [astro-ph]} \BibitemShut {NoStop}%
\bibitem [{\citenamefont {{Yagi}}\ and\ \citenamefont {{Yunes}}(2017)}]{YagiYounes2017}%
  \BibitemOpen
  \bibfield  {author} {\bibinfo {author} {\bibfnamefont {K.}~\bibnamefont {{Yagi}}}\ and\ \bibinfo {author} {\bibfnamefont {N.}~\bibnamefont {{Yunes}}},\ }\bibfield  {title} {\bibinfo {title} {{Approximate universal relations for neutron stars and quark stars}},\ }\href {https://doi.org/10.1016/j.physrep.2017.03.002} {\bibfield  {journal} {\bibinfo  {journal} {\physrep}\ }\textbf {\bibinfo {volume} {681}},\ \bibinfo {pages} {1} (\bibinfo {year} {2017})},\ \Eprint {https://arxiv.org/abs/1608.02582} {arXiv:1608.02582 [gr-qc]} \BibitemShut {NoStop}%
\bibitem [{\citenamefont {{Ofengeim}}(2020)}]{Ofengeim2020}%
  \BibitemOpen
  \bibfield  {author} {\bibinfo {author} {\bibfnamefont {D.~D.}\ \bibnamefont {{Ofengeim}}},\ }\bibfield  {title} {\bibinfo {title} {{Universal properties of maximum-mass neutron stars: A new tool to explore superdense matter}},\ }\href {https://doi.org/10.1103/PhysRevD.101.103029} {\bibfield  {journal} {\bibinfo  {journal} {\prd}\ }\textbf {\bibinfo {volume} {101}},\ \bibinfo {eid} {103029} (\bibinfo {year} {2020})},\ \Eprint {https://arxiv.org/abs/2005.03549} {arXiv:2005.03549 [astro-ph.HE]} \BibitemShut {NoStop}%
\bibitem [{\citenamefont {{Sotani}}\ and\ \citenamefont {{Kumar}}(2021)}]{SotaniKumarPRD2021}%
  \BibitemOpen
  \bibfield  {author} {\bibinfo {author} {\bibfnamefont {H.}~\bibnamefont {{Sotani}}}\ and\ \bibinfo {author} {\bibfnamefont {B.}~\bibnamefont {{Kumar}}},\ }\bibfield  {title} {\bibinfo {title} {{Universal relations between the quasinormal modes of neutron star and tidal deformability}},\ }\href {https://doi.org/10.1103/PhysRevD.104.123002} {\bibfield  {journal} {\bibinfo  {journal} {\prd}\ }\textbf {\bibinfo {volume} {104}},\ \bibinfo {eid} {123002} (\bibinfo {year} {2021})}\BibitemShut {NoStop}%
\bibitem [{\citenamefont {{Saes}}\ and\ \citenamefont {{Mendes}}(2022)}]{SaesMendesPRD2022}%
  \BibitemOpen
  \bibfield  {author} {\bibinfo {author} {\bibfnamefont {J.~A.}\ \bibnamefont {{Saes}}}\ and\ \bibinfo {author} {\bibfnamefont {R.~F.~P.}\ \bibnamefont {{Mendes}}},\ }\bibfield  {title} {\bibinfo {title} {{Equation-of-state-insensitive measure of neutron star stiffness}},\ }\href {https://doi.org/10.1103/PhysRevD.106.043027} {\bibfield  {journal} {\bibinfo  {journal} {\prd}\ }\textbf {\bibinfo {volume} {106}},\ \bibinfo {eid} {043027} (\bibinfo {year} {2022})},\ \Eprint {https://arxiv.org/abs/2109.11571} {arXiv:2109.11571 [gr-qc]} \BibitemShut {NoStop}%
\bibitem [{\citenamefont {{Konstantinou}}\ and\ \citenamefont {{Morsink}}(2022)}]{KonstantinouMorsinkApJ2022}%
  \BibitemOpen
  \bibfield  {author} {\bibinfo {author} {\bibfnamefont {A.}~\bibnamefont {{Konstantinou}}}\ and\ \bibinfo {author} {\bibfnamefont {S.~M.}\ \bibnamefont {{Morsink}}},\ }\bibfield  {title} {\bibinfo {title} {{Universal Relations for the Increase in the Mass and Radius of a Rotating Neutron Star}},\ }\href {https://doi.org/10.3847/1538-4357/ac7b86} {\bibfield  {journal} {\bibinfo  {journal} {\apj}\ }\textbf {\bibinfo {volume} {934}},\ \bibinfo {eid} {139} (\bibinfo {year} {2022})},\ \Eprint {https://arxiv.org/abs/2206.12515} {arXiv:2206.12515 [astro-ph.HE]} \BibitemShut {NoStop}%
\bibitem [{\citenamefont {{Cai}}\ \emph {et~al.}(2023)\citenamefont {{Cai}}, \citenamefont {{Li}},\ and\ \citenamefont {{Zhang}}}]{Cai2023ApJ}%
  \BibitemOpen
  \bibfield  {author} {\bibinfo {author} {\bibfnamefont {B.-J.}\ \bibnamefont {{Cai}}}, \bibinfo {author} {\bibfnamefont {B.-A.}\ \bibnamefont {{Li}}},\ and\ \bibinfo {author} {\bibfnamefont {Z.}~\bibnamefont {{Zhang}}},\ }\bibfield  {title} {\bibinfo {title} {{Core States of Neutron Stars from Anatomizing Their Scaled Structure Equations}},\ }\href {https://doi.org/10.3847/1538-4357/acdef0} {\bibfield  {journal} {\bibinfo  {journal} {\apj}\ }\textbf {\bibinfo {volume} {952}},\ \bibinfo {eid} {147} (\bibinfo {year} {2023})},\ \Eprint {https://arxiv.org/abs/2306.08202} {arXiv:2306.08202 [nucl-th]} \BibitemShut {NoStop}%
\bibitem [{\citenamefont {{Yagi}}\ and\ \citenamefont {{Yunes}}(2013)}]{YagiYounesSci2013}%
  \BibitemOpen
  \bibfield  {author} {\bibinfo {author} {\bibfnamefont {K.}~\bibnamefont {{Yagi}}}\ and\ \bibinfo {author} {\bibfnamefont {N.}~\bibnamefont {{Yunes}}},\ }\bibfield  {title} {\bibinfo {title} {{I-Love-Q: Unexpected Universal Relations for Neutron Stars and Quark Stars}},\ }\href {https://doi.org/10.1126/science.1236462} {\bibfield  {journal} {\bibinfo  {journal} {Science}\ }\textbf {\bibinfo {volume} {341}},\ \bibinfo {pages} {365} (\bibinfo {year} {2013})},\ \Eprint {https://arxiv.org/abs/1302.4499} {arXiv:1302.4499 [gr-qc]} \BibitemShut {NoStop}%
\bibitem [{\citenamefont {{Yagi}}\ and\ \citenamefont {{Yunes}}(2016)}]{YagiYunesCQG2016}%
  \BibitemOpen
  \bibfield  {author} {\bibinfo {author} {\bibfnamefont {K.}~\bibnamefont {{Yagi}}}\ and\ \bibinfo {author} {\bibfnamefont {N.}~\bibnamefont {{Yunes}}},\ }\bibfield  {title} {\bibinfo {title} {{Binary Love relations}},\ }\href {https://doi.org/10.1088/0264-9381/33/13/13LT01} {\bibfield  {journal} {\bibinfo  {journal} {Classical and Quantum Gravity}\ }\textbf {\bibinfo {volume} {33}},\ \bibinfo {eid} {13LT01} (\bibinfo {year} {2016})},\ \Eprint {https://arxiv.org/abs/1512.02639} {arXiv:1512.02639 [gr-qc]} \BibitemShut {NoStop}%
\bibitem [{\citenamefont {{Legred}}\ \emph {et~al.}(2024)\citenamefont {{Legred}}, \citenamefont {{Sy-Garcia}}, \citenamefont {{Chatziioannou}},\ and\ \citenamefont {{Essick}}}]{Legred+PRD2024}%
  \BibitemOpen
  \bibfield  {author} {\bibinfo {author} {\bibfnamefont {I.}~\bibnamefont {{Legred}}}, \bibinfo {author} {\bibfnamefont {B.~O.}\ \bibnamefont {{Sy-Garcia}}}, \bibinfo {author} {\bibfnamefont {K.}~\bibnamefont {{Chatziioannou}}},\ and\ \bibinfo {author} {\bibfnamefont {R.}~\bibnamefont {{Essick}}},\ }\bibfield  {title} {\bibinfo {title} {{Assessing equation of state-independent relations for neutron stars with nonparametric models}},\ }\href {https://doi.org/10.1103/PhysRevD.109.023020} {\bibfield  {journal} {\bibinfo  {journal} {\prd}\ }\textbf {\bibinfo {volume} {109}},\ \bibinfo {eid} {023020} (\bibinfo {year} {2024})},\ \Eprint {https://arxiv.org/abs/2310.10854} {arXiv:2310.10854 [astro-ph.HE]} \BibitemShut {NoStop}%
\bibitem [{\citenamefont {{Choudhury}}\ \emph {et~al.}(2024)\citenamefont {{Choudhury}}, \citenamefont {{Salmi}}, \citenamefont {{Vinciguerra}}, \citenamefont {{Riley}}, \citenamefont {{Kini}}, \citenamefont {{Watts}}, \citenamefont {{Dorsman}}, \citenamefont {{Bogdanov}}, \citenamefont {{Guillot}}, \citenamefont {{Ray}}, \citenamefont {{Reardon}}, \citenamefont {{Remillard}}, \citenamefont {{Bilous}}, \citenamefont {{Huppenkothen}}, \citenamefont {{Lattimer}}, \citenamefont {{Rutherford}}, \citenamefont {{Arzoumanian}}, \citenamefont {{Gendreau}}, \citenamefont {{Morsink}},\ and\ \citenamefont {{Ho}}}]{Choudhury2024}%
  \BibitemOpen
  \bibfield  {author} {\bibinfo {author} {\bibfnamefont {D.}~\bibnamefont {{Choudhury}}}, \bibinfo {author} {\bibfnamefont {T.}~\bibnamefont {{Salmi}}}, \bibinfo {author} {\bibfnamefont {S.}~\bibnamefont {{Vinciguerra}}}, \bibinfo {author} {\bibfnamefont {T.~E.}\ \bibnamefont {{Riley}}}, \bibinfo {author} {\bibfnamefont {Y.}~\bibnamefont {{Kini}}}, \bibinfo {author} {\bibfnamefont {A.~L.}\ \bibnamefont {{Watts}}}, \bibinfo {author} {\bibfnamefont {B.}~\bibnamefont {{Dorsman}}}, \bibinfo {author} {\bibfnamefont {S.}~\bibnamefont {{Bogdanov}}}, \bibinfo {author} {\bibfnamefont {S.}~\bibnamefont {{Guillot}}}, \bibinfo {author} {\bibfnamefont {P.~S.}\ \bibnamefont {{Ray}}}, \bibinfo {author} {\bibfnamefont {D.~J.}\ \bibnamefont {{Reardon}}}, \bibinfo {author} {\bibfnamefont {R.~A.}\ \bibnamefont {{Remillard}}}, \bibinfo {author} {\bibfnamefont {A.~V.}\ \bibnamefont {{Bilous}}}, \bibinfo {author} {\bibfnamefont {D.}~\bibnamefont {{Huppenkothen}}}, \bibinfo {author} {\bibfnamefont {J.~M.}\ \bibnamefont
  {{Lattimer}}}, \bibinfo {author} {\bibfnamefont {N.}~\bibnamefont {{Rutherford}}}, \bibinfo {author} {\bibfnamefont {Z.}~\bibnamefont {{Arzoumanian}}}, \bibinfo {author} {\bibfnamefont {K.~C.}\ \bibnamefont {{Gendreau}}}, \bibinfo {author} {\bibfnamefont {S.~M.}\ \bibnamefont {{Morsink}}},\ and\ \bibinfo {author} {\bibfnamefont {W.~C.~G.}\ \bibnamefont {{Ho}}},\ }\bibfield  {title} {\bibinfo {title} {{A NICER View of the Nearest and Brightest Millisecond Pulsar: PSR J0437$-$4715}},\ }\href {https://doi.org/10.48550/arXiv.2407.06789} {\bibfield  {journal} {\bibinfo  {journal} {arXiv e-prints}\ ,\ \bibinfo {eid} {arXiv:2407.06789}} (\bibinfo {year} {2024})},\ \Eprint {https://arxiv.org/abs/2407.06789} {arXiv:2407.06789 [astro-ph.HE]} \BibitemShut {NoStop}%
\bibitem [{\citenamefont {{Antoniadis}}\ \emph {et~al.}(2013)\citenamefont {{Antoniadis}}, \citenamefont {{Freire}}, \citenamefont {{Wex}}, \citenamefont {{Tauris}}, \citenamefont {{Lynch}},\ and\ \citenamefont {et~al.}}]{0348paper2013}%
  \BibitemOpen
  \bibfield  {author} {\bibinfo {author} {\bibfnamefont {J.}~\bibnamefont {{Antoniadis}}}, \bibinfo {author} {\bibfnamefont {P.~C.~C.}\ \bibnamefont {{Freire}}}, \bibinfo {author} {\bibfnamefont {N.}~\bibnamefont {{Wex}}}, \bibinfo {author} {\bibfnamefont {T.~M.}\ \bibnamefont {{Tauris}}}, \bibinfo {author} {\bibfnamefont {R.~S.}\ \bibnamefont {{Lynch}}},\ and\ \bibinfo {author} {\bibnamefont {et~al.}},\ }\bibfield  {title} {\bibinfo {title} {{A Massive Pulsar in a Compact Relativistic Binary}},\ }\href {https://doi.org/10.1126/science.1233232} {\bibfield  {journal} {\bibinfo  {journal} {Science}\ }\textbf {\bibinfo {volume} {340}},\ \bibinfo {pages} {448} (\bibinfo {year} {2013})},\ \Eprint {https://arxiv.org/abs/1304.6875} {arXiv:1304.6875 [astro-ph.HE]} \BibitemShut {NoStop}%
\bibitem [{\citenamefont {{Fonseca}}\ \emph {et~al.}(2021)\citenamefont {{Fonseca}}, \citenamefont {{Cromartie}}, \citenamefont {{Pennucci}}, \citenamefont {{Ray}}, \citenamefont {{Kirichenko}},\ and\ \citenamefont {et~al.}}]{0740paper2021}%
  \BibitemOpen
  \bibfield  {author} {\bibinfo {author} {\bibfnamefont {E.}~\bibnamefont {{Fonseca}}}, \bibinfo {author} {\bibfnamefont {H.~T.}\ \bibnamefont {{Cromartie}}}, \bibinfo {author} {\bibfnamefont {T.~T.}\ \bibnamefont {{Pennucci}}}, \bibinfo {author} {\bibfnamefont {P.~S.}\ \bibnamefont {{Ray}}}, \bibinfo {author} {\bibfnamefont {A.~Y.}\ \bibnamefont {{Kirichenko}}},\ and\ \bibinfo {author} {\bibnamefont {et~al.}},\ }\bibfield  {title} {\bibinfo {title} {{Refined Mass and Geometric Measurements of the High-mass PSR J0740+6620}},\ }\href {https://doi.org/10.3847/2041-8213/ac03b8} {\bibfield  {journal} {\bibinfo  {journal} {\apjl}\ }\textbf {\bibinfo {volume} {915}},\ \bibinfo {eid} {L12} (\bibinfo {year} {2021})},\ \Eprint {https://arxiv.org/abs/2104.00880} {arXiv:2104.00880 [astro-ph.HE]} \BibitemShut {NoStop}%
\bibitem [{\citenamefont {{Kandel}}\ and\ \citenamefont {{Romani}}(2023)}]{Kandel2023ApJ}%
  \BibitemOpen
  \bibfield  {author} {\bibinfo {author} {\bibfnamefont {D.}~\bibnamefont {{Kandel}}}\ and\ \bibinfo {author} {\bibfnamefont {R.~W.}\ \bibnamefont {{Romani}}},\ }\bibfield  {title} {\bibinfo {title} {{An Optical Study of the Black Widow Population}},\ }\href {https://doi.org/10.3847/1538-4357/aca524} {\bibfield  {journal} {\bibinfo  {journal} {\apj}\ }\textbf {\bibinfo {volume} {942}},\ \bibinfo {eid} {6} (\bibinfo {year} {2023})}\BibitemShut {NoStop}%
\bibitem [{\citenamefont {{Ofengeim}}\ \emph {et~al.}(2023)\citenamefont {{Ofengeim}}, \citenamefont {{Shternin}},\ and\ \citenamefont {{Piran}}}]{OfShtPir2023}%
  \BibitemOpen
  \bibfield  {author} {\bibinfo {author} {\bibfnamefont {D.~D.}\ \bibnamefont {{Ofengeim}}}, \bibinfo {author} {\bibfnamefont {P.~S.}\ \bibnamefont {{Shternin}}},\ and\ \bibinfo {author} {\bibfnamefont {T.}~\bibnamefont {{Piran}}},\ }\bibfield  {title} {\bibinfo {title} {{Maximal Mass Neutron Star as a Key to Superdense Matter Physics}},\ }\href {https://doi.org/10.1134/S1063773723100055} {\bibfield  {journal} {\bibinfo  {journal} {Astronomy Letters}\ }\textbf {\bibinfo {volume} {49}},\ \bibinfo {pages} {567} (\bibinfo {year} {2023})},\ \Eprint {https://arxiv.org/abs/2310.16847} {arXiv:2310.16847 [astro-ph.HE]} \BibitemShut {NoStop}%
\bibitem [{\citenamefont {{{\"O}zel}}\ and\ \citenamefont {{Psaltis}}(2009)}]{OzelPsaltis2009PhRvD}%
  \BibitemOpen
  \bibfield  {author} {\bibinfo {author} {\bibfnamefont {F.}~\bibnamefont {{{\"O}zel}}}\ and\ \bibinfo {author} {\bibfnamefont {D.}~\bibnamefont {{Psaltis}}},\ }\bibfield  {title} {\bibinfo {title} {{Reconstructing the neutron-star equation of state from astrophysical measurements}},\ }\href {https://doi.org/10.1103/PhysRevD.80.103003} {\bibfield  {journal} {\bibinfo  {journal} {\prd}\ }\textbf {\bibinfo {volume} {80}},\ \bibinfo {eid} {103003} (\bibinfo {year} {2009})},\ \Eprint {https://arxiv.org/abs/0905.1959} {arXiv:0905.1959 [astro-ph.HE]} \BibitemShut {NoStop}%
\bibitem [{\citenamefont {{Riley}}\ \emph {et~al.}(2019)\citenamefont {{Riley}}, \citenamefont {{Watts}}, \citenamefont {{Bogdanov}}, \citenamefont {{Ray}}, \citenamefont {{Ludlam}},\ and\ \citenamefont {et~al.}}]{Riley+ApJL2019}%
  \BibitemOpen
  \bibfield  {author} {\bibinfo {author} {\bibfnamefont {T.~E.}\ \bibnamefont {{Riley}}}, \bibinfo {author} {\bibfnamefont {A.~L.}\ \bibnamefont {{Watts}}}, \bibinfo {author} {\bibfnamefont {S.}~\bibnamefont {{Bogdanov}}}, \bibinfo {author} {\bibfnamefont {P.~S.}\ \bibnamefont {{Ray}}}, \bibinfo {author} {\bibfnamefont {R.~M.}\ \bibnamefont {{Ludlam}}},\ and\ \bibinfo {author} {\bibnamefont {et~al.}},\ }\bibfield  {title} {\bibinfo {title} {{A NICER View of PSR J0030+0451: Millisecond Pulsar Parameter Estimation}},\ }\href {https://doi.org/10.3847/2041-8213/ab481c} {\bibfield  {journal} {\bibinfo  {journal} {\apjl}\ }\textbf {\bibinfo {volume} {887}},\ \bibinfo {eid} {L21} (\bibinfo {year} {2019})},\ \Eprint {https://arxiv.org/abs/1912.05702} {arXiv:1912.05702 [astro-ph.HE]} \BibitemShut {NoStop}%
\bibitem [{\citenamefont {{Miller}}\ \emph {et~al.}(2019)\citenamefont {{Miller}}, \citenamefont {{Lamb}}, \citenamefont {{Dittmann}}, \citenamefont {{Bogdanov}}, \citenamefont {{Arzoumanian}}, \citenamefont {{Gendreau}}, \citenamefont {{Guillot}}, \citenamefont {{Harding}}, \citenamefont {{Ho}}, \citenamefont {{Lattimer}}, \citenamefont {{Ludlam}}, \citenamefont {{Mahmoodifar}}, \citenamefont {{Morsink}}, \citenamefont {{Ray}}, \citenamefont {{Strohmayer}}, \citenamefont {{Wood}}, \citenamefont {{Enoto}}, \citenamefont {{Foster}}, \citenamefont {{Okajima}}, \citenamefont {{Prigozhin}},\ and\ \citenamefont {{Soong}}}]{Miller2019ApJ0030}%
  \BibitemOpen
  \bibfield  {author} {\bibinfo {author} {\bibfnamefont {M.~C.}\ \bibnamefont {{Miller}}}, \bibinfo {author} {\bibfnamefont {F.~K.}\ \bibnamefont {{Lamb}}}, \bibinfo {author} {\bibfnamefont {A.~J.}\ \bibnamefont {{Dittmann}}}, \bibinfo {author} {\bibfnamefont {S.}~\bibnamefont {{Bogdanov}}}, \bibinfo {author} {\bibfnamefont {Z.}~\bibnamefont {{Arzoumanian}}}, \bibinfo {author} {\bibfnamefont {K.~C.}\ \bibnamefont {{Gendreau}}}, \bibinfo {author} {\bibfnamefont {S.}~\bibnamefont {{Guillot}}}, \bibinfo {author} {\bibfnamefont {A.~K.}\ \bibnamefont {{Harding}}}, \bibinfo {author} {\bibfnamefont {W.~C.~G.}\ \bibnamefont {{Ho}}}, \bibinfo {author} {\bibfnamefont {J.~M.}\ \bibnamefont {{Lattimer}}}, \bibinfo {author} {\bibfnamefont {R.~M.}\ \bibnamefont {{Ludlam}}}, \bibinfo {author} {\bibfnamefont {S.}~\bibnamefont {{Mahmoodifar}}}, \bibinfo {author} {\bibfnamefont {S.~M.}\ \bibnamefont {{Morsink}}}, \bibinfo {author} {\bibfnamefont {P.~S.}\ \bibnamefont {{Ray}}}, \bibinfo {author} {\bibfnamefont {T.~E.}\
  \bibnamefont {{Strohmayer}}}, \bibinfo {author} {\bibfnamefont {K.~S.}\ \bibnamefont {{Wood}}}, \bibinfo {author} {\bibfnamefont {T.}~\bibnamefont {{Enoto}}}, \bibinfo {author} {\bibfnamefont {R.}~\bibnamefont {{Foster}}}, \bibinfo {author} {\bibfnamefont {T.}~\bibnamefont {{Okajima}}}, \bibinfo {author} {\bibfnamefont {G.}~\bibnamefont {{Prigozhin}}},\ and\ \bibinfo {author} {\bibfnamefont {Y.}~\bibnamefont {{Soong}}},\ }\bibfield  {title} {\bibinfo {title} {{PSR J0030+0451 Mass and Radius from NICER Data and Implications for the Properties of Neutron Star Matter}},\ }\href {https://doi.org/10.3847/2041-8213/ab50c5} {\bibfield  {journal} {\bibinfo  {journal} {\apjl}\ }\textbf {\bibinfo {volume} {887}},\ \bibinfo {eid} {L24} (\bibinfo {year} {2019})},\ \Eprint {https://arxiv.org/abs/1912.05705} {arXiv:1912.05705 [astro-ph.HE]} \BibitemShut {NoStop}%
\bibitem [{\citenamefont {{Raaijmakers}}\ \emph {et~al.}(2019)\citenamefont {{Raaijmakers}}, \citenamefont {{Riley}}, \citenamefont {{Watts}}, \citenamefont {{Greif}}, \citenamefont {{Morsink}},\ and\ \citenamefont {et~al.}}]{Raaijmakers+2019}%
  \BibitemOpen
  \bibfield  {author} {\bibinfo {author} {\bibfnamefont {G.}~\bibnamefont {{Raaijmakers}}}, \bibinfo {author} {\bibfnamefont {T.~E.}\ \bibnamefont {{Riley}}}, \bibinfo {author} {\bibfnamefont {A.~L.}\ \bibnamefont {{Watts}}}, \bibinfo {author} {\bibfnamefont {S.~K.}\ \bibnamefont {{Greif}}}, \bibinfo {author} {\bibfnamefont {S.~M.}\ \bibnamefont {{Morsink}}},\ and\ \bibinfo {author} {\bibnamefont {et~al.}},\ }\bibfield  {title} {\bibinfo {title} {{A Nicer View of PSR J0030+0451: Implications for the Dense Matter Equation of State}},\ }\href {https://doi.org/10.3847/2041-8213/ab451a} {\bibfield  {journal} {\bibinfo  {journal} {\apjl}\ }\textbf {\bibinfo {volume} {887}},\ \bibinfo {eid} {L22} (\bibinfo {year} {2019})},\ \Eprint {https://arxiv.org/abs/1912.05703} {arXiv:1912.05703 [astro-ph.HE]} \BibitemShut {NoStop}%
\bibitem [{\citenamefont {{Vinciguerra}}\ \emph {et~al.}(2024)\citenamefont {{Vinciguerra}}, \citenamefont {{Salmi}}, \citenamefont {{Watts}}, \citenamefont {{Choudhury}}, \citenamefont {{Riley}},\ and\ \citenamefont {et~al.}}]{Vinciguerra+ApJ2024}%
  \BibitemOpen
  \bibfield  {author} {\bibinfo {author} {\bibfnamefont {S.}~\bibnamefont {{Vinciguerra}}}, \bibinfo {author} {\bibfnamefont {T.}~\bibnamefont {{Salmi}}}, \bibinfo {author} {\bibfnamefont {A.~L.}\ \bibnamefont {{Watts}}}, \bibinfo {author} {\bibfnamefont {D.}~\bibnamefont {{Choudhury}}}, \bibinfo {author} {\bibfnamefont {T.~E.}\ \bibnamefont {{Riley}}},\ and\ \bibinfo {author} {\bibnamefont {et~al.}},\ }\bibfield  {title} {\bibinfo {title} {{An Updated Mass-Radius Analysis of the 2017-2018 NICER Data Set of PSR J0030+0451}},\ }\href {https://doi.org/10.3847/1538-4357/acfb83} {\bibfield  {journal} {\bibinfo  {journal} {\apj}\ }\textbf {\bibinfo {volume} {961}},\ \bibinfo {eid} {62} (\bibinfo {year} {2024})},\ \Eprint {https://arxiv.org/abs/2308.09469} {arXiv:2308.09469 [astro-ph.HE]} \BibitemShut {NoStop}%
\bibitem [{\citenamefont {{Riley}}\ \emph {et~al.}(2021)\citenamefont {{Riley}}, \citenamefont {{Watts}}, \citenamefont {{Ray}}, \citenamefont {{Bogdanov}}, \citenamefont {{Guillot}},\ and\ \citenamefont {et~al.}}]{Riley+ApJL2021}%
  \BibitemOpen
  \bibfield  {author} {\bibinfo {author} {\bibfnamefont {T.~E.}\ \bibnamefont {{Riley}}}, \bibinfo {author} {\bibfnamefont {A.~L.}\ \bibnamefont {{Watts}}}, \bibinfo {author} {\bibfnamefont {P.~S.}\ \bibnamefont {{Ray}}}, \bibinfo {author} {\bibfnamefont {S.}~\bibnamefont {{Bogdanov}}}, \bibinfo {author} {\bibfnamefont {S.}~\bibnamefont {{Guillot}}},\ and\ \bibinfo {author} {\bibnamefont {et~al.}},\ }\bibfield  {title} {\bibinfo {title} {{A NICER View of the Massive Pulsar PSR J0740+6620 Informed by Radio Timing and XMM-Newton Spectroscopy}},\ }\href {https://doi.org/10.3847/2041-8213/ac0a81} {\bibfield  {journal} {\bibinfo  {journal} {\apjl}\ }\textbf {\bibinfo {volume} {918}},\ \bibinfo {eid} {L27} (\bibinfo {year} {2021})},\ \Eprint {https://arxiv.org/abs/2105.06980} {arXiv:2105.06980 [astro-ph.HE]} \BibitemShut {NoStop}%
\bibitem [{\citenamefont {{Raaijmakers}}\ \emph {et~al.}(2021)\citenamefont {{Raaijmakers}}, \citenamefont {{Greif}}, \citenamefont {{Hebeler}}, \citenamefont {{Hinderer}}, \citenamefont {{Nissanke}},\ and\ \citenamefont {et~al.}}]{Raaijmakers+2021}%
  \BibitemOpen
  \bibfield  {author} {\bibinfo {author} {\bibfnamefont {G.}~\bibnamefont {{Raaijmakers}}}, \bibinfo {author} {\bibfnamefont {S.~K.}\ \bibnamefont {{Greif}}}, \bibinfo {author} {\bibfnamefont {K.}~\bibnamefont {{Hebeler}}}, \bibinfo {author} {\bibfnamefont {T.}~\bibnamefont {{Hinderer}}}, \bibinfo {author} {\bibfnamefont {S.}~\bibnamefont {{Nissanke}}},\ and\ \bibinfo {author} {\bibnamefont {et~al.}},\ }\bibfield  {title} {\bibinfo {title} {{Constraints on the Dense Matter Equation of State and Neutron Star Properties from NICER's Mass-Radius Estimate of PSR J0740+6620 and Multimessenger Observations}},\ }\href {https://doi.org/10.3847/2041-8213/ac089a} {\bibfield  {journal} {\bibinfo  {journal} {\apjl}\ }\textbf {\bibinfo {volume} {918}},\ \bibinfo {eid} {L29} (\bibinfo {year} {2021})},\ \Eprint {https://arxiv.org/abs/2105.06981} {arXiv:2105.06981 [astro-ph.HE]} \BibitemShut {NoStop}%
\bibitem [{\citenamefont {{Salmi}}\ \emph {et~al.}(2022)\citenamefont {{Salmi}}, \citenamefont {{Vinciguerra}}, \citenamefont {{Choudhury}}, \citenamefont {{Riley}}, \citenamefont {{Watts}}, \citenamefont {{Remillard}}, \citenamefont {{Ray}}, \citenamefont {{Bogdanov}}, \citenamefont {{Guillot}}, \citenamefont {{Arzoumanian}}, \citenamefont {{Chirenti}}, \citenamefont {{Dittmann}}, \citenamefont {{Gendreau}}, \citenamefont {{Ho}}, \citenamefont {{Miller}}, \citenamefont {{Morsink}}, \citenamefont {{Wadiasingh}},\ and\ \citenamefont {{Wolff}}}]{Salmi2022ApJ}%
  \BibitemOpen
  \bibfield  {author} {\bibinfo {author} {\bibfnamefont {T.}~\bibnamefont {{Salmi}}}, \bibinfo {author} {\bibfnamefont {S.}~\bibnamefont {{Vinciguerra}}}, \bibinfo {author} {\bibfnamefont {D.}~\bibnamefont {{Choudhury}}}, \bibinfo {author} {\bibfnamefont {T.~E.}\ \bibnamefont {{Riley}}}, \bibinfo {author} {\bibfnamefont {A.~L.}\ \bibnamefont {{Watts}}}, \bibinfo {author} {\bibfnamefont {R.~A.}\ \bibnamefont {{Remillard}}}, \bibinfo {author} {\bibfnamefont {P.~S.}\ \bibnamefont {{Ray}}}, \bibinfo {author} {\bibfnamefont {S.}~\bibnamefont {{Bogdanov}}}, \bibinfo {author} {\bibfnamefont {S.}~\bibnamefont {{Guillot}}}, \bibinfo {author} {\bibfnamefont {Z.}~\bibnamefont {{Arzoumanian}}}, \bibinfo {author} {\bibfnamefont {C.}~\bibnamefont {{Chirenti}}}, \bibinfo {author} {\bibfnamefont {A.~J.}\ \bibnamefont {{Dittmann}}}, \bibinfo {author} {\bibfnamefont {K.~C.}\ \bibnamefont {{Gendreau}}}, \bibinfo {author} {\bibfnamefont {W.~C.~G.}\ \bibnamefont {{Ho}}}, \bibinfo {author} {\bibfnamefont {M.~C.}\ \bibnamefont
  {{Miller}}}, \bibinfo {author} {\bibfnamefont {S.~M.}\ \bibnamefont {{Morsink}}}, \bibinfo {author} {\bibfnamefont {Z.}~\bibnamefont {{Wadiasingh}}},\ and\ \bibinfo {author} {\bibfnamefont {M.~T.}\ \bibnamefont {{Wolff}}},\ }\bibfield  {title} {\bibinfo {title} {{The Radius of PSR J0740+6620 from NICER with NICER Background Estimates}},\ }\href {https://doi.org/10.3847/1538-4357/ac983d} {\bibfield  {journal} {\bibinfo  {journal} {\apj}\ }\textbf {\bibinfo {volume} {941}},\ \bibinfo {eid} {150} (\bibinfo {year} {2022})},\ \Eprint {https://arxiv.org/abs/2209.12840} {arXiv:2209.12840 [astro-ph.HE]} \BibitemShut {NoStop}%
\bibitem [{\citenamefont {{Salmi}}\ \emph {et~al.}(2024)\citenamefont {{Salmi}}, \citenamefont {{Choudhury}}, \citenamefont {{Kini}}, \citenamefont {{Riley}}, \citenamefont {{Vinciguerra}}, \citenamefont {{Watts}}, \citenamefont {{Wolff}}, \citenamefont {{Arzoumanian}}, \citenamefont {{Bogdanov}}, \citenamefont {{Chakrabarty}}, \citenamefont {{Gendreau}}, \citenamefont {{Guillot}}, \citenamefont {{Ho}}, \citenamefont {{Huppenkothen}}, \citenamefont {{Ludlam}}, \citenamefont {{Morsink}},\ and\ \citenamefont {{Ray}}}]{Salmi2024jun}%
  \BibitemOpen
  \bibfield  {author} {\bibinfo {author} {\bibfnamefont {T.}~\bibnamefont {{Salmi}}}, \bibinfo {author} {\bibfnamefont {D.}~\bibnamefont {{Choudhury}}}, \bibinfo {author} {\bibfnamefont {Y.}~\bibnamefont {{Kini}}}, \bibinfo {author} {\bibfnamefont {T.~E.}\ \bibnamefont {{Riley}}}, \bibinfo {author} {\bibfnamefont {S.}~\bibnamefont {{Vinciguerra}}}, \bibinfo {author} {\bibfnamefont {A.~L.}\ \bibnamefont {{Watts}}}, \bibinfo {author} {\bibfnamefont {M.~T.}\ \bibnamefont {{Wolff}}}, \bibinfo {author} {\bibfnamefont {Z.}~\bibnamefont {{Arzoumanian}}}, \bibinfo {author} {\bibfnamefont {S.}~\bibnamefont {{Bogdanov}}}, \bibinfo {author} {\bibfnamefont {D.}~\bibnamefont {{Chakrabarty}}}, \bibinfo {author} {\bibfnamefont {K.}~\bibnamefont {{Gendreau}}}, \bibinfo {author} {\bibfnamefont {S.}~\bibnamefont {{Guillot}}}, \bibinfo {author} {\bibfnamefont {W.~C.~G.}\ \bibnamefont {{Ho}}}, \bibinfo {author} {\bibfnamefont {D.}~\bibnamefont {{Huppenkothen}}}, \bibinfo {author} {\bibfnamefont {R.~M.}\ \bibnamefont {{Ludlam}}},
  \bibinfo {author} {\bibfnamefont {S.~M.}\ \bibnamefont {{Morsink}}},\ and\ \bibinfo {author} {\bibfnamefont {P.~S.}\ \bibnamefont {{Ray}}},\ }\bibfield  {title} {\bibinfo {title} {{The Radius of the High Mass Pulsar PSR J0740+6620 With 3.6 Years of NICER Data}},\ }\href {https://doi.org/10.48550/arXiv.2406.14466} {\bibfield  {journal} {\bibinfo  {journal} {arXiv e-prints}\ ,\ \bibinfo {eid} {arXiv:2406.14466}} (\bibinfo {year} {2024})},\ \Eprint {https://arxiv.org/abs/2406.14466} {arXiv:2406.14466 [astro-ph.HE]} \BibitemShut {NoStop}%
\bibitem [{\citenamefont {{Shternin}}\ \emph {et~al.}(2023)\citenamefont {{Shternin}}, \citenamefont {{Ofengeim}}, \citenamefont {{Heinke}},\ and\ \citenamefont {{Ho}}}]{Shternin2023MNRAS}%
  \BibitemOpen
  \bibfield  {author} {\bibinfo {author} {\bibfnamefont {P.~S.}\ \bibnamefont {{Shternin}}}, \bibinfo {author} {\bibfnamefont {D.~D.}\ \bibnamefont {{Ofengeim}}}, \bibinfo {author} {\bibfnamefont {C.~O.}\ \bibnamefont {{Heinke}}},\ and\ \bibinfo {author} {\bibfnamefont {W.~C.~G.}\ \bibnamefont {{Ho}}},\ }\bibfield  {title} {\bibinfo {title} {{Constraints on neutron star superfluidity from the cooling neutron star in Cassiopeia A using all Chandra ACIS-S observations}},\ }\href {https://doi.org/10.1093/mnras/stac3226} {\bibfield  {journal} {\bibinfo  {journal} {\mnras}\ }\textbf {\bibinfo {volume} {518}},\ \bibinfo {pages} {2775} (\bibinfo {year} {2023})}\BibitemShut {NoStop}%
\bibitem [{\citenamefont {{N{\"a}ttil{\"a}}}\ \emph {et~al.}(2017)\citenamefont {{N{\"a}ttil{\"a}}}, \citenamefont {{Miller}}, \citenamefont {{Steiner}}, \citenamefont {{Kajava}}, \citenamefont {{Suleimanov}},\ and\ \citenamefont {et~al.}}]{Nattila+AA2017}%
  \BibitemOpen
  \bibfield  {author} {\bibinfo {author} {\bibfnamefont {J.}~\bibnamefont {{N{\"a}ttil{\"a}}}}, \bibinfo {author} {\bibfnamefont {M.~C.}\ \bibnamefont {{Miller}}}, \bibinfo {author} {\bibfnamefont {A.~W.}\ \bibnamefont {{Steiner}}}, \bibinfo {author} {\bibfnamefont {J.~J.~E.}\ \bibnamefont {{Kajava}}}, \bibinfo {author} {\bibfnamefont {V.~F.}\ \bibnamefont {{Suleimanov}}},\ and\ \bibinfo {author} {\bibnamefont {et~al.}},\ }\bibfield  {title} {\bibinfo {title} {{Neutron star mass and radius measurements from atmospheric model fits to X-ray burst cooling tail spectra}},\ }\href {https://doi.org/10.1051/0004-6361/201731082} {\bibfield  {journal} {\bibinfo  {journal} {\aap}\ }\textbf {\bibinfo {volume} {608}},\ \bibinfo {eid} {A31} (\bibinfo {year} {2017})},\ \Eprint {https://arxiv.org/abs/1709.09120} {arXiv:1709.09120 [astro-ph.HE]} \BibitemShut {NoStop}%
\bibitem [{\citenamefont {{N{\"a}ttil{\"a}}}\ \emph {et~al.}(2016)\citenamefont {{N{\"a}ttil{\"a}}}, \citenamefont {{Steiner}}, \citenamefont {{Kajava}}, \citenamefont {{Suleimanov}},\ and\ \citenamefont {{Poutanen}}}]{Nattila+AA2016}%
  \BibitemOpen
  \bibfield  {author} {\bibinfo {author} {\bibfnamefont {J.}~\bibnamefont {{N{\"a}ttil{\"a}}}}, \bibinfo {author} {\bibfnamefont {A.~W.}\ \bibnamefont {{Steiner}}}, \bibinfo {author} {\bibfnamefont {J.~J.~E.}\ \bibnamefont {{Kajava}}}, \bibinfo {author} {\bibfnamefont {V.~F.}\ \bibnamefont {{Suleimanov}}},\ and\ \bibinfo {author} {\bibfnamefont {J.}~\bibnamefont {{Poutanen}}},\ }\bibfield  {title} {\bibinfo {title} {{Equation of state constraints for the cold dense matter inside neutron stars using the cooling tail method}},\ }\href {https://doi.org/10.1051/0004-6361/201527416} {\bibfield  {journal} {\bibinfo  {journal} {\aap}\ }\textbf {\bibinfo {volume} {591}},\ \bibinfo {eid} {A25} (\bibinfo {year} {2016})},\ \Eprint {https://arxiv.org/abs/1509.06561} {arXiv:1509.06561 [astro-ph.HE]} \BibitemShut {NoStop}%
\bibitem [{\citenamefont {{Abbott}}\ \emph {et~al.}(2018)\citenamefont {{Abbott}}, \citenamefont {{Abbott}}, \citenamefont {{Abbott}} \emph {et~al.}}]{GW170817-EoS}%
  \BibitemOpen
  \bibfield  {author} {\bibinfo {author} {\bibfnamefont {B.~P.}\ \bibnamefont {{Abbott}}}, \bibinfo {author} {\bibfnamefont {R.}~\bibnamefont {{Abbott}}}, \bibinfo {author} {\bibfnamefont {T.~D.}\ \bibnamefont {{Abbott}}}, \emph {et~al.},\ }\bibfield  {title} {\bibinfo {title} {{GW170817: Measurements of Neutron Star Radii and Equation of State}},\ }\href {https://doi.org/10.1103/PhysRevLett.121.161101} {\bibfield  {journal} {\bibinfo  {journal} {\prl}\ }\textbf {\bibinfo {volume} {121}},\ \bibinfo {eid} {161101} (\bibinfo {year} {2018})},\ \Eprint {https://arxiv.org/abs/1805.11581} {arXiv:1805.11581 [gr-qc]} \BibitemShut {NoStop}%
\bibitem [{\citenamefont {{Demorest}}\ \emph {et~al.}(2010)\citenamefont {{Demorest}}, \citenamefont {{Pennucci}}, \citenamefont {{Ransom}}, \citenamefont {{Roberts}},\ and\ \citenamefont {{Hessels}}}]{1614paper2010}%
  \BibitemOpen
  \bibfield  {author} {\bibinfo {author} {\bibfnamefont {P.~B.}\ \bibnamefont {{Demorest}}}, \bibinfo {author} {\bibfnamefont {T.}~\bibnamefont {{Pennucci}}}, \bibinfo {author} {\bibfnamefont {S.~M.}\ \bibnamefont {{Ransom}}}, \bibinfo {author} {\bibfnamefont {M.~S.~E.}\ \bibnamefont {{Roberts}}},\ and\ \bibinfo {author} {\bibfnamefont {J.~W.~T.}\ \bibnamefont {{Hessels}}},\ }\bibfield  {title} {\bibinfo {title} {{A two-solar-mass neutron star measured using Shapiro delay}},\ }\href {https://doi.org/10.1038/nature09466} {\bibfield  {journal} {\bibinfo  {journal} {\nat}\ }\textbf {\bibinfo {volume} {467}},\ \bibinfo {pages} {1081} (\bibinfo {year} {2010})}\BibitemShut {NoStop}%
\bibitem [{\citenamefont {{Kandel}}\ and\ \citenamefont {{Romani}}(2020)}]{Kandel2020ApJ}%
  \BibitemOpen
  \bibfield  {author} {\bibinfo {author} {\bibfnamefont {D.}~\bibnamefont {{Kandel}}}\ and\ \bibinfo {author} {\bibfnamefont {R.~W.}\ \bibnamefont {{Romani}}},\ }\bibfield  {title} {\bibinfo {title} {{Atmospheric Circulation on Black Widow Companions}},\ }\href {https://doi.org/10.3847/1538-4357/ab7b62} {\bibfield  {journal} {\bibinfo  {journal} {\apj}\ }\textbf {\bibinfo {volume} {892}},\ \bibinfo {eid} {101} (\bibinfo {year} {2020})}\BibitemShut {NoStop}%
\bibitem [{\citenamefont {{Rezzolla}}\ \emph {et~al.}(2018)\citenamefont {{Rezzolla}}, \citenamefont {{Most}},\ and\ \citenamefont {{Weih}}}]{Rezzolla2018ApJ}%
  \BibitemOpen
  \bibfield  {author} {\bibinfo {author} {\bibfnamefont {L.}~\bibnamefont {{Rezzolla}}}, \bibinfo {author} {\bibfnamefont {E.~R.}\ \bibnamefont {{Most}}},\ and\ \bibinfo {author} {\bibfnamefont {L.~R.}\ \bibnamefont {{Weih}}},\ }\bibfield  {title} {\bibinfo {title} {{Using Gravitational-wave Observations and Quasi-universal Relations to Constrain the Maximum Mass of Neutron Stars}},\ }\href {https://doi.org/10.3847/2041-8213/aaa401} {\bibfield  {journal} {\bibinfo  {journal} {\apjl}\ }\textbf {\bibinfo {volume} {852}},\ \bibinfo {eid} {L25} (\bibinfo {year} {2018})}\BibitemShut {NoStop}%
\bibitem [{\citenamefont {{Dietrich}}\ \emph {et~al.}(2020)\citenamefont {{Dietrich}}, \citenamefont {{Coughlin}}, \citenamefont {{Pang}}, \citenamefont {{Bulla}}, \citenamefont {{Heinzel}}, \citenamefont {{Issa}}, \citenamefont {{Tews}},\ and\ \citenamefont {{Antier}}}]{Dietrich2020Sci}%
  \BibitemOpen
  \bibfield  {author} {\bibinfo {author} {\bibfnamefont {T.}~\bibnamefont {{Dietrich}}}, \bibinfo {author} {\bibfnamefont {M.~W.}\ \bibnamefont {{Coughlin}}}, \bibinfo {author} {\bibfnamefont {P.~T.~H.}\ \bibnamefont {{Pang}}}, \bibinfo {author} {\bibfnamefont {M.}~\bibnamefont {{Bulla}}}, \bibinfo {author} {\bibfnamefont {J.}~\bibnamefont {{Heinzel}}}, \bibinfo {author} {\bibfnamefont {L.}~\bibnamefont {{Issa}}}, \bibinfo {author} {\bibfnamefont {I.}~\bibnamefont {{Tews}}},\ and\ \bibinfo {author} {\bibfnamefont {S.}~\bibnamefont {{Antier}}},\ }\bibfield  {title} {\bibinfo {title} {{Multimessenger constraints on the neutron-star equation of state and the Hubble constant}},\ }\href {https://doi.org/10.1126/science.abb4317} {\bibfield  {journal} {\bibinfo  {journal} {Science}\ }\textbf {\bibinfo {volume} {370}},\ \bibinfo {pages} {1450} (\bibinfo {year} {2020})},\ \Eprint {https://arxiv.org/abs/2002.11355} {arXiv:2002.11355 [astro-ph.HE]} \BibitemShut {NoStop}%
\bibitem [{\citenamefont {{Ayriyan}}\ \emph {et~al.}(2021)\citenamefont {{Ayriyan}}, \citenamefont {{Blaschke}}, \citenamefont {{Grunfeld}}, \citenamefont {{Alvarez-Castillo}}, \citenamefont {{Grigorian}},\ and\ \citenamefont {{Abgaryan}}}]{Ayriyan2021EPJA}%
  \BibitemOpen
  \bibfield  {author} {\bibinfo {author} {\bibfnamefont {A.}~\bibnamefont {{Ayriyan}}}, \bibinfo {author} {\bibfnamefont {D.}~\bibnamefont {{Blaschke}}}, \bibinfo {author} {\bibfnamefont {A.~G.}\ \bibnamefont {{Grunfeld}}}, \bibinfo {author} {\bibfnamefont {D.}~\bibnamefont {{Alvarez-Castillo}}}, \bibinfo {author} {\bibfnamefont {H.}~\bibnamefont {{Grigorian}}},\ and\ \bibinfo {author} {\bibfnamefont {V.}~\bibnamefont {{Abgaryan}}},\ }\bibfield  {title} {\bibinfo {title} {{Bayesian analysis of multimessenger M-R data with interpolated hybrid EoS}},\ }\href {https://doi.org/10.1140/epja/s10050-021-00619-0} {\bibfield  {journal} {\bibinfo  {journal} {European Physical Journal A}\ }\textbf {\bibinfo {volume} {57}},\ \bibinfo {eid} {318} (\bibinfo {year} {2021})},\ \Eprint {https://arxiv.org/abs/2102.13485} {arXiv:2102.13485 [astro-ph.HE]} \BibitemShut {NoStop}%
\bibitem [{\citenamefont {{Ascenzi}}\ \emph {et~al.}(2024)\citenamefont {{Ascenzi}}, \citenamefont {{Graber}},\ and\ \citenamefont {{Rea}}}]{Ascenzi2024arXiv}%
  \BibitemOpen
  \bibfield  {author} {\bibinfo {author} {\bibfnamefont {S.}~\bibnamefont {{Ascenzi}}}, \bibinfo {author} {\bibfnamefont {V.}~\bibnamefont {{Graber}}},\ and\ \bibinfo {author} {\bibfnamefont {N.}~\bibnamefont {{Rea}}},\ }\bibfield  {title} {\bibinfo {title} {{Neutron-star measurements in the multi-messenger Era}},\ }\href {https://doi.org/10.1016/j.astropartphys.2024.102935} {\bibfield  {journal} {\bibinfo  {journal} {Astroparticle Physics}\ }\textbf {\bibinfo {volume} {158}},\ \bibinfo {eid} {102935} (\bibinfo {year} {2024})},\ \Eprint {https://arxiv.org/abs/2401.14930} {arXiv:2401.14930 [astro-ph.HE]} \BibitemShut {NoStop}%
\bibitem [{\citenamefont {{Blinnikov}}\ \emph {et~al.}(2022)\citenamefont {{Blinnikov}}, \citenamefont {{Yudin}}, \citenamefont {{Kramarev}},\ and\ \citenamefont {{Potashov}}}]{Blinnikov+P2022}%
  \BibitemOpen
  \bibfield  {author} {\bibinfo {author} {\bibfnamefont {S.}~\bibnamefont {{Blinnikov}}}, \bibinfo {author} {\bibfnamefont {A.}~\bibnamefont {{Yudin}}}, \bibinfo {author} {\bibfnamefont {N.}~\bibnamefont {{Kramarev}}},\ and\ \bibinfo {author} {\bibfnamefont {M.}~\bibnamefont {{Potashov}}},\ }\bibfield  {title} {\bibinfo {title} {{Stripping Model for Short Gamma-Ray Bursts in Neutron Star Mergers}},\ }\href {https://doi.org/10.3390/particles5020018} {\bibfield  {journal} {\bibinfo  {journal} {Particles}\ }\textbf {\bibinfo {volume} {5}},\ \bibinfo {pages} {198} (\bibinfo {year} {2022})}\BibitemShut {NoStop}%
\bibitem [{\citenamefont {{Potekhin}}\ \emph {et~al.}(2020)\citenamefont {{Potekhin}}, \citenamefont {{Zyuzin}}, \citenamefont {{Yakovlev}}, \citenamefont {{Beznogov}},\ and\ \citenamefont {{Shibanov}}}]{Potekhin2020MNRAS}%
  \BibitemOpen
  \bibfield  {author} {\bibinfo {author} {\bibfnamefont {A.~Y.}\ \bibnamefont {{Potekhin}}}, \bibinfo {author} {\bibfnamefont {D.~A.}\ \bibnamefont {{Zyuzin}}}, \bibinfo {author} {\bibfnamefont {D.~G.}\ \bibnamefont {{Yakovlev}}}, \bibinfo {author} {\bibfnamefont {M.~V.}\ \bibnamefont {{Beznogov}}},\ and\ \bibinfo {author} {\bibfnamefont {Y.~A.}\ \bibnamefont {{Shibanov}}},\ }\bibfield  {title} {\bibinfo {title} {{Thermal luminosities of cooling neutron stars}},\ }\href {https://doi.org/10.1093/mnras/staa1871} {\bibfield  {journal} {\bibinfo  {journal} {\mnras}\ }\textbf {\bibinfo {volume} {496}},\ \bibinfo {pages} {5052} (\bibinfo {year} {2020})}\BibitemShut {NoStop}%
\bibitem [{\citenamefont {{Doroshenko}}\ \emph {et~al.}(2022)\citenamefont {{Doroshenko}}, \citenamefont {{Suleimanov}}, \citenamefont {{P{\"u}hlhofer}},\ and\ \citenamefont {{Santangelo}}}]{Doroshenko2022NatAs}%
  \BibitemOpen
  \bibfield  {author} {\bibinfo {author} {\bibfnamefont {V.}~\bibnamefont {{Doroshenko}}}, \bibinfo {author} {\bibfnamefont {V.}~\bibnamefont {{Suleimanov}}}, \bibinfo {author} {\bibfnamefont {G.}~\bibnamefont {{P{\"u}hlhofer}}},\ and\ \bibinfo {author} {\bibfnamefont {A.}~\bibnamefont {{Santangelo}}},\ }\bibfield  {title} {\bibinfo {title} {{A strangely light neutron star within a supernova remnant}},\ }\href {https://doi.org/10.1038/s41550-022-01800-1} {\bibfield  {journal} {\bibinfo  {journal} {Nature Astronomy}\ }\textbf {\bibinfo {volume} {6}},\ \bibinfo {pages} {1444} (\bibinfo {year} {2022})}\BibitemShut {NoStop}%
\bibitem [{\citenamefont {{Alford}}\ and\ \citenamefont {{Halpern}}(2023)}]{Alford2023ApJ}%
  \BibitemOpen
  \bibfield  {author} {\bibinfo {author} {\bibfnamefont {J.~A.~J.}\ \bibnamefont {{Alford}}}\ and\ \bibinfo {author} {\bibfnamefont {J.~P.}\ \bibnamefont {{Halpern}}},\ }\bibfield  {title} {\bibinfo {title} {{Do Central Compact Objects have Carbon Atmospheres?}},\ }\href {https://doi.org/10.3847/1538-4357/acaf55} {\bibfield  {journal} {\bibinfo  {journal} {\apj}\ }\textbf {\bibinfo {volume} {944}},\ \bibinfo {eid} {36} (\bibinfo {year} {2023})},\ \Eprint {https://arxiv.org/abs/2302.05893} {arXiv:2302.05893 [astro-ph.HE]} \BibitemShut {NoStop}%
\bibitem [{\citenamefont {{Steiner}}\ \emph {et~al.}(2018)\citenamefont {{Steiner}}, \citenamefont {{Heinke}}, \citenamefont {{Bogdanov}}, \citenamefont {{Li}}, \citenamefont {{Ho}}, \citenamefont {{Bahramian}},\ and\ \citenamefont {{Han}}}]{Steiner2018MNRAS}%
  \BibitemOpen
  \bibfield  {author} {\bibinfo {author} {\bibfnamefont {A.~W.}\ \bibnamefont {{Steiner}}}, \bibinfo {author} {\bibfnamefont {C.~O.}\ \bibnamefont {{Heinke}}}, \bibinfo {author} {\bibfnamefont {S.}~\bibnamefont {{Bogdanov}}}, \bibinfo {author} {\bibfnamefont {C.~K.}\ \bibnamefont {{Li}}}, \bibinfo {author} {\bibfnamefont {W.~C.~G.}\ \bibnamefont {{Ho}}}, \bibinfo {author} {\bibfnamefont {A.}~\bibnamefont {{Bahramian}}},\ and\ \bibinfo {author} {\bibfnamefont {S.}~\bibnamefont {{Han}}},\ }\bibfield  {title} {\bibinfo {title} {{Constraining the mass and radius of neutron stars in globular clusters}},\ }\href {https://doi.org/10.1093/mnras/sty215} {\bibfield  {journal} {\bibinfo  {journal} {\mnras}\ }\textbf {\bibinfo {volume} {476}},\ \bibinfo {pages} {421} (\bibinfo {year} {2018})},\ \Eprint {https://arxiv.org/abs/1709.05013} {arXiv:1709.05013 [astro-ph.HE]} \BibitemShut {NoStop}%
\bibitem [{\citenamefont {{Marino}}\ \emph {et~al.}(2018)\citenamefont {{Marino}}, \citenamefont {{Degenaar}}, \citenamefont {{Di Salvo}}, \citenamefont {{Wijnands}}, \citenamefont {{Burderi}},\ and\ \citenamefont {{Iaria}}}]{Marino2018MNRAS}%
  \BibitemOpen
  \bibfield  {author} {\bibinfo {author} {\bibfnamefont {A.}~\bibnamefont {{Marino}}}, \bibinfo {author} {\bibfnamefont {N.}~\bibnamefont {{Degenaar}}}, \bibinfo {author} {\bibfnamefont {T.}~\bibnamefont {{Di Salvo}}}, \bibinfo {author} {\bibfnamefont {R.}~\bibnamefont {{Wijnands}}}, \bibinfo {author} {\bibfnamefont {L.}~\bibnamefont {{Burderi}}},\ and\ \bibinfo {author} {\bibfnamefont {R.}~\bibnamefont {{Iaria}}},\ }\bibfield  {title} {\bibinfo {title} {{On obtaining neutron star mass and radius constraints from quiescent low-mass X-ray binaries in the Galactic plane}},\ }\href {https://doi.org/10.1093/mnras/sty1585} {\bibfield  {journal} {\bibinfo  {journal} {\mnras}\ }\textbf {\bibinfo {volume} {479}},\ \bibinfo {pages} {3634} (\bibinfo {year} {2018})},\ \Eprint {https://arxiv.org/abs/1806.04557} {arXiv:1806.04557 [astro-ph.HE]} \BibitemShut {NoStop}%
\bibitem [{\citenamefont {{Lattimer}}(2019)}]{LattimerU2019}%
  \BibitemOpen
  \bibfield  {author} {\bibinfo {author} {\bibfnamefont {J.~M.}\ \bibnamefont {{Lattimer}}},\ }\bibfield  {title} {\bibinfo {title} {{Neutron Star Mass and Radius Measurements}},\ }\href {https://doi.org/10.3390/universe5070159} {\bibfield  {journal} {\bibinfo  {journal} {Universe}\ }\textbf {\bibinfo {volume} {5}},\ \bibinfo {eid} {159} (\bibinfo {year} {2019})}\BibitemShut {NoStop}%
\bibitem [{\citenamefont {{Barr}}\ \emph {et~al.}(2024)\citenamefont {{Barr}}, \citenamefont {{Dutta}}, \citenamefont {{Freire}}, \citenamefont {{Cadelano}}, \citenamefont {{Gautam}}, \citenamefont {{Kramer}}, \citenamefont {{Pallanca}}, \citenamefont {{Ransom}}, \citenamefont {{Ridolfi}}, \citenamefont {{Stappers}}, \citenamefont {{Tauris}}, \citenamefont {{Venkatraman Krishnan}}, \citenamefont {{Wex}}, \citenamefont {{Bailes}}, \citenamefont {{Behrend}}, \citenamefont {{Buchner}}, \citenamefont {{Burgay}}, \citenamefont {{Chen}}, \citenamefont {{Champion}}, \citenamefont {{Chen}}, \citenamefont {{Corongiu}}, \citenamefont {{Geyer}}, \citenamefont {{Men}}, \citenamefont {{Padmanabh}},\ and\ \citenamefont {{Possenti}}}]{Barr2024Sci}%
  \BibitemOpen
  \bibfield  {author} {\bibinfo {author} {\bibfnamefont {E.~D.}\ \bibnamefont {{Barr}}}, \bibinfo {author} {\bibfnamefont {A.}~\bibnamefont {{Dutta}}}, \bibinfo {author} {\bibfnamefont {P.~C.~C.}\ \bibnamefont {{Freire}}}, \bibinfo {author} {\bibfnamefont {M.}~\bibnamefont {{Cadelano}}}, \bibinfo {author} {\bibfnamefont {T.}~\bibnamefont {{Gautam}}}, \bibinfo {author} {\bibfnamefont {M.}~\bibnamefont {{Kramer}}}, \bibinfo {author} {\bibfnamefont {C.}~\bibnamefont {{Pallanca}}}, \bibinfo {author} {\bibfnamefont {S.~M.}\ \bibnamefont {{Ransom}}}, \bibinfo {author} {\bibfnamefont {A.}~\bibnamefont {{Ridolfi}}}, \bibinfo {author} {\bibfnamefont {B.~W.}\ \bibnamefont {{Stappers}}}, \bibinfo {author} {\bibfnamefont {T.~M.}\ \bibnamefont {{Tauris}}}, \bibinfo {author} {\bibfnamefont {V.}~\bibnamefont {{Venkatraman Krishnan}}}, \bibinfo {author} {\bibfnamefont {N.}~\bibnamefont {{Wex}}}, \bibinfo {author} {\bibfnamefont {M.}~\bibnamefont {{Bailes}}}, \bibinfo {author} {\bibfnamefont {J.}~\bibnamefont {{Behrend}}},
  \bibinfo {author} {\bibfnamefont {S.}~\bibnamefont {{Buchner}}}, \bibinfo {author} {\bibfnamefont {M.}~\bibnamefont {{Burgay}}}, \bibinfo {author} {\bibfnamefont {W.}~\bibnamefont {{Chen}}}, \bibinfo {author} {\bibfnamefont {D.~J.}\ \bibnamefont {{Champion}}}, \bibinfo {author} {\bibfnamefont {C.~H.~R.}\ \bibnamefont {{Chen}}}, \bibinfo {author} {\bibfnamefont {A.}~\bibnamefont {{Corongiu}}}, \bibinfo {author} {\bibfnamefont {M.}~\bibnamefont {{Geyer}}}, \bibinfo {author} {\bibfnamefont {Y.~P.}\ \bibnamefont {{Men}}}, \bibinfo {author} {\bibfnamefont {P.~V.}\ \bibnamefont {{Padmanabh}}},\ and\ \bibinfo {author} {\bibfnamefont {A.}~\bibnamefont {{Possenti}}},\ }\bibfield  {title} {\bibinfo {title} {{A pulsar in a binary with a compact object in the mass gap between neutron stars and black holes}},\ }\href {https://doi.org/10.1126/science.adg3005} {\bibfield  {journal} {\bibinfo  {journal} {Science}\ }\textbf {\bibinfo {volume} {383}},\ \bibinfo {pages} {275} (\bibinfo {year} {2024})},\ \Eprint
  {https://arxiv.org/abs/2401.09872} {arXiv:2401.09872 [astro-ph.HE]} \BibitemShut {NoStop}%
\bibitem [{\citenamefont {{Annala}}\ \emph {et~al.}(2023)\citenamefont {{Annala}}, \citenamefont {{Gorda}}, \citenamefont {{Hirvonen}}, \citenamefont {{Komoltsev}}, \citenamefont {{Kurkela}},\ and\ \citenamefont {et~al.}}]{Annala+Nat2023}%
  \BibitemOpen
  \bibfield  {author} {\bibinfo {author} {\bibfnamefont {E.}~\bibnamefont {{Annala}}}, \bibinfo {author} {\bibfnamefont {T.}~\bibnamefont {{Gorda}}}, \bibinfo {author} {\bibfnamefont {J.}~\bibnamefont {{Hirvonen}}}, \bibinfo {author} {\bibfnamefont {O.}~\bibnamefont {{Komoltsev}}}, \bibinfo {author} {\bibfnamefont {A.}~\bibnamefont {{Kurkela}}},\ and\ \bibinfo {author} {\bibnamefont {et~al.}},\ }\bibfield  {title} {\bibinfo {title} {{Strongly interacting matter exhibits deconfined behavior in massive neutron stars}},\ }\href {https://doi.org/https://doi.org/10.1038/s41467-023-44051-y} {\bibfield  {journal} {\bibinfo  {journal} {Nat. Commun.}\ }\textbf {\bibinfo {volume} {14}},\ \bibinfo {eid} {arXiv:2303.11356} (\bibinfo {year} {2023})}\BibitemShut {NoStop}%
\bibitem [{\citenamefont {{Brandes}}\ \emph {et~al.}(2023{\natexlab{a}})\citenamefont {{Brandes}}, \citenamefont {{Weise}},\ and\ \citenamefont {{Kaiser}}}]{Brandes+PRD2023}%
  \BibitemOpen
  \bibfield  {author} {\bibinfo {author} {\bibfnamefont {L.}~\bibnamefont {{Brandes}}}, \bibinfo {author} {\bibfnamefont {W.}~\bibnamefont {{Weise}}},\ and\ \bibinfo {author} {\bibfnamefont {N.}~\bibnamefont {{Kaiser}}},\ }\bibfield  {title} {\bibinfo {title} {{Evidence against a strong first-order phase transition in neutron star cores: Impact of new data}},\ }\href {https://doi.org/10.1103/PhysRevD.108.094014} {\bibfield  {journal} {\bibinfo  {journal} {\prd}\ }\textbf {\bibinfo {volume} {108}},\ \bibinfo {eid} {094014} (\bibinfo {year} {2023}{\natexlab{a}})},\ \Eprint {https://arxiv.org/abs/2306.06218} {arXiv:2306.06218 [nucl-th]} \BibitemShut {NoStop}%
\bibitem [{\citenamefont {{Brandes}}\ \emph {et~al.}(2023{\natexlab{b}})\citenamefont {{Brandes}}, \citenamefont {{Weise}},\ and\ \citenamefont {{Kaiser}}}]{Brandes2023PhRvDa}%
  \BibitemOpen
  \bibfield  {author} {\bibinfo {author} {\bibfnamefont {L.}~\bibnamefont {{Brandes}}}, \bibinfo {author} {\bibfnamefont {W.}~\bibnamefont {{Weise}}},\ and\ \bibinfo {author} {\bibfnamefont {N.}~\bibnamefont {{Kaiser}}},\ }\bibfield  {title} {\bibinfo {title} {{Inference of the sound speed and related properties of neutron stars}},\ }\href {https://doi.org/10.1103/PhysRevD.107.014011} {\bibfield  {journal} {\bibinfo  {journal} {\prd}\ }\textbf {\bibinfo {volume} {107}},\ \bibinfo {eid} {014011} (\bibinfo {year} {2023}{\natexlab{b}})},\ \Eprint {https://arxiv.org/abs/2208.03026} {arXiv:2208.03026 [nucl-th]} \BibitemShut {NoStop}%
\bibitem [{\citenamefont {Ofengeim}\ \emph {et~al.}(2024)\citenamefont {Ofengeim}, \citenamefont {Shternin},\ and\ \citenamefont {Piran}}]{PaperData}%
  \BibitemOpen
  \bibfield  {author} {\bibinfo {author} {\bibfnamefont {D.}~\bibnamefont {Ofengeim}}, \bibinfo {author} {\bibfnamefont {P.}~\bibnamefont {Shternin}},\ and\ \bibinfo {author} {\bibfnamefont {T.}~\bibnamefont {Piran}},\ }\bibfield  {title} {\bibinfo {title} {{A three-parameter characterization of neutron stars' mass-radius relation and equation of state [Data Set]}},\ }\href {https://doi.org/10.5281/zenodo.12819983} {10.5281/zenodo.12819983} (\bibinfo {year} {2024})\BibitemShut {NoStop}%
\bibitem [{\citenamefont {{Lattimer}}\ and\ \citenamefont {{Prakash}}(2016)}]{LattPrak2016}%
  \BibitemOpen
  \bibfield  {author} {\bibinfo {author} {\bibfnamefont {J.~M.}\ \bibnamefont {{Lattimer}}}\ and\ \bibinfo {author} {\bibfnamefont {M.}~\bibnamefont {{Prakash}}},\ }\bibfield  {title} {\bibinfo {title} {{The equation of state of hot, dense matter and neutron stars}},\ }\href {https://doi.org/10.1016/j.physrep.2015.12.005} {\bibfield  {journal} {\bibinfo  {journal} {\physrep}\ }\textbf {\bibinfo {volume} {621}},\ \bibinfo {pages} {127} (\bibinfo {year} {2016})},\ \Eprint {https://arxiv.org/abs/1512.07820} {arXiv:1512.07820 [astro-ph.SR]} \BibitemShut {NoStop}%
\bibitem [{\citenamefont {{Komoltsev}}\ and\ \citenamefont {{Kurkela}}(2022)}]{KomoltsevKurkelaPRL2022}%
  \BibitemOpen
  \bibfield  {author} {\bibinfo {author} {\bibfnamefont {O.}~\bibnamefont {{Komoltsev}}}\ and\ \bibinfo {author} {\bibfnamefont {A.}~\bibnamefont {{Kurkela}}},\ }\bibfield  {title} {\bibinfo {title} {{How Perturbative QCD Constrains the Equation of State at Neutron-Star Densities}},\ }\href {https://doi.org/10.1103/PhysRevLett.128.202701} {\bibfield  {journal} {\bibinfo  {journal} {\prl}\ }\textbf {\bibinfo {volume} {128}},\ \bibinfo {eid} {202701} (\bibinfo {year} {2022})},\ \Eprint {https://arxiv.org/abs/2111.05350} {arXiv:2111.05350 [nucl-th]} \BibitemShut {NoStop}%
\bibitem [{\citenamefont {{Hebeler}}\ \emph {et~al.}(2013)\citenamefont {{Hebeler}}, \citenamefont {{Lattimer}}, \citenamefont {{Pethick}},\ and\ \citenamefont {{Schwenk}}}]{Hebeler+ApJ2013}%
  \BibitemOpen
  \bibfield  {author} {\bibinfo {author} {\bibfnamefont {K.}~\bibnamefont {{Hebeler}}}, \bibinfo {author} {\bibfnamefont {J.~M.}\ \bibnamefont {{Lattimer}}}, \bibinfo {author} {\bibfnamefont {C.~J.}\ \bibnamefont {{Pethick}}},\ and\ \bibinfo {author} {\bibfnamefont {A.}~\bibnamefont {{Schwenk}}},\ }\bibfield  {title} {\bibinfo {title} {{Equation of State and Neutron Star Properties Constrained by Nuclear Physics and Observation}},\ }\href {https://doi.org/10.1088/0004-637X/773/1/11} {\bibfield  {journal} {\bibinfo  {journal} {\apj}\ }\textbf {\bibinfo {volume} {773}},\ \bibinfo {eid} {11} (\bibinfo {year} {2013})},\ \Eprint {https://arxiv.org/abs/1303.4662} {arXiv:1303.4662 [astro-ph.SR]} \BibitemShut {NoStop}%
\bibitem [{\citenamefont {{Lattimer}}(2021)}]{Lattimer2021}%
  \BibitemOpen
  \bibfield  {author} {\bibinfo {author} {\bibfnamefont {J.~M.}\ \bibnamefont {{Lattimer}}},\ }\bibfield  {title} {\bibinfo {title} {{Neutron Stars and the Nuclear Matter Equation of State}},\ }\href {https://doi.org/10.1146/annurev-nucl-102419-124827} {\bibfield  {journal} {\bibinfo  {journal} {Annual Review of Nuclear and Particle Science}\ }\textbf {\bibinfo {volume} {71}},\ \bibinfo {pages} {433} (\bibinfo {year} {2021})}\BibitemShut {NoStop}%
\bibitem [{\citenamefont {{Lindblom}}\ and\ \citenamefont {{Indik}}(2012)}]{LindInd2012}%
  \BibitemOpen
  \bibfield  {author} {\bibinfo {author} {\bibfnamefont {L.}~\bibnamefont {{Lindblom}}}\ and\ \bibinfo {author} {\bibfnamefont {N.~M.}\ \bibnamefont {{Indik}}},\ }\bibfield  {title} {\bibinfo {title} {{Spectral approach to the relativistic inverse stellar structure problem}},\ }\href {https://doi.org/10.1103/PhysRevD.86.084003} {\bibfield  {journal} {\bibinfo  {journal} {\prd}\ }\textbf {\bibinfo {volume} {86}},\ \bibinfo {eid} {084003} (\bibinfo {year} {2012})},\ \Eprint {https://arxiv.org/abs/1207.3744} {arXiv:1207.3744 [astro-ph.HE]} \BibitemShut {NoStop}%
\bibitem [{\citenamefont {{Soma}}\ \emph {et~al.}(2022)\citenamefont {{Soma}}, \citenamefont {{Wang}}, \citenamefont {{Shi}}, \citenamefont {{St{\"o}cker}}, \citenamefont {{Zhou}},\ and\ \citenamefont {et~al.}}]{Soma+2022}%
  \BibitemOpen
  \bibfield  {author} {\bibinfo {author} {\bibfnamefont {S.}~\bibnamefont {{Soma}}}, \bibinfo {author} {\bibfnamefont {L.}~\bibnamefont {{Wang}}}, \bibinfo {author} {\bibfnamefont {S.}~\bibnamefont {{Shi}}}, \bibinfo {author} {\bibfnamefont {H.}~\bibnamefont {{St{\"o}cker}}}, \bibinfo {author} {\bibfnamefont {K.}~\bibnamefont {{Zhou}}},\ and\ \bibinfo {author} {\bibnamefont {et~al.}},\ }\bibfield  {title} {\bibinfo {title} {{Neural network reconstruction of the dense matter equation of state from neutron star observables}},\ }\href {https://doi.org/10.1088/1475-7516/2022/08/071} {\bibfield  {journal} {\bibinfo  {journal} {\jcap}\ }\textbf {\bibinfo {volume} {2022}},\ \bibinfo {eid} {071} (\bibinfo {year} {2022})},\ \Eprint {https://arxiv.org/abs/2201.01756} {arXiv:2201.01756 [hep-ph]} \BibitemShut {NoStop}%
\bibitem [{\citenamefont {{Lindblom}}(2010)}]{Lindblom2010}%
  \BibitemOpen
  \bibfield  {author} {\bibinfo {author} {\bibfnamefont {L.}~\bibnamefont {{Lindblom}}},\ }\bibfield  {title} {\bibinfo {title} {{Spectral representations of neutron-star equations of state}},\ }\href {https://doi.org/10.1103/PhysRevD.82.103011} {\bibfield  {journal} {\bibinfo  {journal} {\prd}\ }\textbf {\bibinfo {volume} {82}},\ \bibinfo {eid} {103011} (\bibinfo {year} {2010})},\ \Eprint {https://arxiv.org/abs/1009.0738} {arXiv:1009.0738 [astro-ph.HE]} \BibitemShut {NoStop}%
\bibitem [{\citenamefont {{Typel}}\ \emph {et~al.}(2015)\citenamefont {{Typel}}, \citenamefont {{Oertel}},\ and\ \citenamefont {{Kl{\"a}hn}}}]{CompOSE2015}%
  \BibitemOpen
  \bibfield  {author} {\bibinfo {author} {\bibfnamefont {S.}~\bibnamefont {{Typel}}}, \bibinfo {author} {\bibfnamefont {M.}~\bibnamefont {{Oertel}}},\ and\ \bibinfo {author} {\bibfnamefont {T.}~\bibnamefont {{Kl{\"a}hn}}},\ }\bibfield  {title} {\bibinfo {title} {{CompOSE CompStar online supernova equations of state harmonising the concert of nuclear physics and astrophysics compose.obspm.fr}},\ }\href {https://doi.org/10.1134/S1063779615040061} {\bibfield  {journal} {\bibinfo  {journal} {Physics of Particles and Nuclei}\ }\textbf {\bibinfo {volume} {46}},\ \bibinfo {pages} {633} (\bibinfo {year} {2015})}\BibitemShut {NoStop}%
\bibitem [{\citenamefont {{Read}}\ \emph {et~al.}(2009)\citenamefont {{Read}}, \citenamefont {{Lackey}}, \citenamefont {{Owen}},\ and\ \citenamefont {{Friedman}}}]{Read+2009}%
  \BibitemOpen
  \bibfield  {author} {\bibinfo {author} {\bibfnamefont {J.~S.}\ \bibnamefont {{Read}}}, \bibinfo {author} {\bibfnamefont {B.~D.}\ \bibnamefont {{Lackey}}}, \bibinfo {author} {\bibfnamefont {B.~J.}\ \bibnamefont {{Owen}}},\ and\ \bibinfo {author} {\bibfnamefont {J.~L.}\ \bibnamefont {{Friedman}}},\ }\bibfield  {title} {\bibinfo {title} {{Constraints on a phenomenologically parametrized neutron-star equation of state}},\ }\href {https://doi.org/10.1103/PhysRevD.79.124032} {\bibfield  {journal} {\bibinfo  {journal} {\prd}\ }\textbf {\bibinfo {volume} {79}},\ \bibinfo {eid} {124032} (\bibinfo {year} {2009})},\ \Eprint {https://arxiv.org/abs/0812.2163} {arXiv:0812.2163 [astro-ph]} \BibitemShut {NoStop}%
\bibitem [{\citenamefont {{{\"O}zel}}\ and\ \citenamefont {{Freire}}(2016)}]{OzelFreire2016}%
  \BibitemOpen
  \bibfield  {author} {\bibinfo {author} {\bibfnamefont {F.}~\bibnamefont {{{\"O}zel}}}\ and\ \bibinfo {author} {\bibfnamefont {P.}~\bibnamefont {{Freire}}},\ }\bibfield  {title} {\bibinfo {title} {{Masses, Radii, and the Equation of State of Neutron Stars}},\ }\href {https://doi.org/10.1146/annurev-astro-081915-023322} {\bibfield  {journal} {\bibinfo  {journal} {\araa}\ }\textbf {\bibinfo {volume} {54}},\ \bibinfo {pages} {401} (\bibinfo {year} {2016})},\ \Eprint {https://arxiv.org/abs/1603.02698} {arXiv:1603.02698 [astro-ph.HE]} \BibitemShut {NoStop}%
\bibitem [{\citenamefont {{Maslov}}\ \emph {et~al.}(2016)\citenamefont {{Maslov}}, \citenamefont {{Kolomeitsev}},\ and\ \citenamefont {{Voskresensky}}}]{MKV2016}%
  \BibitemOpen
  \bibfield  {author} {\bibinfo {author} {\bibfnamefont {K.~A.}\ \bibnamefont {{Maslov}}}, \bibinfo {author} {\bibfnamefont {E.~E.}\ \bibnamefont {{Kolomeitsev}}},\ and\ \bibinfo {author} {\bibfnamefont {D.~N.}\ \bibnamefont {{Voskresensky}}},\ }\bibfield  {title} {\bibinfo {title} {{Relativistic mean-field models with scaled hadron masses and couplings: Hyperons and maximum neutron star mass}},\ }\href {https://doi.org/10.1016/j.nuclphysa.2016.03.011} {\bibfield  {journal} {\bibinfo  {journal} {\nphysa}\ }\textbf {\bibinfo {volume} {950}},\ \bibinfo {pages} {64} (\bibinfo {year} {2016})},\ \Eprint {https://arxiv.org/abs/1509.02538} {arXiv:1509.02538 [astro-ph.HE]} \BibitemShut {NoStop}%
\bibitem [{\citenamefont {{Pearson}}\ \emph {et~al.}(2018)\citenamefont {{Pearson}}, \citenamefont {{Chamel}}, \citenamefont {{Potekhin}}, \citenamefont {{Fantina}}, \citenamefont {{Ducoin}}, \citenamefont {{Dutta}},\ and\ \citenamefont {{Goriely}}}]{BSk2018}%
  \BibitemOpen
  \bibfield  {author} {\bibinfo {author} {\bibfnamefont {J.~M.}\ \bibnamefont {{Pearson}}}, \bibinfo {author} {\bibfnamefont {N.}~\bibnamefont {{Chamel}}}, \bibinfo {author} {\bibfnamefont {A.~Y.}\ \bibnamefont {{Potekhin}}}, \bibinfo {author} {\bibfnamefont {A.~F.}\ \bibnamefont {{Fantina}}}, \bibinfo {author} {\bibfnamefont {C.}~\bibnamefont {{Ducoin}}}, \bibinfo {author} {\bibfnamefont {A.~K.}\ \bibnamefont {{Dutta}}},\ and\ \bibinfo {author} {\bibfnamefont {S.}~\bibnamefont {{Goriely}}},\ }\bibfield  {title} {\bibinfo {title} {{Unified equations of state for cold non-accreting neutron stars with Brussels-Montreal functionals - I. Role of symmetry energy}},\ }\href {https://doi.org/10.1093/mnras/sty2413} {\bibfield  {journal} {\bibinfo  {journal} {\mnras}\ }\textbf {\bibinfo {volume} {481}},\ \bibinfo {pages} {2994} (\bibinfo {year} {2018})},\ \Eprint {https://arxiv.org/abs/1903.04981} {arXiv:1903.04981 [astro-ph.HE]} \BibitemShut {NoStop}%
\bibitem [{\citenamefont {{J{\"a}rvinen}}(2022)}]{JarvinenEPJC2022}%
  \BibitemOpen
  \bibfield  {author} {\bibinfo {author} {\bibfnamefont {M.}~\bibnamefont {{J{\"a}rvinen}}},\ }\bibfield  {title} {\bibinfo {title} {{Holographic modeling of nuclear matter and neutron stars}},\ }\href {https://doi.org/10.1140/epjc/s10052-022-10227-x} {\bibfield  {journal} {\bibinfo  {journal} {European Physical Journal C}\ }\textbf {\bibinfo {volume} {82}},\ \bibinfo {eid} {282} (\bibinfo {year} {2022})},\ \Eprint {https://arxiv.org/abs/2110.08281} {arXiv:2110.08281 [hep-ph]} \BibitemShut {NoStop}%
\bibitem [{\citenamefont {{Haensel}}\ \emph {et~al.}(2007)\citenamefont {{Haensel}}, \citenamefont {{Potekhin}},\ and\ \citenamefont {{Yakovlev}}}]{HPY2007}%
  \BibitemOpen
  \bibfield  {author} {\bibinfo {author} {\bibfnamefont {P.}~\bibnamefont {{Haensel}}}, \bibinfo {author} {\bibfnamefont {A.~Y.}\ \bibnamefont {{Potekhin}}},\ and\ \bibinfo {author} {\bibfnamefont {D.~G.}\ \bibnamefont {{Yakovlev}}},\ }\href@noop {} {\emph {\bibinfo {title} {{Neutron Stars 1 : Equation of State and Structure}}}},\ \bibinfo {series} {Astrophysics and Space Science Library}, Vol.\ \bibinfo {volume} {326}\ (\bibinfo  {publisher} {{Springer}},\ \bibinfo {address} {{New York}},\ \bibinfo {year} {2007})\BibitemShut {NoStop}%
\bibitem [{\citenamefont {{Gusakov}}\ \emph {et~al.}(2005)\citenamefont {{Gusakov}}, \citenamefont {{Kaminker}}, \citenamefont {{Yakovlev}},\ and\ \citenamefont {{Gnedin}}}]{Gusakov+2005}%
  \BibitemOpen
  \bibfield  {author} {\bibinfo {author} {\bibfnamefont {M.~E.}\ \bibnamefont {{Gusakov}}}, \bibinfo {author} {\bibfnamefont {A.~D.}\ \bibnamefont {{Kaminker}}}, \bibinfo {author} {\bibfnamefont {D.~G.}\ \bibnamefont {{Yakovlev}}},\ and\ \bibinfo {author} {\bibfnamefont {O.~Y.}\ \bibnamefont {{Gnedin}}},\ }\bibfield  {title} {\bibinfo {title} {{The cooling of Akmal-Pandharipande-Ravenhall neutron star models}},\ }\href {https://doi.org/10.1111/j.1365-2966.2005.09459.x} {\bibfield  {journal} {\bibinfo  {journal} {\mnras}\ }\textbf {\bibinfo {volume} {363}},\ \bibinfo {pages} {555} (\bibinfo {year} {2005})},\ \Eprint {https://arxiv.org/abs/astro-ph/0507560} {arXiv:astro-ph/0507560 [astro-ph]} \BibitemShut {NoStop}%
\bibitem [{\citenamefont {{Kaminker}}\ \emph {et~al.}(2014)\citenamefont {{Kaminker}}, \citenamefont {{Kaurov}}, \citenamefont {{Potekhin}},\ and\ \citenamefont {{Yakovlev}}}]{Kaurov+2014}%
  \BibitemOpen
  \bibfield  {author} {\bibinfo {author} {\bibfnamefont {A.~D.}\ \bibnamefont {{Kaminker}}}, \bibinfo {author} {\bibfnamefont {A.~A.}\ \bibnamefont {{Kaurov}}}, \bibinfo {author} {\bibfnamefont {A.~Y.}\ \bibnamefont {{Potekhin}}},\ and\ \bibinfo {author} {\bibfnamefont {D.~G.}\ \bibnamefont {{Yakovlev}}},\ }\bibfield  {title} {\bibinfo {title} {{Thermal emission of neutron stars with internal heaters}},\ }\href {https://doi.org/10.1093/mnras/stu1102} {\bibfield  {journal} {\bibinfo  {journal} {\mnras}\ }\textbf {\bibinfo {volume} {442}},\ \bibinfo {pages} {3484} (\bibinfo {year} {2014})},\ \Eprint {https://arxiv.org/abs/1406.0723} {arXiv:1406.0723 [astro-ph.HE]} \BibitemShut {NoStop}%
\bibitem [{\citenamefont {{Potekhin}}\ \emph {et~al.}(2013)\citenamefont {{Potekhin}}, \citenamefont {{Fantina}}, \citenamefont {{Chamel}}, \citenamefont {{Pearson}},\ and\ \citenamefont {{Goriely}}}]{BSk2013}%
  \BibitemOpen
  \bibfield  {author} {\bibinfo {author} {\bibfnamefont {A.~Y.}\ \bibnamefont {{Potekhin}}}, \bibinfo {author} {\bibfnamefont {A.~F.}\ \bibnamefont {{Fantina}}}, \bibinfo {author} {\bibfnamefont {N.}~\bibnamefont {{Chamel}}}, \bibinfo {author} {\bibfnamefont {J.~M.}\ \bibnamefont {{Pearson}}},\ and\ \bibinfo {author} {\bibfnamefont {S.}~\bibnamefont {{Goriely}}},\ }\bibfield  {title} {\bibinfo {title} {{Analytical representations of unified equations of state for neutron-star matter}},\ }\href {https://doi.org/10.1051/0004-6361/201321697} {\bibfield  {journal} {\bibinfo  {journal} {\aap}\ }\textbf {\bibinfo {volume} {560}},\ \bibinfo {eid} {A48} (\bibinfo {year} {2013})},\ \Eprint {https://arxiv.org/abs/1310.0049} {arXiv:1310.0049 [astro-ph.SR]} \BibitemShut {NoStop}%
\bibitem [{\citenamefont {{Yakovlev}}\ \emph {et~al.}(2011)\citenamefont {{Yakovlev}}, \citenamefont {{Ho}}, \citenamefont {{Shternin}}, \citenamefont {{Heinke}},\ and\ \citenamefont {{Potekhin}}}]{Yakovlev+2011}%
  \BibitemOpen
  \bibfield  {author} {\bibinfo {author} {\bibfnamefont {D.~G.}\ \bibnamefont {{Yakovlev}}}, \bibinfo {author} {\bibfnamefont {W.~C.~G.}\ \bibnamefont {{Ho}}}, \bibinfo {author} {\bibfnamefont {P.~S.}\ \bibnamefont {{Shternin}}}, \bibinfo {author} {\bibfnamefont {C.~O.}\ \bibnamefont {{Heinke}}},\ and\ \bibinfo {author} {\bibfnamefont {A.~Y.}\ \bibnamefont {{Potekhin}}},\ }\bibfield  {title} {\bibinfo {title} {{Cooling rates of neutron stars and the young neutron star in the Cassiopeia A supernova remnant}},\ }\href {https://doi.org/10.1111/j.1365-2966.2010.17827.x} {\bibfield  {journal} {\bibinfo  {journal} {\mnras}\ }\textbf {\bibinfo {volume} {411}},\ \bibinfo {pages} {1977} (\bibinfo {year} {2011})},\ \Eprint {https://arxiv.org/abs/1010.1154} {arXiv:1010.1154 [astro-ph.HE]} \BibitemShut {NoStop}%
\bibitem [{\citenamefont {{Ofengeim}}\ \emph {et~al.}(2019)\citenamefont {{Ofengeim}}, \citenamefont {{Gusakov}}, \citenamefont {{Haensel}},\ and\ \citenamefont {{Fortin}}}]{Ofengeim+2019}%
  \BibitemOpen
  \bibfield  {author} {\bibinfo {author} {\bibfnamefont {D.~D.}\ \bibnamefont {{Ofengeim}}}, \bibinfo {author} {\bibfnamefont {M.~E.}\ \bibnamefont {{Gusakov}}}, \bibinfo {author} {\bibfnamefont {P.}~\bibnamefont {{Haensel}}},\ and\ \bibinfo {author} {\bibfnamefont {M.}~\bibnamefont {{Fortin}}},\ }\bibfield  {title} {\bibinfo {title} {{Bulk viscosity in neutron stars with hyperon cores}},\ }\href {https://doi.org/10.1103/PhysRevD.100.103017} {\bibfield  {journal} {\bibinfo  {journal} {\prd}\ }\textbf {\bibinfo {volume} {100}},\ \bibinfo {eid} {103017} (\bibinfo {year} {2019})},\ \Eprint {https://arxiv.org/abs/1911.08407} {arXiv:1911.08407 [astro-ph.HE]} \BibitemShut {NoStop}%
\bibitem [{\citenamefont {{Gusakov}}\ \emph {et~al.}(2014)\citenamefont {{Gusakov}}, \citenamefont {{Haensel}},\ and\ \citenamefont {{Kantor}}}]{Gusakov+2014}%
  \BibitemOpen
  \bibfield  {author} {\bibinfo {author} {\bibfnamefont {M.~E.}\ \bibnamefont {{Gusakov}}}, \bibinfo {author} {\bibfnamefont {P.}~\bibnamefont {{Haensel}}},\ and\ \bibinfo {author} {\bibfnamefont {E.~M.}\ \bibnamefont {{Kantor}}},\ }\bibfield  {title} {\bibinfo {title} {{Physics input for modelling superfluid neutron stars with hyperon cores}},\ }\href {https://doi.org/10.1093/mnras/stt2438} {\bibfield  {journal} {\bibinfo  {journal} {\mnras}\ }\textbf {\bibinfo {volume} {439}},\ \bibinfo {pages} {318} (\bibinfo {year} {2014})},\ \Eprint {https://arxiv.org/abs/1401.2827} {arXiv:1401.2827 [astro-ph.HE]} \BibitemShut {NoStop}%
\bibitem [{\citenamefont {{Brandes}}\ \emph {et~al.}(2024)\citenamefont {{Brandes}}, \citenamefont {{Modi}}, \citenamefont {{Ghosh}}, \citenamefont {{Farrell}}, \citenamefont {{Lindblom}},\ and\ \citenamefont {et~al.}}]{Brandes+2024mar}%
  \BibitemOpen
  \bibfield  {author} {\bibinfo {author} {\bibfnamefont {L.}~\bibnamefont {{Brandes}}}, \bibinfo {author} {\bibfnamefont {C.}~\bibnamefont {{Modi}}}, \bibinfo {author} {\bibfnamefont {A.}~\bibnamefont {{Ghosh}}}, \bibinfo {author} {\bibfnamefont {D.}~\bibnamefont {{Farrell}}}, \bibinfo {author} {\bibfnamefont {L.}~\bibnamefont {{Lindblom}}},\ and\ \bibinfo {author} {\bibnamefont {et~al.}},\ }\bibfield  {title} {\bibinfo {title} {{Neural Simulation-Based Inference of the Neutron Star Equation of State directly from Telescope Spectra}},\ }\href@noop {} {\bibfield  {journal} {\bibinfo  {journal} {arXiv e-prints}\ ,\ \bibinfo {eid} {arXiv:2403.00287}} (\bibinfo {year} {2024})},\ \Eprint {https://arxiv.org/abs/2403.00287} {arXiv:2403.00287 [astro-ph.HE]} \BibitemShut {NoStop}%
\bibitem [{\citenamefont {{Fan}}\ \emph {et~al.}(2024)\citenamefont {{Fan}}, \citenamefont {{Han}}, \citenamefont {{Jiang}}, \citenamefont {{Shao}}, \citenamefont {{Tang}},\ and\ \citenamefont {et~al.}}]{Fan+PRD2024}%
  \BibitemOpen
  \bibfield  {author} {\bibinfo {author} {\bibfnamefont {Y.-Z.}\ \bibnamefont {{Fan}}}, \bibinfo {author} {\bibfnamefont {M.-Z.}\ \bibnamefont {{Han}}}, \bibinfo {author} {\bibfnamefont {J.-L.}\ \bibnamefont {{Jiang}}}, \bibinfo {author} {\bibfnamefont {D.-S.}\ \bibnamefont {{Shao}}}, \bibinfo {author} {\bibfnamefont {S.-P.}\ \bibnamefont {{Tang}}},\ and\ \bibinfo {author} {\bibnamefont {et~al.}},\ }\bibfield  {title} {\bibinfo {title} {{Maximum gravitational mass M$_{TOV}$=2.2 5$_{-0.07}$$^{+0.08}$M$_{{\ensuremath{\odot}}}$ inferred at about 3\% precision with multimessenger data of neutron stars}},\ }\href {https://doi.org/10.1103/PhysRevD.109.043052} {\bibfield  {journal} {\bibinfo  {journal} {\prd}\ }\textbf {\bibinfo {volume} {109}},\ \bibinfo {eid} {043052} (\bibinfo {year} {2024})},\ \Eprint {https://arxiv.org/abs/2309.12644} {arXiv:2309.12644 [astro-ph.HE]} \BibitemShut {NoStop}%
\bibitem [{\citenamefont {Schweder}\ and\ \citenamefont {Hjort}(2016)}]{Schweder_Hjort_2016}%
  \BibitemOpen
  \bibfield  {author} {\bibinfo {author} {\bibfnamefont {T.}~\bibnamefont {Schweder}}\ and\ \bibinfo {author} {\bibfnamefont {N.~L.}\ \bibnamefont {Hjort}},\ }\href@noop {} {\emph {\bibinfo {title} {Confidence, Likelihood, Probability: Statistical Inference with Confidence Distributions}}},\ Cambridge Series in Statistical and Probabilistic Mathematics\ (\bibinfo  {publisher} {Cambridge University Press},\ \bibinfo {year} {2016})\BibitemShut {NoStop}%
\bibitem [{\citenamefont {{Hernandez Vivanco}}\ \emph {et~al.}(2020)\citenamefont {{Hernandez Vivanco}}, \citenamefont {{Smith}}, \citenamefont {{Thrane}},\ and\ \citenamefont {{Lasky}}}]{Hernandez2020MNRAS}%
  \BibitemOpen
  \bibfield  {author} {\bibinfo {author} {\bibfnamefont {F.}~\bibnamefont {{Hernandez Vivanco}}}, \bibinfo {author} {\bibfnamefont {R.}~\bibnamefont {{Smith}}}, \bibinfo {author} {\bibfnamefont {E.}~\bibnamefont {{Thrane}}},\ and\ \bibinfo {author} {\bibfnamefont {P.~D.}\ \bibnamefont {{Lasky}}},\ }\bibfield  {title} {\bibinfo {title} {{A scalable random forest regressor for combining neutron-star equation of state measurements: a case study with GW170817 and GW190425}},\ }\href {https://doi.org/10.1093/mnras/staa3243} {\bibfield  {journal} {\bibinfo  {journal} {\mnras}\ }\textbf {\bibinfo {volume} {499}},\ \bibinfo {pages} {5972} (\bibinfo {year} {2020})}\BibitemShut {NoStop}%
\bibitem [{\citenamefont {{Bungert}}\ and\ \citenamefont {{Wacker}}(2020)}]{Bungert2020arXiv}%
  \BibitemOpen
  \bibfield  {author} {\bibinfo {author} {\bibfnamefont {L.}~\bibnamefont {{Bungert}}}\ and\ \bibinfo {author} {\bibfnamefont {P.}~\bibnamefont {{Wacker}}},\ }\bibfield  {title} {\bibinfo {title} {{The lion in the attic -- A resolution of the Borel--Kolmogorov paradox}},\ }\href {https://doi.org/10.48550/arXiv.2009.04778} {\bibfield  {journal} {\bibinfo  {journal} {arXiv e-prints}\ ,\ \bibinfo {eid} {arXiv:2009.04778}} (\bibinfo {year} {2020})},\ \Eprint {https://arxiv.org/abs/2009.04778} {arXiv:2009.04778 [math.PR]} \BibitemShut {NoStop}%
\bibitem [{\citenamefont {{Poole}}\ and\ \citenamefont {{Raftery}}(2000)}]{Poole2000}%
  \BibitemOpen
  \bibfield  {author} {\bibinfo {author} {\bibfnamefont {D.}~\bibnamefont {{Poole}}}\ and\ \bibinfo {author} {\bibfnamefont {A.~E.}\ \bibnamefont {{Raftery}}},\ }\bibfield  {title} {\bibinfo {title} {{Inference for Deterministic Simulation Models: The Bayesian Melding Approach}},\ }\href {https://doi.org/10.2307/2669764} {\bibfield  {journal} {\bibinfo  {journal} {Journal of the American Statistical Association}\ }\textbf {\bibinfo {volume} {95}},\ \bibinfo {pages} {1244} (\bibinfo {year} {2000})}\BibitemShut {NoStop}%
\bibitem [{\citenamefont {{Steiner}}(2018)}]{Steiner2018arXiv}%
  \BibitemOpen
  \bibfield  {author} {\bibinfo {author} {\bibfnamefont {A.~W.}\ \bibnamefont {{Steiner}}},\ }\bibfield  {title} {\bibinfo {title} {{Two- and Multi-dimensional Curve Fitting using Bayesian Inference}},\ }\href {https://doi.org/10.48550/arXiv.1802.05339} {\bibfield  {journal} {\bibinfo  {journal} {arXiv e-prints}\ ,\ \bibinfo {eid} {arXiv:1802.05339}} (\bibinfo {year} {2018})},\ \Eprint {https://arxiv.org/abs/1802.05339} {arXiv:1802.05339 [physics.data-an]} \BibitemShut {NoStop}%
\bibitem [{\citenamefont {{Pihajoki}}(2017)}]{Pihajoki2017MNRAS}%
  \BibitemOpen
  \bibfield  {author} {\bibinfo {author} {\bibfnamefont {P.}~\bibnamefont {{Pihajoki}}},\ }\bibfield  {title} {\bibinfo {title} {{A geometric approach to non-linear correlations with intrinsic scatter}},\ }\href {https://doi.org/10.1093/mnras/stx2179} {\bibfield  {journal} {\bibinfo  {journal} {\mnras}\ }\textbf {\bibinfo {volume} {472}},\ \bibinfo {pages} {3407} (\bibinfo {year} {2017})},\ \Eprint {https://arxiv.org/abs/1704.05466} {arXiv:1704.05466 [astro-ph.IM]} \BibitemShut {NoStop}%
\bibitem [{\citenamefont {{Foreman-Mackey}}\ \emph {et~al.}(2013)\citenamefont {{Foreman-Mackey}}, \citenamefont {{Hogg}}, \citenamefont {{Lang}},\ and\ \citenamefont {{Goodman}}}]{Foreman-Mackey2013}%
  \BibitemOpen
  \bibfield  {author} {\bibinfo {author} {\bibfnamefont {D.}~\bibnamefont {{Foreman-Mackey}}}, \bibinfo {author} {\bibfnamefont {D.~W.}\ \bibnamefont {{Hogg}}}, \bibinfo {author} {\bibfnamefont {D.}~\bibnamefont {{Lang}}},\ and\ \bibinfo {author} {\bibfnamefont {J.}~\bibnamefont {{Goodman}}},\ }\bibfield  {title} {\bibinfo {title} {{emcee: The MCMC Hammer}},\ }\href {https://doi.org/10.1086/670067} {\bibfield  {journal} {\bibinfo  {journal} {Publications of the Astronomical Society of the Pacific}\ }\textbf {\bibinfo {volume} {125}},\ \bibinfo {pages} {306} (\bibinfo {year} {2013})}\BibitemShut {NoStop}%
\end{thebibliography}
\end{document}